\documentclass{article}
\usepackage[utf8]{inputenc}
\usepackage{verbatim}
\usepackage[dvipsnames]{xcolor}
\usepackage[export]{adjustbox}
\usepackage{subcaption}
\usepackage{dirtytalk}
\usepackage{float}
\usepackage{setspace}
\usepackage{url}
\usepackage{multirow}
\usepackage{geometry}
\usepackage{lscape}
\usepackage{amsmath}
\usepackage{amsfonts}
\usepackage{booktabs}
\usepackage{titling}
\usepackage{textcomp}
\usepackage{lineno}
\graphicspath{ {./images/} }

\usepackage{amssymb}
\newcommand{\R}{\mathbb{R}}

\DeclareFontFamily{U}{mathx}{\hyphenchar\font45}
\DeclareFontShape{U}{mathx}{m}{n}{<-> mathx10}{}
\DeclareSymbolFont{mathx}{U}{mathx}{m}{n}
\DeclareMathAccent{\widebar}{0}{mathx}{"73}

\title{Multivariable Fractional Polynomials for lithium-ion batteries degradation models under dynamic conditions}
\author{Clara Bertinelli Salucci\thanks{Corresponding author. \newline  \hspace*{1.9em}  Department of Mathematics, University of Oslo, 0851 Oslo, Norway, clarabe$@$math.uio.no}  $\,$, Azzeddine Bakdi\thanks{
Department of Mathematics, University of Oslo, 0851 Oslo, Norway, azzeddib$@$math.uio.no} $\,$, \\ Ingrid K. Glad\thanks{
Department of Mathematics, University of Oslo, 0851 Oslo, Norway, glad$@$math.uio.no} $\,$, Erik Vanem\thanks{Department of Mathematics, University of Oslo, 0851 Oslo, Norway; \newline  \hspace*{1.9em} DNV GL Group Technology and Research, 1322 Høvik, Norway, erik.vanem$@$dnvgl.com} $\,$, Riccardo De Bin\thanks{Department of Mathematics, University of Oslo, 0851 Oslo, Norway, debin$@$math.uio.no}}
\date{November 11, 2021}

\thanksmarkseries{arabic}

\DeclareMathOperator*{\argmin}{argmin}
\DeclareMathOperator*{\E}{\mathbb{E}}

\newcommand\m[1]{%
\renewcommand\arraystretch{1}%
\begin{tabular}{@{}l@{}}#1\end{tabular}}

\newcommand\mm[1]{%
\renewcommand\arraystretch{1}%
\begin{tabular}{@{}l@{}l@{}}#1\end{tabular}}

\newcommand\mmm[1]{%
\renewcommand\arraystretch{1}%
\begin{tabular}{@{}l@{}l@{}l@{}}#1\end{tabular}}

\newcommand\s[1]{%
\renewcommand\arraystretch{1}%
\begin{tabular}{@{}l@{}l@{}l@{}l@{}l@{}}#1\end{tabular}}

\newcommand{\beginsupplement}{%
        \setcounter{table}{0}
        \renewcommand{\thetable}{S\arabic{table}}%
        \setcounter{figure}{0}
        \renewcommand{\thefigure}{S\arabic{figure}}%
     }

\begin{document}

\maketitle

\section*{Abstract}

Longevity and safety of lithium-ion batteries are facilitated by efficient monitoring and adjustment of the battery operating conditions. Hence, it is crucial to implement fast and accurate algorithms for State of Health (SoH) monitoring on the Battery Management System. The task is challenging due to the complexity and multitude of the factors contributing to the battery capacity degradation, especially because the different degradation processes occur at various timescales and their interactions play an important role. Data-driven methods bypass this issue by approximating the complex processes with statistical or machine learning models: they rely solely on the available cycling data, while remaining agnostic to the underlying real physical processes. This paper proposes a data-driven approach which is understudied in the context of battery degradation, despite being characterised by simplicity and ease of computation: the Multivariable Fractional Polynomial (MFP) regression. Models are trained from historical data of one exhausted cell and used to predict the SoH of other cells. The data are provided by the NASA Ames Prognostics Center of Excellence, and are characterised by varying loads which simulate dynamic operating conditions. Two hypothetical scenarios are considered: one assumes that a recent observed capacity measurement is known, the other is based only on the nominal capacity of the cell. It was shown that the degradation behaviour of the batteries under examination is influenced by their historical data, as supported by the low prediction errors achieved (root mean squared errors ranging from 1.2\% to 7.22\% when considering data up to the battery End of Life). Moreover, we offer a multi-factor perspective where the degree of impact of each different factor to ageing acceleration is analysed. Finally, we compare with a Long Short-Term Memory Neural Network and other works from the literature on the same dataset. We conclude that the MFP regression is effective and competitive with contemporary works, and provides several additional advantages e.g. in terms of interpretability, generalisability, and implementability.

\section*{Keywords}

Lithium-ion batteries;  State of Health prediction; Data-driven approach; Multivariable Fractional Polynomial regression; NASA Randomized Battery Usage Data Set; History-based prediction.

\section{Introduction}

Transports highly contribute to greenhouse gas emissions \cite{ghg}, hence it is fundamental to optimise sustainable and low-emission solutions such as electric batteries. Lithium-ion batteries (LIBs) are the most popular battery technology, as they offer important advantages compared to other battery types such as lead-acid, nickel-cadmium or nickel-metal-hydride \cite{li_ion_advantages, sbarufatti}. The performance of LIBs is nevertheless destined to deteriorate over time (calendar ageing) and usage (cycle ageing): in fact, they are complex electrochemical systems sensitive to operating conditions, and their nonlinear characteristics are time-varying due to ageing. To cope with the increasing demand for high-performance and durability of rechargeable batteries, Prognostics and Health Management (PHM) received tremendous attention over the recent years. PHM plays a crucial role: it enables the operators to monitor the State of Health (SoH) of the battery and take actions to maintain availability and reliability. The LIB prognostics phase includes five steps \cite{berecibar}: measurement, feature extraction, SoH estimation, SoH prediction, and Remaining Useful Life (RUL) estimation. 

We define the battery SoH in Section \ref{Prob_formulation_dataset}, and throughout the article we distinguish between estimation, prediction and forecast according to the following: \emph{estimation} involves evaluating SoH at a given cycle $k$ using features pertaining only to the $k^{\mbox{\small th}}$ cycle; \emph{prediction} evaluates SoH at cycle $k$ using the whole battery historical sequence, i.e. features from cycles $1, ..., k$; \emph{forecast} involves evaluation of SoH in the future ageing pattern of the battery.

Various approaches are proposed in the literature for SoH estimation and prediction of LIBs, which may be categorised as: (i) experimental approaches, including direct SoH testing and experiment-based static models; and (ii) adaptive data-based methods, including filters, observers, expert systems, statistical and machine learning (ML) methods. This article focuses on the latter, particularly on statistical methods as compared to ML methods.

Experimental direct testing methods \cite{D} include internal resistance and impedance measurement, battery energy level, and incremental capacity analysis methods \cite{E}. Common measurement-based models for SoH estimation include Coulomb counting, destructive tests \cite{berecibar}, and other data-fitting models obtained only from test measurements. The experimental methods can only be conducted offline and are highly time-consuming: hence they are inappropriate for the Battery Management System (BMS). Adaptive models are based on online parameter estimation and include: physical models such as equivalent circuits \cite{C}, and electrochemical models \cite{electrochemical_models}; purely data-driven models; domain knowledge; hybrid models. Among the approximate physical models, the widest classes of SoH estimation methods are filters and state observers: especially, Kalman filter and its unscented and extended versions are widely adopted for the estimation of State of Charge (SoC), defined e.g. in \cite{review_ng}, and have their applications successfully extended to SoH estimation \cite{F} and sliding-mode observers \cite{G}. Pure knowledge-based SoH estimation approaches comprise expert systems such as fuzzy logic \cite{H} and Bayesian networks \cite{I} with structures designed by experts: these are however limited, though widely integrated with the other approaches \cite{J}.

Data-driven SoH estimation models are abundant in the literature, due to their favourable capability of modelling the battery degradation while being agnostic to the underlying complex physical and electrochemical phenomena: they are usually based on partial segments of charging/discharging curves, and mainly rely on ML methods. 
For example, \cite{L} developed an energy-based segmentation called \say{energy of equal distance voltage difference} to estimate SoH using deep neural network (NN). A gate recurrent unit-convolutional NN was developed in \cite{M}, and deep convolutional NN (CNN) in \cite{shen2020} to estimate SoH from full trajectories of fixed charge curves; whereas \cite{O} extended the idea to multiple cells SoH estimation from their voltage-current-temperature trajectories over a sliding window using Long Short-Term Memory (LSTM). A more comprehensive overview on data-driven models can be found in \cite{review_ng} and \cite{US}. 

While exhibiting good prediction abilities, ML approaches lack interpretability and require large amounts of data: the complexity of ML models, which are difficult to be handled and checked, risks replacing the complexity of the physical problem that was to be avoided by considering a data-driven approach. For instance, a NN acts as a black box: it provides results based on transformations unavailable to the user. As an example, \cite{venugopal} uses a transformation of the response among the inputs, and yet the prediction error is not 0: this shows that the NN may be forced to find spurious relationships, since the predictions would have been all equal to the true values, had the right relationship been identified. This case also points out that recognizing potential issues can be difficult when dealing with black-box methods: in contrast, a linear regression with the same response transformation among the inputs would have provided a clear indication of the one-to-one relationship between the input and the outcome. 

Among the studies considering regression in connection to LIB capacity estimation or prediction we mention \cite{LR1, LR2, LR3, LR4, LR5}. Different statistical methods have also been proposed. For example, \cite{K} used support vector machines to develop a curve-similarity factor to estimate SoH from charging voltage segments; \cite{richardson2017} estimate SoH using Gaussian process regression based on another form of partial segments designed as $n$ equispaced voltage points. Further methods require the segment to be a full cycle, such as approximate weighted total least squares \cite{Q} to estimate the rated discharge capacity from arbitrary total capacity. Due to their common idea of partial segments, the accuracy of these methods is subject to the availability of long deep monotonic segments, and they cannot be applied to SoH estimation for LIBs under dynamic conditions and calendar ageing. 

Conversely, history-based SoH prediction approaches can explain the degradation of battery health as influenced by the whole LIB history, and they are crucial in PHM for their simultaneous advantages of predicting the future performance and optimising the present operating conditions. Different operating conditions have different effects on the LIB ageing behaviour; a SoH prediction model with capabilities of predictive and prescriptive analytics, such as those considered in this article, enables the BMS to adjust temperature and charge/discharge currents to increase longevity and facilitate safe, high-performance operation. 

The LIB degradation emerges from a complex interplay between many influencing elements, degradation mechanisms (internal side reactions), degradation modes, and observed degradation effects \cite{A}. The influencing elements include: cell and pack design factors; production factors; and application (stress) factors; they influence internal side reactions through complex irreversible physical and chemical processes, which in turn lead to the various degradation modes of lithium depletion, active material loss, electrolyte decomposition, and increase of internal resistance \cite{A}. As a result, the observed LIB degradation effects are capacity fade and power fade. While the design and production factors of influence are fixed and depend on the monitored LIB, the stress factors are dynamic features that may accelerate the LIB degradation behaviour. They can be extracted from measured variables or estimated states and they include: exposure to elevated and low voltages; Depth of Discharge (DoD); cycle bandwidth; cycling frequency; high and low temperatures; high discharge rates (Section \ref{features_section}). 

A critical review \cite{berecibar} emphasises that the degradation effects originate from various processes and their interactions: studying the ageing mechanism is challenging as these processes occur simultaneously, they have different time scales, and it should be avoided to analyse them independently. However, very few multi-factor SoH degradation analyses are reported in the literature: \cite{R} studied the effects of current, cycling limits, and temperature on battery ageing using four dependent models of these factors; \cite{S} conducted a weight analysis to study the influence of voltage, capacity and internal resistance inconsistency on module capacity; \cite{T} conducted orthogonal experiments to study the impact degree of single and multiple stress factors on capacity loss. Unfortunately, these factors are considered independently or in subgroups. Thus, this paper also aims at exploring how various combinations of stress factors affect the LIB degradation, and the degree of impact of different factors to ageing acceleration.

Different reviews, e.g. \cite{berecibar, US}, agree that there is not a perfect method for SoH estimation. As we have briefly outlined, a variety of different approaches are available, all presenting strengths and weaknesses: the best option depends strongly on the available data. In this paper, using data simulating realistic operating conditions, we present a perspective in which simple statistical models are identified via the Multivariable Fractional Polynomial (MFP) approach. We promote the MFP as a valid solution to overcome many of the downsides reported so far:

\begin{enumerate}
    \item The \say{black-box} feature of ML methods, and their common inability of providing uncertainty estimates around the point estimates or predictions; 
    \item The unsuitableness of complex models for the BMS;
    \item The need for researching meaningful features or specific health indicators carrying electrochemical meaning;
    \item The need for long deep monotonic charge/discharge segments which are not likely to be available under realistic operating conditions;
    \item The limitation of considering stress factors for LIBs independently from each other or in subgroups.
\end{enumerate}

\noindent The advantages of MFP have never been exploited, to the knowledge of the authors, for modelling the LIB capacity degradation. We research to which extent MFP models can be used to model the battery capacity reduction under dynamic conditions. Besides, by comparing different MFP models we examine the relationship between the most relevant features and the battery capacity degradation itself. Finally, we compare our method to a ML approach based on a Deep LSTM Regression Network (D–LSTM–RN), and other contemporary works from the literature.

The remainder of this article is organised as follows: Section \ref{Prob_formulation_dataset} introduces the problem of SoH degradation in relation to the battery capacity fade and presents the experimental data used for the analysis; Section \ref{statistics_approach} describes the MFP algorithm in the regression context, overviews the stress factors for LIBs and the related features, presents three MFP models and their results, with a particular focus on the interpretation of the estimated coefficients; Section \ref{discussion} compares the MFP regression with a Long Short-Term Memory Neural Network, as well as with recently published methodologies applied on the same dataset; Section \ref{conclusion} summarises the main results and provides concluding remarks. 

\section{Dataset description and problem formulation} \label{Prob_formulation_dataset}

The dataset we use is part of the \say{Randomized Battery Usage Data Set} by the NASA Ames Prognostics Center of Excellence (PCoE) \cite{dataset_cit}, which comprises ageing data for 18650 LIBs cycled under randomly generated current profiles. The aim of the experiment was to mimic the dynamic operating condition of batteries used in real-life. In fact, despite their important role in the study of LIB deterioration, laboratory data are typically gathered under particular and unrealistic conditions, for example with small temperature variation and constant current. This randomised dataset constitutes an effort towards a better approximation of real-life conditions. It is a well-known benchmark that is widely used for training, testing, and comparing various methods in the literature, hence we adopt it to validate our work. The data considered in this study pertain to 11 out of the 28 battery cells that were made available by the NASA PCoE. The remaining 17 cells were discarded due to unrealistic values in the temperature data (cells 2, 3, 17 and 18) and/or in the cycle duration (negative values in cells 6, 16, 17, and 18), or due to the presence of too few observations of the response variable (cells 13, 14, 15 and 21 to 28). The considered cells belong to four groups that were operated in a controlled environment in different ways, as reported in Table \ref{cell_groups}. All the discharge processes were randomised to recreate real operation, and for all groups the two modalities relevant for this analysis are:

\begin{itemize}
    \item \textbf{Reference discharge:} a controlled full discharge cycle, occurring immediately after a controlled full charge cycle, which allows to compute the cell capacity periodically. During a reference charge cycle, the cell is initially charged at a constant current of 2 A, until the battery reaches the maximum voltage of 4.2 V; then, the voltage is kept constant until the charging current drops to 0.01 A. During the subsequent reference discharge cycle, the cell is discharged at 1 A until the voltage reaches the threshold of 3.2 V.
    
    \item \textbf{Random walk (RW) steps:} a charge and/or discharge process where the current load or the duration is selected at random from a pre-defined set; see Table \ref{cell_groups} for details. 
    \end{itemize}

\noindent Temperature, voltage and current were recorded for each operational mode and are the only data used for this study.

\begin{table}[H] 
\renewcommand{\arraystretch}{2}
\centering
\begin{tabular}{p{0.09\textwidth}p{0.12\textwidth}p{0.34\textwidth}p{0.34\textwidth}}
\toprule
& \hspace{.2cm}Cells & RW procedure & Reference cycles \\[-.1em]
\midrule
& & & \\[\dimexpr-\normalbaselineskip-2em] \\
  Group 1 & 1, 7, 8 &  \m{Charge: random duration \\ (0.5 to 3 h, or until full). \\[.5em] Discharge: random current \\ until 3.2 V. \vspace{.5cm} } & \mm{Every 50 RW steps \\[.5em] Rest after charge and discharge \\ (introduced only halfway). \\ \\ } \\
  \hline
    & & & \\[\dimexpr-\normalbaselineskip-2em] \\
    Group 2 & 4,5 & \m{Charge: not random, i.e. \\ constant current (2 A) \\ to 4.2 V. \\[.5em] Discharge: random current \\ for 300 s or until 4.2 V. \vspace{.5cm}} &  \mm{Every 50 charge-discharge processes \\[.5em] Rest after charge and discharge \\ (introduced only halfway). \\ \\ } \\
  \hline
& & & \\[\dimexpr-\normalbaselineskip-2em] \\
  Group 3 & 9, 10, 11, 12 &  \mmm{Charge: random current \\ for 300 s or until 4.2 V. \\[.5ex] Discharge: random current \\for 300 s or until 3.2 V. \\[.5em]  Whether the RW step is \\ charge or discharge \\ is also random. \vspace{.5cm}} & \mm{ Every 1500 RW steps \\[.5em] Rest after charge and discharge \\ (introduced only halfway). \\ \\ } \\
    \hline
& & & \\[\dimexpr-\normalbaselineskip-2em] \\
  Group 4 & 19, 20 &  \mm{Same as group 2, but each step \\ lasts 60 s and the probability  \\distribution is skewed towards \\ higher currents. \vspace{.5cm}} & \mm{Every 50 charge/discharge \\[.5em] Rest after charge and discharge. \\ \\ } \\
  
\bottomrule
\end{tabular}
\caption{Brief description of the RW procedure for each group of batteries in the NASA Randomized Data Set.}
\label{cell_groups}
\end{table}

The degradation of the battery SoH entails a decrease in the battery capacity and an increase in the battery impedance \cite{berecibar, vetter}: hence, both these phenomena can be alternatively used to define SoH and analyse the LIB health deterioration. In this study, we consider the reduction in the battery capacity. The remaining capacity before and after every RW phase can be computed through integration of the discharging current over the elapsed time during the reference discharge \cite{kirchev:2015},

\begin{equation}
C_d \, = \,  \int_0^{t_{cutoff}} I_d \, dt.
\label{int_capacity}
\end{equation}

\noindent In particular, since the discharging current for the reference cycles throughout the experiment is always $I_d = 1$ A for all the considered cell groups, the magnitude of the discharging capacity corresponds to the discharging time expressed in hours. It should be noted, however, that the batteries are not likely to be at equilibrium at the beginning of each reference cycle, since there were too short resting periods (or no resting periods at all) to allow reaching the steady state. Consequently, the initial voltage of the cycles is uncertain, and generally different from the desired value of 4.2 V. This translates into an uncertainty on the benchmarked capacity \cite{equilibrium}, the determination of which would involve an accurate study of the battery transient dynamics, possibly complemented by gathering of data from cycles interspersed with longer resting periods of the cell. An adjustment for the transient effects is introduced in Section \ref{section_models}.

\begin{figure}
\centering
\renewcommand{\arraystretch}{.9}
    \begin{tabular}{c c}
        Group 1 & Group 2 \\
        \includegraphics[width=.49\textwidth]{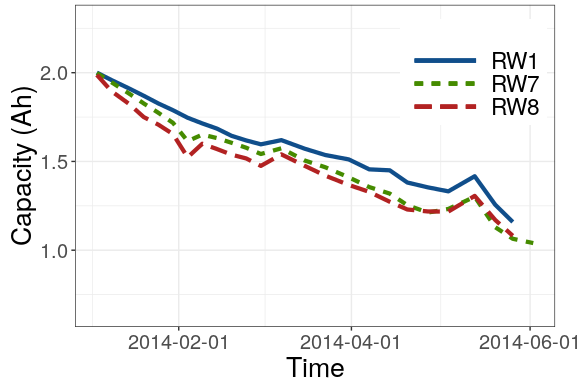} & \includegraphics[width=.49\textwidth]{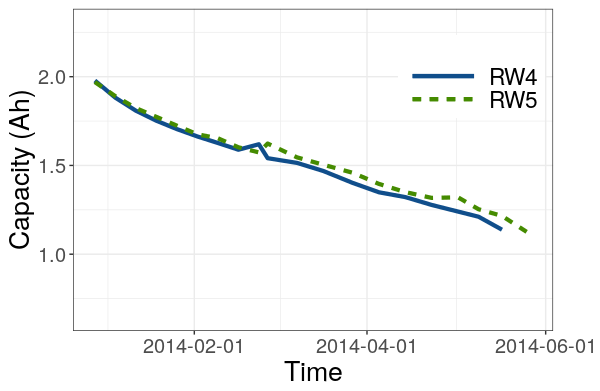} \\[1ex]
        Group 3 & Group 5 \\
        \includegraphics[width=.49\textwidth]{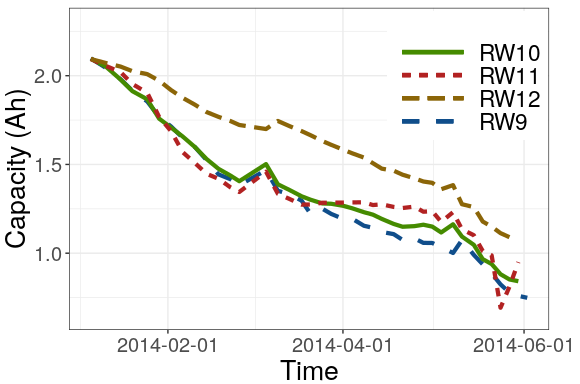} & \includegraphics[width=.49\textwidth]{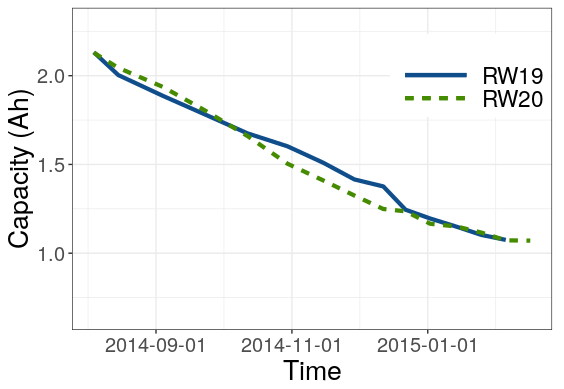} \\
    \end{tabular}
\caption{Observed capacity values for the cells used in this study. The jumps in the capacity curves are likely to be caused by a prolonged rest which is accounted for in our models. Note the dissimilarities in the degradation behaviours of cells pertaining to the same experiment in the groups 3 and 4.}
\label{capacity_all_batteries}
\end{figure}

The observed capacities of all the cells considered in the study, grouped in the four different subsets, are shown in Figure \ref{capacity_all_batteries}. Note that there are sudden increases in the capacity in almost all the degradation curves: they are likely to be due to a prolonged rest taking place before the reference cycle, as it is confirmed by studies such as \cite{rest_time1, rest_time2}. When modelling the capacity degradation, this is taken into account through a feature related to the resting time. 

It should also be noted that in some cases there are dissimilarities in the degradation curves of the cells even when they belong to the same group. This is particularly accentuated in the third group: the cell RW11 follows a similar path to RW9 and RW10 for the first part of the experiment, but develops differently in the second half, while the cell RW12 has a completely different behaviour. These differences are presumably caused by differences both in the production phase and in the operational history of the cells; it is reasonable to expect that such differences will be reflected in less precise predictions.

\begin{figure}
\centering
\renewcommand{\arraystretch}{.9}
    \begin{tabular}{c}
        \includegraphics[width=.9\textwidth]{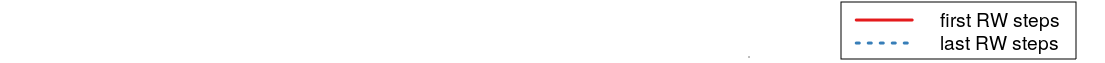} \\[-.4em] 
        \includegraphics[width=.9\textwidth]{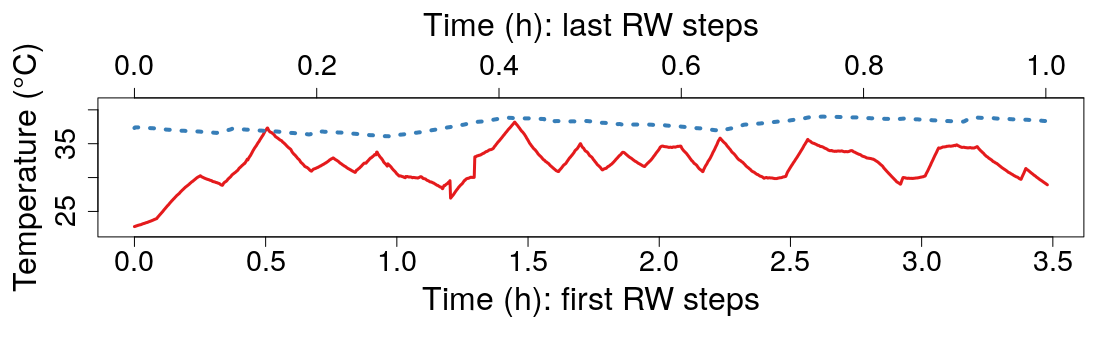} \\[-1em] \includegraphics[width=.9\textwidth]{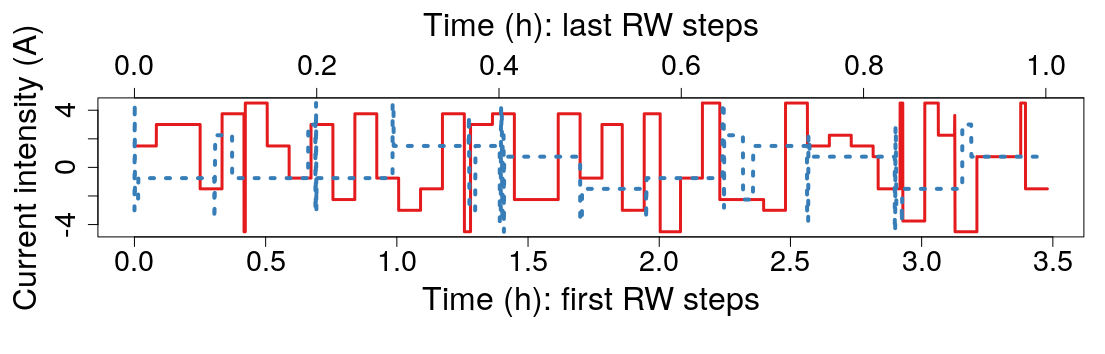} \\[-1em]
        \includegraphics[width=.9\textwidth]{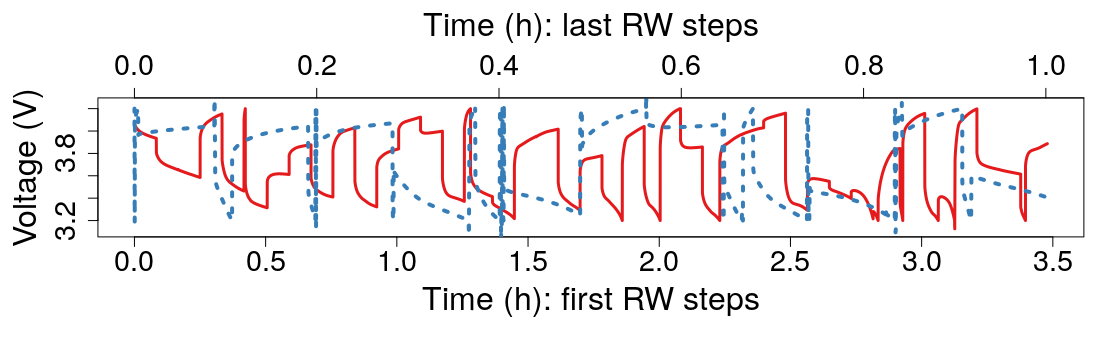}
    \end{tabular}
\caption{Comparison between the measurements of temperature (upper plot), current intensity (middle plot) and voltage (lower plot) during the first 50 RW steps (red lines) and the last 50 RW steps (blue lines) of cell RW9. Note the differences between the initial and final steps: the temperature is consistently higher in the final phase, which takes a significantly shorter time (upper x-axis) and hits the voltage extremes (3.2 V and 4.2 V) more frequently than the initial phase.}
    \label{rw_steps}
\end{figure}

As illustrative of the RW cycling phase, Figure \ref{rw_steps} shows the temperature, current and voltage sensed data for the cell RW9 during the first and the last 50 RW steps (solid red lines and dashed blue lines, respectively). Note that there is a significant difference between the temperatures, which are consistently higher in the last RW phase, as well as in the voltage curves: in the first steps, the voltage rarely hits the range boundaries of 3.2 and 4.2 V, while it happens more frequently in the last steps. As a result, the last RW phase is also much shorter, since many of the steps last for less than the default duration of 5 minutes. All these effects are a clear indication of the cell health degradation, and are to be considered when modelling the capacity reduction.

\newpage

\section{Multivariable Fractional Polynomials} \label{statistics_approach}

\subsection{Algorithm Description} \label{mfp}

The Multivariable Fractional Polynomial (MFP) approach of Sauerbrei and Royston \cite{sauerbrei_mfp} is based on an algorithm to find the best input transformations in a multivariable linear regression setting. A multivariable linear regression assumes a linear relationship between the response (or target variable or outcome) $y$ and a set of inputs (also called features or covariates) $x_1, ..., x_p$,

\begin{equation}
y \, = \,  \E [\, y \, \vert \, x_1, ..., x_p \,] + \varepsilon  \, = \, \beta_0 + \beta_1 \, x_1 + ... + \beta_p \, x_p \, + \, \varepsilon 
\end{equation}

\noindent where $\beta_0, ..., \beta_p$ are the regression coefficients, $\varepsilon$ is an error term, and $\E [\, y \, \vert \, x_1, ... x_p \,] = \beta_0 + \beta_1 \, x_1 + ... + \beta_p \, x_p$ is the expected value of $y$ conditioned on $x_1, ..., x_p$. Note that linearity is assumed with respect to the regression coefficients $\beta_0, ..., \beta_p$, not necessarily to $x_1, ..., x_p$: on the contrary, it is important to consider possible nonlinear contributions from the inputs, which, if not accounted for, may lead to misspecified final models \cite{sauerbrei_mfp}.  The MFP method, implemented in R with the \texttt{mfp} package \cite{mfp_package}, handles non-linearities by transforming the inputs through the most suitable FP functions. 

Given a unidimensional input $x$, a FP function of first degree is defined as $x^l$, where the power $l$ can be either integer or fraction, positive or negative, from the predefined set 

    \vspace{0.1cm}
  \begin{center} $\mathcal{A}$ = {
    \{-2,-1, -0.5, 0, 0.5, 1, 2, 3\}}, 
    \end{center}
    \vspace{0.1cm}

\noindent where $x^0 \equiv \log x$. The best $l$ in the set $\mathcal{A}$ is considered to be that yielding the lowest deviance $d = -2 \ell (\hat{\beta}_{ML})$, where $\ell(\hat{\beta}_{ML})$ is the maximised log-likelihood function. In a multivariate setting with $p > 1$ covariates, a backfitting-like algorithm is used to fit a MFP model: a linear term for each covariate is initially included; then, the best transformation within the considered FPs is selected with the same procedure as for the univariate case, and the model is updated consequently. Iterations are repeated until all FPs remain constant. The \texttt{mfp} package allows for FPs of degree $m>1$: the second-degree polynomial is preferred over the first-degree on the basis of a $\chi^2$ test, and so on. Further details about the MFP method with FPs of a generic degree $m$ can be found in \cite{sauerbrei_mfp, mfp_paper, mfp_paper2}. For this analysis, we considered FPs with maximum permitted degree $m = 2$, but only first-degree models were eventually selected by the algorithm.

Considering a training set of $n$ labelled pairs ($y$, $\boldsymbol{x}^{\boldsymbol{l}}$), where $\boldsymbol{x}^{\boldsymbol{l}}$ is the $q$-dimensional vector of transformed inputs according to the MFP algorithm and ${\bf l} = (l_1, \dots, l_q)$ the set of selected powers, the linear model can be written as 

\begin{equation}
    \mbox{\textbf{\emph{y}}} \, = \,  \E [\, \mbox{\textbf{\emph{y}}} \, \vert \, \tilde{X} \,] +  \boldsymbol{\varepsilon}  \, = \, \tilde{X} \boldsymbol{\beta} +  \boldsymbol{\varepsilon},
\end{equation}

\noindent where: \textbf{\emph{y}} is a vector of $n$ observations of the target variable; $\tilde{X}$ is a $n \times (q+1)$ matrix assumed to have full rank $q+1$, containing a column of 1's in the first position to account for the intercept term $\beta_0$, and the transformed inputs in the remaining columns; $\boldsymbol{\varepsilon}$ is a vector of $n$ independent error terms that are here assumed to be $N(0, \sigma^2)$-distributed. In general, $q \neq p$ due to the possibility of introducing an $m$-degree polynomial for each input $x_1, ..., x_p$; however, $q=p$ in our case where only first-degree polynomials were selected. Given the training set, one can learn the linear relationship through the least squares criterion, i.e. by solving

\begin{equation}
\hat{\beta} \, = \, \argmin_{\beta \in \mathbb{R}^q} \,  \Vert \mbox{\textbf{\emph{y}}} - \tilde{X} \boldsymbol{\beta} \Vert^2.
\end{equation}

Despite constituting such a simple setup, linear regression proved to work extremely well in many non-trivial contexts and different applications, see e.g. \cite{linear_regression_ex1, linear_regression_ex2, linear_regression_ex3}. An attractive characteristic of multiple linear regression is the interpretation of the regression coefficients $\beta_j$ for $j=1,...,q$, which may be read as the change in $\E [\, y \, \vert \, x_1^{l_1}, ... x_q^{l_q} \,]$ for an increase of one unit in $x_j^{l_j}$, holding all other features constant. Importantly, this enables immediate identification of the most relevant features, and allows studying the effect of one single feature while adjusting for the effects of the others. In addition, the \emph{coefficient of determination} $R^2$ can be interpreted, in the context of linear regression, as the proportion of total variability in the outcomes that is explained by the model, with $0 \leq R^2 \leq 1$.  For this reason, $R^2$ provides an interesting diagnostic measure. However, since $R^2$ increases every time new features are added to the regression equation, one should preferably consider its penalised version, the \emph{adjusted coefficient of determination} $R^2_{adj}$, that only increases when newly added features increase the variability explained by the model. 

Finally, to avoid including features that do not add valuable information, it is important to introduce a variable selection mechanism which enables to keep only the subset of the most significant variables to predict the outcome. In this case, we used a stepback procedure based on the AIC criterion. Variable selection is important
for both theoretical and practical reasons, such as: achieving a reduction of the model variance, thus improving the prediction accuracy; facilitating the model interpretation and providing a cleaner view of the data-generating process; and reducing the computational and usage time of the model, which makes it more portable. Performing variable selection is an important and delicate point in particular for explanatory models such as linear regression, since including or excluding highly correlated features may lead to significantly different interpretations of their effects \cite{var_selection}.

\subsection{Features extraction} \label{features_section}

LIB capacity decrease is attributed to multiple factors and processes and their interactions over various timescales, including: exposure to elevated voltages; Depth of Discharge (DoD); cycle bandwidth; cycling frequency; elevated temperatures; and high discharge rates \cite{BDA_factors_1, BDA_factors_2}. A multitude of stress factors is accounted for in our models; however, since it is a data-driven method that relies only on temperature-current-voltage data gathered in the RW discharging processes, some of the relevant factors need to be considered indirectly. 

\medskip

Temperature has a significant impact on performance, safety, and ageing of LIBs. 
The temperature range $(-20 ^\circ \mbox{C}, +60 ^\circ \mbox{C})$ is considered acceptable \cite{Temp_acceptable}, whereas the desired temperature range is $(+15 ^\circ \mbox{C}, +35 ^\circ \mbox{C})$ \cite{Temp_Desired}.
The effects of temperature on LIBs can be classified into low-temperature and high-temperature effects: in both cases, extreme temperatures affect both calendar and cycle ageing. Low temperatures may cause slow chemical reactions and charge transfer, decreased ionic conductivity, and lithium plating \cite{Temp_low}. Discharging at low temperatures results in power limitation, while low-temperature charge produces a reduced power capability and cold cranking \cite{Temp_low_2}. High temperatures may cause loss of lithium and increase of internal resistance, which in turn produce loss of capacity and of power, respectively. Furthermore, the operating temperature affects the State of Charge (SoC) of the battery, and extreme temperatures may accelerate lithium plating on the anode. Temperature-based features are designed to account for the influence of temperature on the battery ageing behaviour. In particular, for each reference cycle $i$ occurring after a RW phase having $m$ RW discharge steps, we include the average of the minimum and maximum temperatures at each RW discharge step:

\begin{equation}
\begin{split}
\widebar{T}_{\mbox{\footnotesize min}} = \frac{1}{m} \sum_{k=1}^m (T^{\mbox{ \footnotesize min}}_{ik}); \\
\widebar{T}_{\mbox{\footnotesize max}} = \frac{1}{m} \sum_{k=1}^m (T^{ \mbox{\footnotesize max}}_{ik}).
 \end{split}
\end{equation}

\noindent Note that, by construction of the linear regression methods, we cannot input the entire multivariate sequence of data corresponding to each RW phase, but we rather have to summarise the sequence to a single scalar for each input.

\medskip
  
Fast charging is a desired aspect in batteries; however, high charge-discharge rates may cause mechanical-induced damage of active particles in LIBs and accelerate the capacity fade \cite{C_rate}. Lithium plating is also associated with fast charging, particularly in combination with low temperatures. The effects of different current rates on coulombic efficiency and capacity loss are studied in \cite{C_rate_2}, and \cite{c_rate_3} explores the impact on capacity estimation, confirming that the C-rate (or, equivalently, the current intensity) is an important feature for capacity fade prediction. In this work, we consider the average of the current intensities of the RW discharge steps,

\begin{equation}
 \bar{I} = \frac{1}{m} \sum_{k=1}^m (I_{ik}).   
\end{equation}
 
\medskip

The charge–discharge cycling frequency is related to mechanical stress on LIBs: it affects the degradation behaviour, especially when it is extreme \cite{Frequency}, but high frequency may also come as a consequence of the battery ageing. The data at hand are characterised by a varying cycling frequency, which we consider in the analysis through: the cycle duration,

\begin{equation}
\Delta t = \sum_{k=1}^{m+l} (t^{\mbox{ \footnotesize end}}_{ik} - t^{\mbox{ \footnotesize start}}_{ik}),  \\[1ex]
\end{equation}

\noindent where $m$ and $l$ are the discharge and charge RW steps occurring before reference cycle $i$ at which we are doing the prediction, respectively; the proportion of cycles lasting less than the default duration,

\begin{equation}
\lambda = \frac{\# \, \mbox{\small short steps}}{m}; 
\end{equation}

\noindent and the rest time, which is input as a nonlinear smooth saturation function of the time elapsed between the last RW step of the cycle and the beginning of the reference cycle, 

$$ \Delta t_{\mbox{\footnotesize rest}} = \frac{1}{1 + \exp \biggr[-\frac{1}{4}\bigl(\frac{\tilde{t}}{\sigma^2_{ \tilde{t}<20}}\bigr)\biggr]} \quad \mbox{with}  \quad \tilde{t} = t^{\mbox{ \footnotesize start, ref}}_{i} - t^{{\mbox{ \footnotesize end}}, (m+l)}_{i},$$

\noindent where $\sigma^2_{ \tilde{t}<20}$ is the variance of the observations such that $\tilde{t} < 20$ h. Note that the rest time affects the recovery effects and the charge balancing of the battery, and influences its lifetime \cite{Rest_time}. The estimation of SoC is also affected by the rest time, besides temperature and cycle current; however, estimating the SoC is a complex process which often leads to uncertain results, and this work advantageously avoids relying on SoC estimates.

High voltages and overcharge contribute to lithium plating and electrolyte decomposition, which accelerate the battery ageing \cite{OverCharge}. The Depth of Discharge DoD $= \mbox{SoC}_1 - \mbox{SoC}_2$ is also usually considered as a stress factor; however, \cite{DoD} showed that the adopted SoC range (SoC$_1$, SoC$_2$) plays a bigger role than DoD itself: in fact, though batteries cycled with range (100\%, 25\%) degraded faster than those with (100\%, 40\%), it was also observed that (100\%, 40\%) degraded much faster than (85\%, 25\%), despite the same DoD. Furthermore, (100\%, 50\%) showed a faster degradation than (85\%, 25\%), despite the lower DoD. In our work, high voltages, DoD and SoC range are accounted for through the average initial voltage and voltage difference of the RW discharge steps,

\begin{equation}
    \begin{split}
        \widebar{V}_{\mbox{\footnotesize in}}  &= \frac{1}{m} \sum_{k=1}^m v^{\mbox{ \footnotesize in}}_k; \\
        \widebar{\Delta V}  &= \frac{1}{m} \sum_{k=1}^m (v^{\mbox{ \footnotesize end}}_{ik} - v^{\mbox{ \footnotesize start}}_{ik}).
    \end{split}
\end{equation}

Note that there are redundancies in the input feature space, to account for the presence of nonlinear relationships between the features. For example, the cycles duration, voltage difference and cycle current are not independent from each other. Previously these factors were analysed separately in the literature; however, they should not be analysed independently as their interactions play an important role in LIBs ageing. This analysis models the battery capacity loss as a process of all factors simultaneously, and investigates which combinations of features contribute mostly in explaining the capacity fade, measuring their degree of importance. 

In addition to the aforementioned stress factors, two additional features are considered: the observed capacity at the last reference discharge cycle,

$$ C_{\mbox{\footnotesize prev}} = C_{i-1},$$

\noindent and an approximation,

$$C_{\mbox{\footnotesize approx}} = \frac{1}{m} \sum_{k=1}^m (\hat{C}_{ik}),$$

\noindent obtained as an average of rough capacity predictions at each RW discharge step, coming from a preliminary linear regression model which is described in section S.2 in the Supplementary Material. The inclusion or exclusion of these two additional features from the pool of possible inputs will enable the comparison of models representing different scenarios.

\subsection{Structure} \label{section_models}

We considered three regression models:

\begin{enumerate}
\renewcommand{\labelenumi}{(\alph{enumi})}
\item a model including all features presented in Section \ref{features_section}, but $C_{\mbox{\footnotesize prev}}$ and $C_{\mbox{\footnotesize approx}}$;

\item a model including all features presented in Section \ref{features_section}, but $C_{\mbox{\footnotesize prev}}$;

\item a model including all features presented in Section \ref{features_section}, with no exceptions.

\end{enumerate}

\noindent Model ($c$) assumes that $C_{\mbox{\footnotesize prev}}$, the observed capacity measured at the most recent reference cycle, is known and can be used to add important information to the regression model. Model ($b$) reflects the more realistic scenario in which $C_{\mbox{\footnotesize prev}}$ is undetermined, but it includes $C_{\mbox{\footnotesize approx}}$ as an approximate surrogate with the purpose of maximising the information extracted from the data sensed during the RW discharges; model ($a$) relies uniquely on the data directly gathered from the sensors, without additional true or predicted capacity values. \noindent In each case, the target variable for the analysis is the change in the battery capacity at time $t$ compared to its nominal capacity,

\begin{equation}
y \, = \, \Delta C(t) \, = \, C(t_0) - C(t). \label{DeltaC}
\end{equation}

\noindent The observed capacity values in Equation \eqref{DeltaC} are not directly computed from Equation \eqref{int_capacity} in Section \ref{Prob_formulation_dataset}: to account for the batteries not having reached equilibrium, the voltage curves of the reference discharges have been interpolated with a monotone Hermite spline, which allowed to extrapolate how much longer it would have taken if the cycles had started from the threshold voltage value of 4.2 V, instead of the observed ones. This is but a first step towards a more accurate capacity prediction, but it allows to introduce a small correction (Section S.1 in the Supplementary Material).

After having undergone the MFP and variable selection procedures, the three final models trained on the first battery of group 3, cell RW9, are:
 
\bigskip
 
\begin{itemize}
 \setlength\itemsep{2em}
 \setlength{\itemindent}{3em}
 
\item[\textbf{MFPa}] \[\Delta C_{a,i} \, \, = \, \, \alpha_0 \, + \, \alpha_1 \cdot \widebar{T}_{\mbox{\footnotesize min},i} \, + \, \alpha_2 \cdot \widebar{T}_{\mbox{\footnotesize max},i} \, + \, \alpha_3 \cdot \lambda^3_i \, + \, \varepsilon_i \]

\item[\textbf{MFPb}]  \[ \Delta C_{b,i} \, \, = \, \, \beta_0 \, + \,
    \beta_1 \cdot C_{\mbox{\footnotesize approx},i}^\frac{1}{2} \, + \, 
    \beta_2 \cdot \Delta t_{\mbox{\footnotesize rest},i} \, + \,
    \beta_3 \cdot \widebar{T}_{\mbox{\footnotesize min},i} \, + \, \] 
\[     \quad \quad  \beta_4 \cdot \lambda_i \, + \,
    \beta_5 \cdot \widebar{V}_{\mbox{\footnotesize in}, i}  \, + \, \varepsilon_i \]

\item[\textbf{MFPc}]  \[ \Delta C_{c,i} \, \, = \, \, \gamma_0 \, + \, 
    \gamma_1 \cdot C_{\mbox{\footnotesize prev}, i}^{\frac{1}{2}}  \, + \,
    \gamma_2 \cdot \Delta t_{\mbox{\footnotesize rest},i}  \, + \,
    \gamma_3 \cdot C_{\mbox{\footnotesize approx},i}  \, + \, \]
\[    \quad \quad \gamma_4 \cdot \widebar{T}_{\mbox{\footnotesize min},i}  \, + \,  \varepsilon_i  \]

\end{itemize}

\noindent Similarly, for all the other groups of cells considered in this study the whole history of the first cell is used as training data, whereas the remaining cells are used for testing.

\subsection{Results} \label{results_MFP_section}

Results for all models are shown in this section. In particular, subsection \ref{group3} offers a detailed comparison of the models obtained for the third group (the largest), with interpretation and discussion of the selected features; results pertaining all other groups are shown in \ref{other_groups}.

To evaluate accuracy in prediction, we mainly consider the Root Mean Squared Error, 

    \begin{equation}
     \mbox{RMSE}(\hat{C}, C) \, = \, \sqrt{\frac{1}{n} \sum_{i = 1}^n \bigl( C_i - \hat{C}_i \bigr)^2}.   
    \end{equation}

\noindent Note that we compare RMSEs on the capacity scale, despite the target variable being the capacity drop between two consecutive reference cycles. This is done to ease the comparison with other works in the literature using the same data. A normalised version where the deviation is divided by the observed capacity is also provided:
    
    \begin{equation}
     \mbox{RMSE}_{\mbox{\tiny norm}}(\hat{C}, C) \, = \, \sqrt{\frac{1}{n} \sum_{i = 1}^n \biggl( \frac{ C_i - \hat{C}_i}{C_i}\biggr)^2}. 
    \end{equation}
    
\noindent  Moreover, since the data-gathering for the cells in this experiment extended well beyond their End of Life (EoL) (70\% or 80\% of the cell nominal capacity \cite{eol_def, eol_def2}), we present also the counterparts computed on data up to the EoL, RMSE$^{\tiny \mbox{EoL}}$ and RMSE$^{\tiny \mbox{EoL}}_{\tiny \mbox{norm}}$. This metric is meaningful since SoH prediction after EoL is practically less important, and generally ignored in many works in the literature. The EoL is here defined to occur when the capacity is 80\% of the nominal capacity. Other error measures can be found in section S.5 the Supplementary Material.

\subsubsection{Group 3: comparison between models and their interpretation} \label{group3}
The trained models for group 3 are reported in Table \ref{summary_mfp}: the estimated regression coefficients are shown together with their standard errors, the corresponding p-values, and the $R^2$ and $R^2_{\mbox{\footnotesize adj}}$ coefficients. 

The table shows that MFPa is including only $\widebar{T}_{\mbox{\footnotesize min}}$ and $\widebar{T}_{\mbox{\footnotesize max}}$, together with the proportion $\lambda$ of steps interrupted due to the voltage reaching the boundaries before the default duration. The temperatures are the most significant variables in the model. Increasing $\widebar{T}_{\mbox{\footnotesize max}}$ by one degree while holding all other features constant implies an increase in the capacity drop of 0.54; this effect is counterbalanced by $\widebar{T}_{\mbox{\footnotesize min}}$ having an opposite coefficient. When they have close values, the effect of the two variables together is close to 0, meaning that a narrow temperature range, hence a controlled temperature variation, does not affect the battery health seriously. Concerning $\lambda$, it also appears strongly significant and it is included as a cubic effect with a positive regression coefficient.

With MFPb we included the average of the approximate capacity for each RW step, $C_{\mbox{\footnotesize approx}}$. This led to increased values of both $R^2$ and $R^2_{\mbox{\footnotesize adj}}$: the square root of $C_{\mbox{\footnotesize approx}}$ is in fact the most significant feature of the model, correctly associated with a negative coefficient: a higher value of the square root of $C_{\mbox{\footnotesize approx}}$ involves a smaller capacity gap. A mild beneficial effect is attributed also to $\widebar{V}_{\mbox{\footnotesize in}}$ and $\widebar{T}_{\mbox{\footnotesize min}}$: the initial voltage is not extremely significant; while the minimum temperature continues to have a very low p-value, which makes it the second most important feature in the model. $\widebar{T}_{\mbox{\footnotesize min}}$ is the only temperature variable selected for MFPb, and its effect is smaller than and opposite to that of MFPa: without $\widebar{T}_{\mbox{\footnotesize max}}$ in the model, $\widebar{T}_{\mbox{\footnotesize min}}$ is left alone to account for the effect of extreme temperatures. $\Delta t_{\mbox{\footnotesize rest}}$ is the third covariate for importance in the model, and it produces a reduction in the capacity variation: in fact, as discussed in Section \ref{Prob_formulation_dataset}, an apparent increase in the capacity of LIBs can be achieved by allowing the cell to rest for some time. The resting time was not selected in MFPa: this might be ascribed to the correlation that exists between $\Delta t_{\mbox{\footnotesize rest}}$ and  $\lambda$, as we also see that $\lambda$ has a reduced effect in MFPb compared to MFPa: its p-value changed from 0.0007 to 0.0146 and the feature is now included as a linear effect with a smaller estimated coefficient. Concurrently, since $\lambda$ is both a cause and a consequence of capacity fade in this experimental setting, it is likely that the reduction in its significance compared to MFPa is strongly connected to the presence of $C_{\mbox{\footnotesize approx}}$.

MFPc is the result of variable selection starting from the full set of inputs described in Section \ref{features_section}, including the most recent observed capacity value, $C_{\mbox{\footnotesize prev}}$. The inclusion of $C_{\mbox{\footnotesize prev}}$ adds a great deal of exact information to the model, which unsurprisingly results in a further increase of both $R^2$ and $R^2_{\mbox{\footnotesize adj}}$, almost reaching their maximum value of 1.  The results of MFPc seem consistent with those of MFPb: the most important feature is now the square root of $C_{\mbox{\footnotesize prev}}$, which also has the larger (in magnitude) estimated coefficient.

\begin{table}[H] 
\renewcommand{\arraystretch}{2}
\centering
\begin{tabular}{p{0.11\textwidth}p{0.11\textwidth}p{0.11\textwidth}p{0.11\textwidth}p{0.11\textwidth}p{0.11\textwidth}p{0.11\textwidth}}
\toprule
& Covariate & Est & Std err & p-value & $R^2$ & $R^2_{\mbox{\footnotesize adj}}$ \\[-.1em]
\midrule
  \multirow{4}{5em}{$\mbox{MFPa}$} & intercept &  0.43 & 0.17 & 0.0142 & \multirow{4}{5em}{\hfil 0.985 } & \multirow{4}{5em}{\hfil 0.984} \\
& $\widebar{T}_{\mbox{\footnotesize min}}$ &  0.54  & 0.08 & 5.35e-08 & & \\ 
& $\widebar{T}_{\mbox{\footnotesize max}}$  & -0.53  & 0.08 & 9.13e-08 & & \\
& $\lambda^3$  & 0.79  & 0.21 & 0.0007 & &  \\
\hline
 \multirow{6}{5em}{$\mbox{MFPb}$} & intercept  & 7.40 & 0.90 & 1.72e-09 & \multirow{6}{5em}{\hfil 0.992} & \multirow{6}{5em}{\hfil 0.991} \\
& $C_{\mbox{\footnotesize approx}}^\frac{1}{2}$  & -3.99 & 0.40 & 1.83e-11 & & \\ 
& $\Delta t_{\mbox{\footnotesize rest}}$  & -0.26  & 0.07 & 0.60e-03 & & \\ 
& $\widebar{T}_{\mbox{\footnotesize min}}$  & -0.06  & 0.01 & 8.39e-09 & & \\
& $\lambda$  & 0.62  & 0.24 & 0.0146 & &  \\
& $\widebar{V}_{\mbox{\footnotesize in}}$   & -0.10  & 0.04 & 0.0230 & & \\
\hline
\multirow{5}{5em}{$\mbox{MFPc}$} & intercept  & 4.63  & 0.21  & $<$2e-16 & \multirow{5}{5em}{\hfil 0.997} & \multirow{5}{5em}{\hfil 0.997} \\
& $C_{\mbox{\footnotesize prev}}^{\frac{1}{2}}$ &  -1.36  &  0.13 & 4.87e-12  & &  \\
& $\Delta t_{\mbox{\footnotesize rest}}$  & -0.34 & 0.04 & 5.23e-10  & &  \\
& $C_{\mbox{\footnotesize approx}}$  & -0.87 & 0.13 & 7.83e-08 & &  \\
& $\widebar{T}_{\mbox{\footnotesize min}}$  & -0.03 & 0.005 & 1.05e-05 & & \\
\bottomrule
\end{tabular}
\caption{Results for the three models MFPa, MFPb and MFPc for group 3: estimated regression coefficients (Est), standard errors (Std err) and corresponding $p$-values, $R^2$ and adjusted $R^2$ coefficients.}
\label{summary_mfp}
\end{table}

\noindent Interestingly, the second most significant covariate is now $\Delta t_{\mbox{\footnotesize rest}}$, with a stronger effect also in its coefficient: this could again be explained with its relation to $\lambda$, which is not present in this model. However, once again, the absence of $\lambda$ should be also related to its connection with $C_{\mbox{\footnotesize prev}}$ and $C_{\mbox{\footnotesize approx}}$: in fact, $\lambda$ was strongly significant in MFPa, where neither $C_{\mbox{\footnotesize prev}}$ nor $C_{\mbox{\footnotesize approx}}$ were considered; less important in MFPb where $C_{\mbox{\footnotesize approx}}$ was included; not present at all in MFPc where both the capacity measures are part of the model. $C_{\mbox{\footnotesize approx}}$ continues to be extremely significant, but it is now included linearly and it is less important than in MFPb. Finally, $\widebar{T}_{\mbox{\footnotesize min}}$ has an even smaller effect than in MFPb, but persists in being an important input to predict the change in capacity.

For the sake of interpretability and explanation, MFPc seems the best choice: it has the highest coefficients of determination, it is reasonably sparse and all the included variables are extremely significant. However, it assumes that the observed capacity is known at every previous cycle, which is hardly the case in real conditions: then, MFPb constitutes a valid alternative as it takes advantage of approximate evaluations of the capacity derived directly from the data gathered during the RW steps. However, it is noticeable that MFPa also reaches high $R^2$ and $R^2_{\mbox{\footnotesize adj}}$ while comprising only three features which can be very easily obtained in practice. 

\begin{figure}[H]
\centering
        \textbf{\hspace{.6cm} MFPa \hspace{3.5cm} MFPb \hspace{3.5cm} MFPc}
        
        {\rotatebox{90}{\textbf{RW9}}} \hspace{0cm} \hfill\includegraphics[width=0.315\textwidth, valign=c]{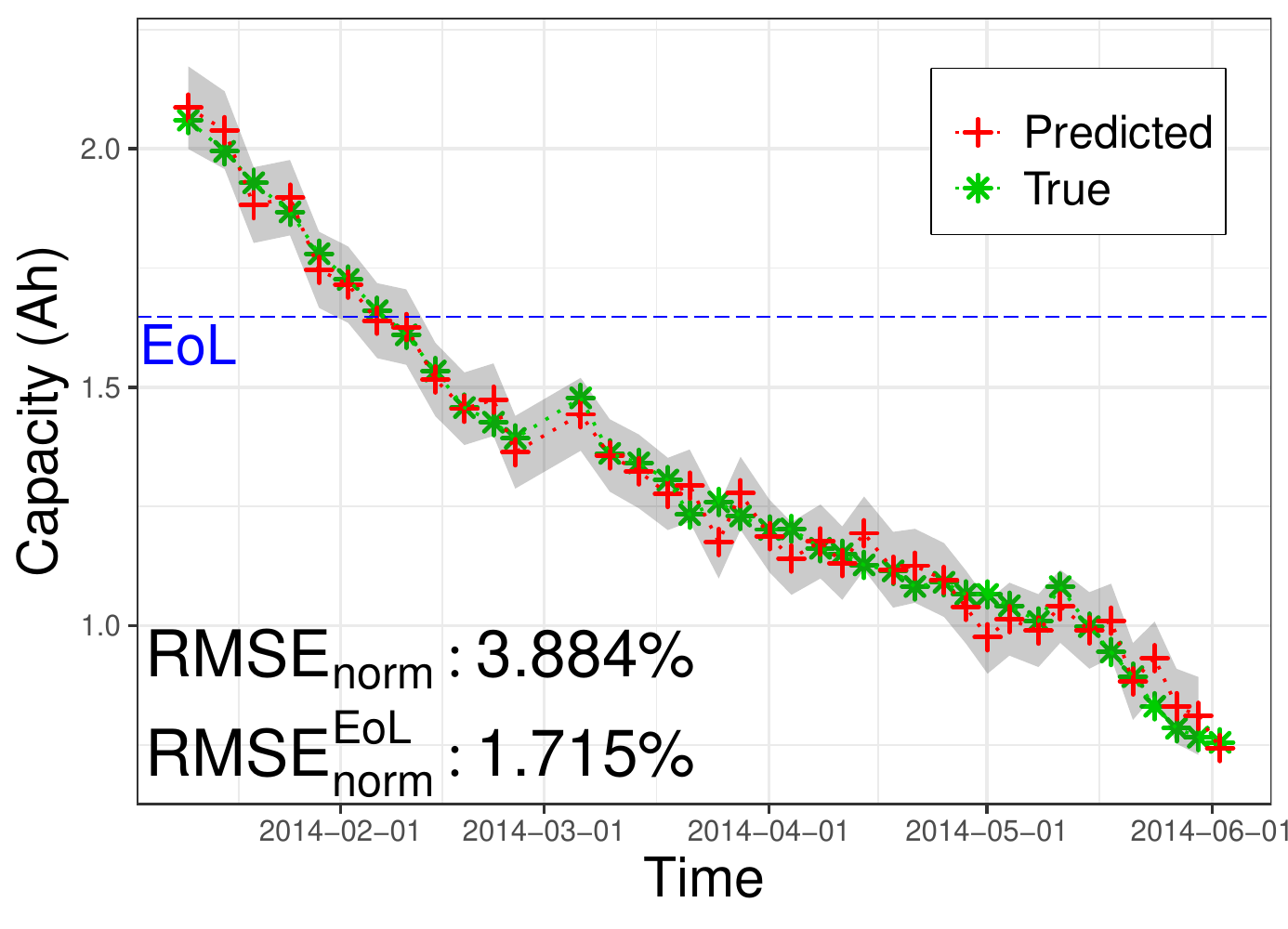}
        \hfill\includegraphics[width=0.315\textwidth, valign=c]{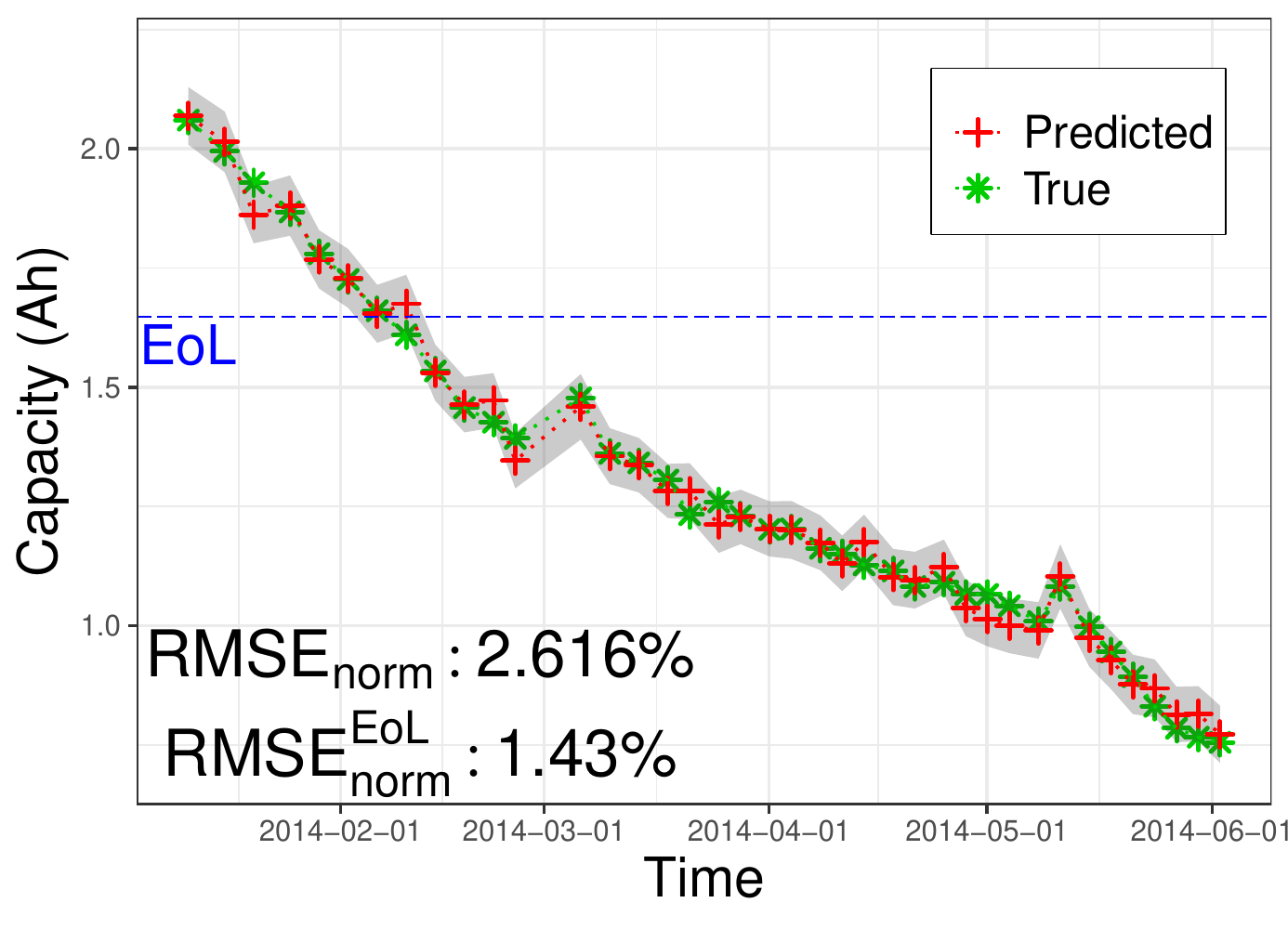}
        \hfill\includegraphics[width=0.315\textwidth, valign=c]{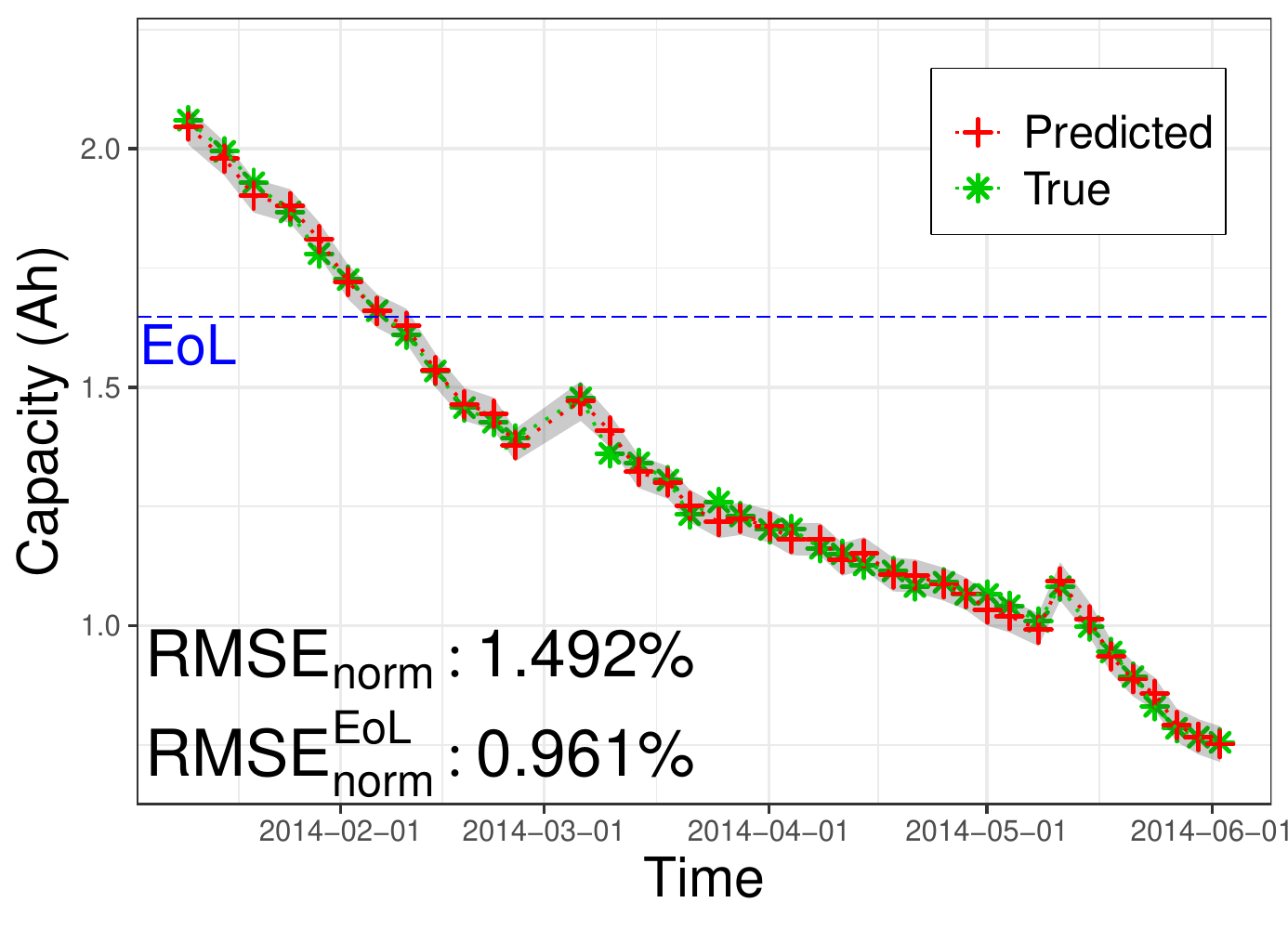}
        
        {\rotatebox{90}{\textbf{RW10}}} \hspace{0cm} \hfill\includegraphics[width=0.315\textwidth, valign=c]{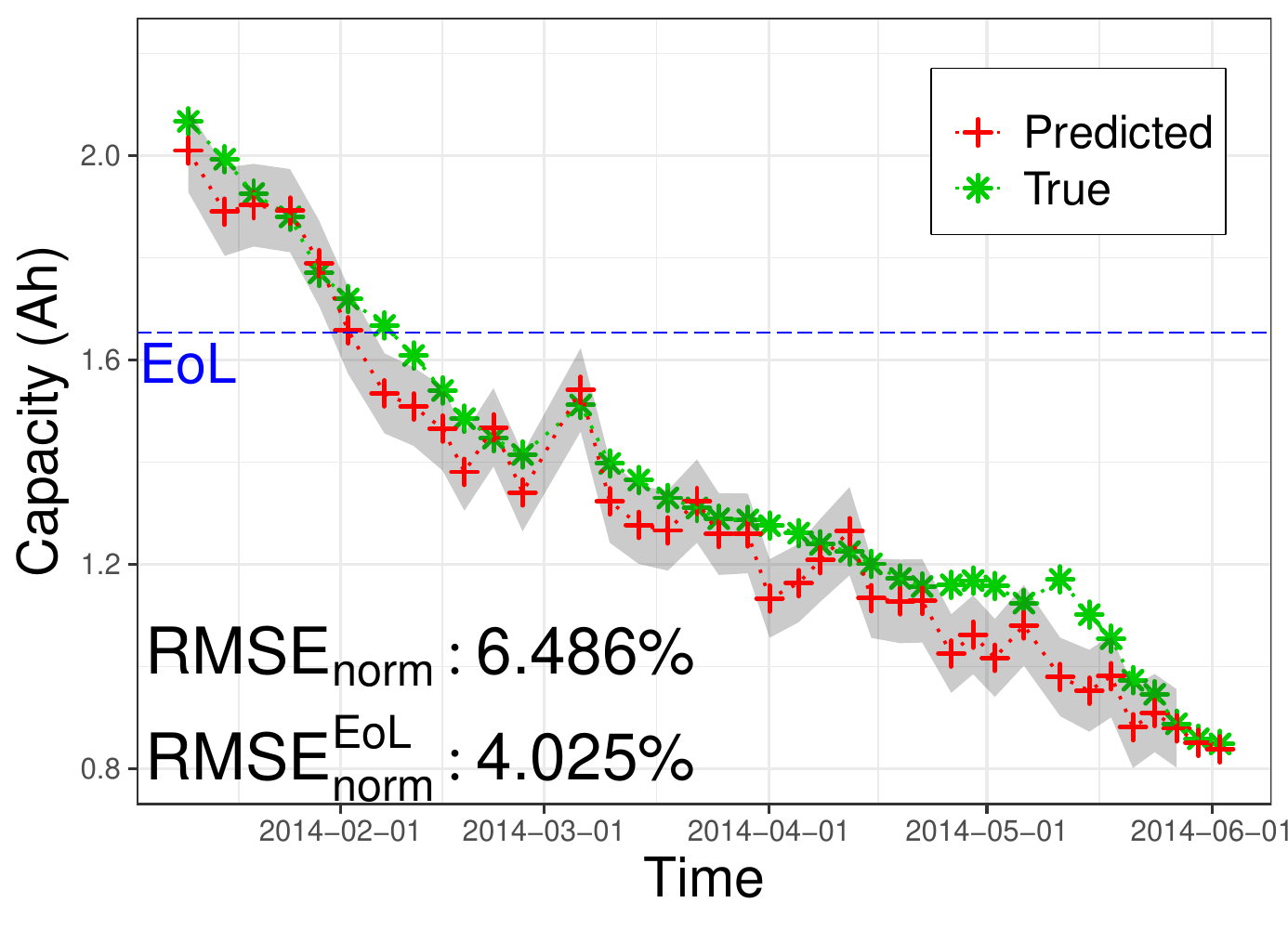}
        \hfill\includegraphics[width=0.315\textwidth, valign=c]{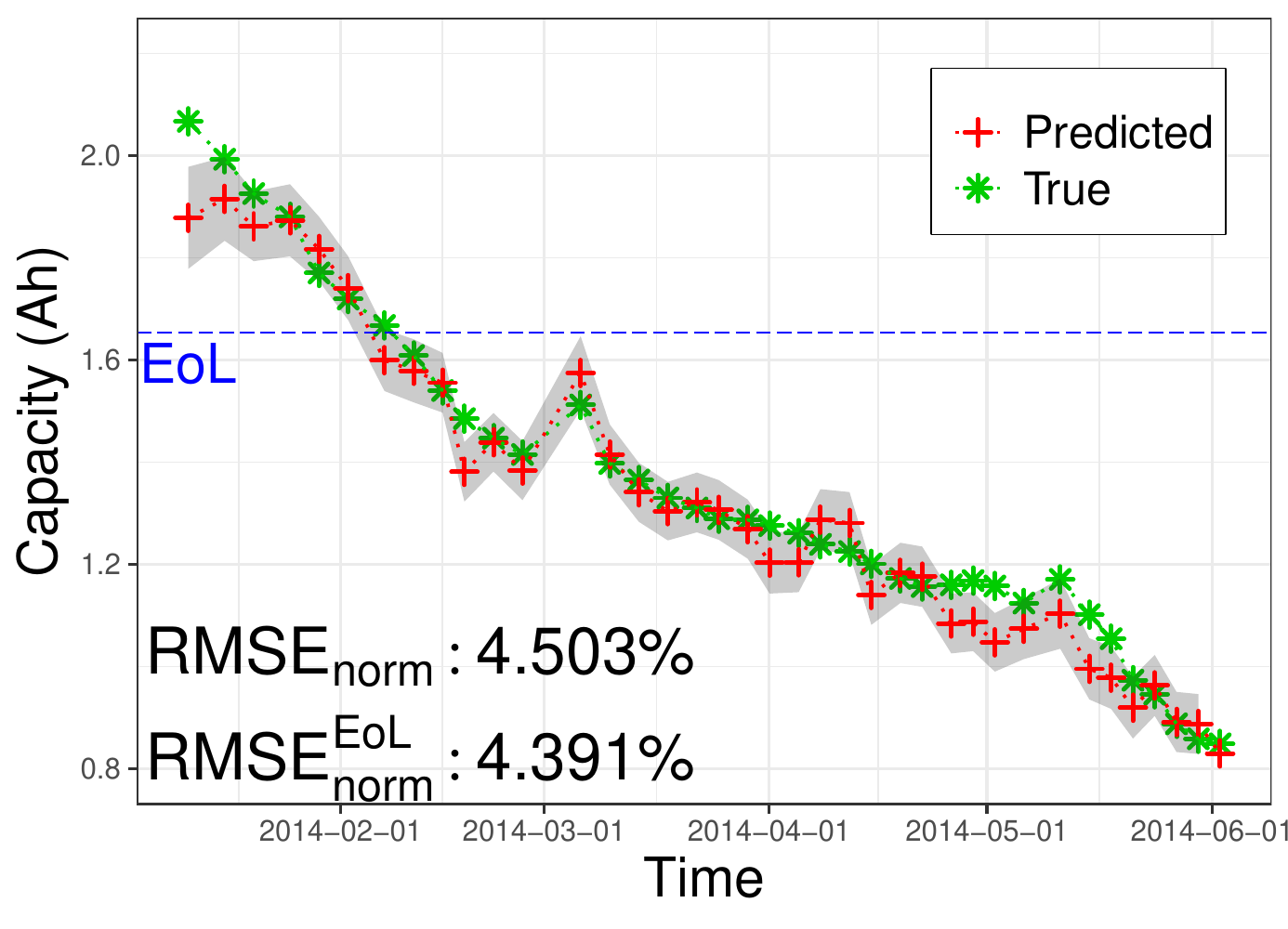}
        \hfill\includegraphics[width=0.315\textwidth, valign=c]{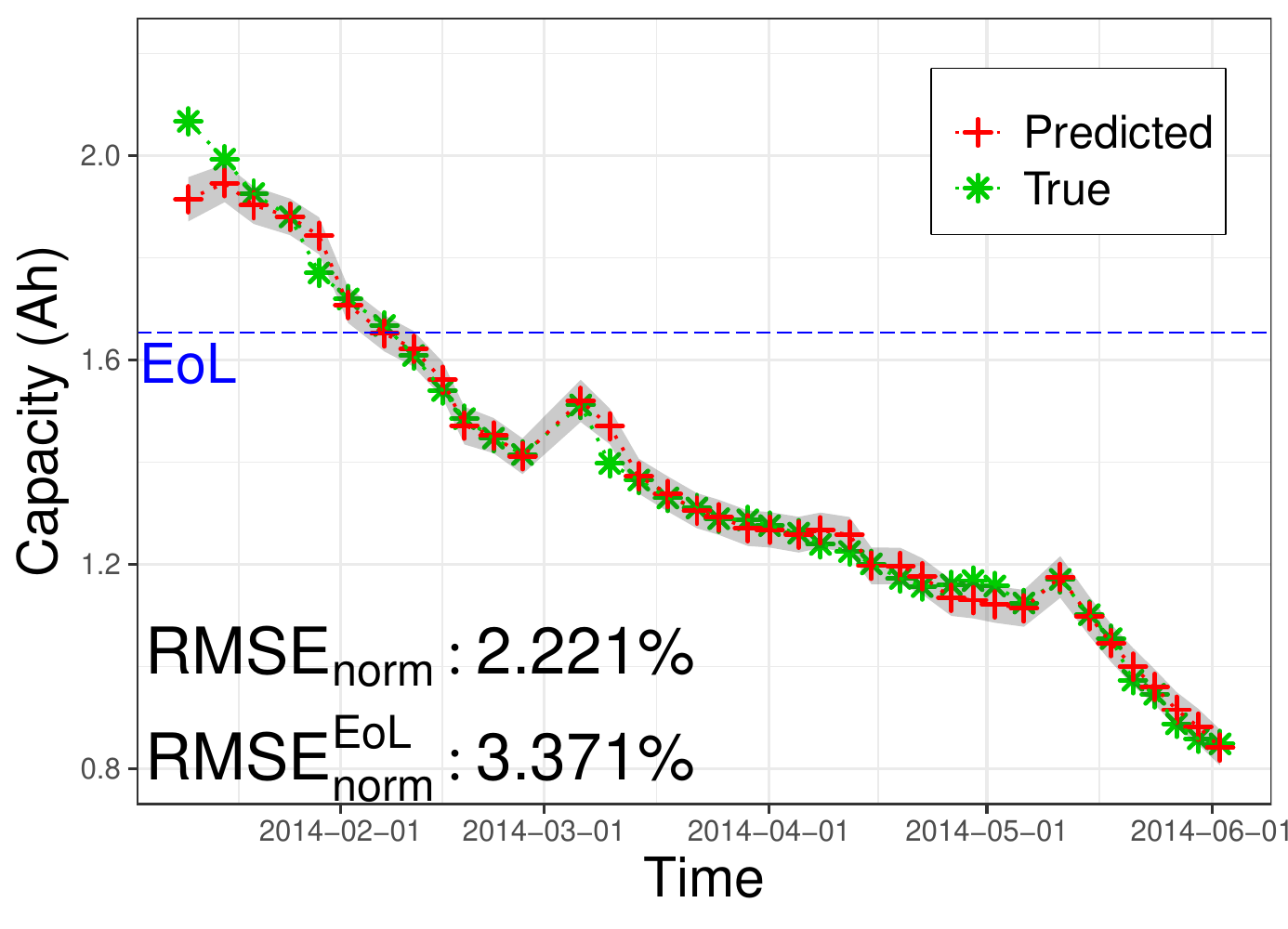}
        
        {\rotatebox{90}{\textbf{RW11}}} \hspace{0cm} \hfill\includegraphics[width=0.315\textwidth, valign=c]{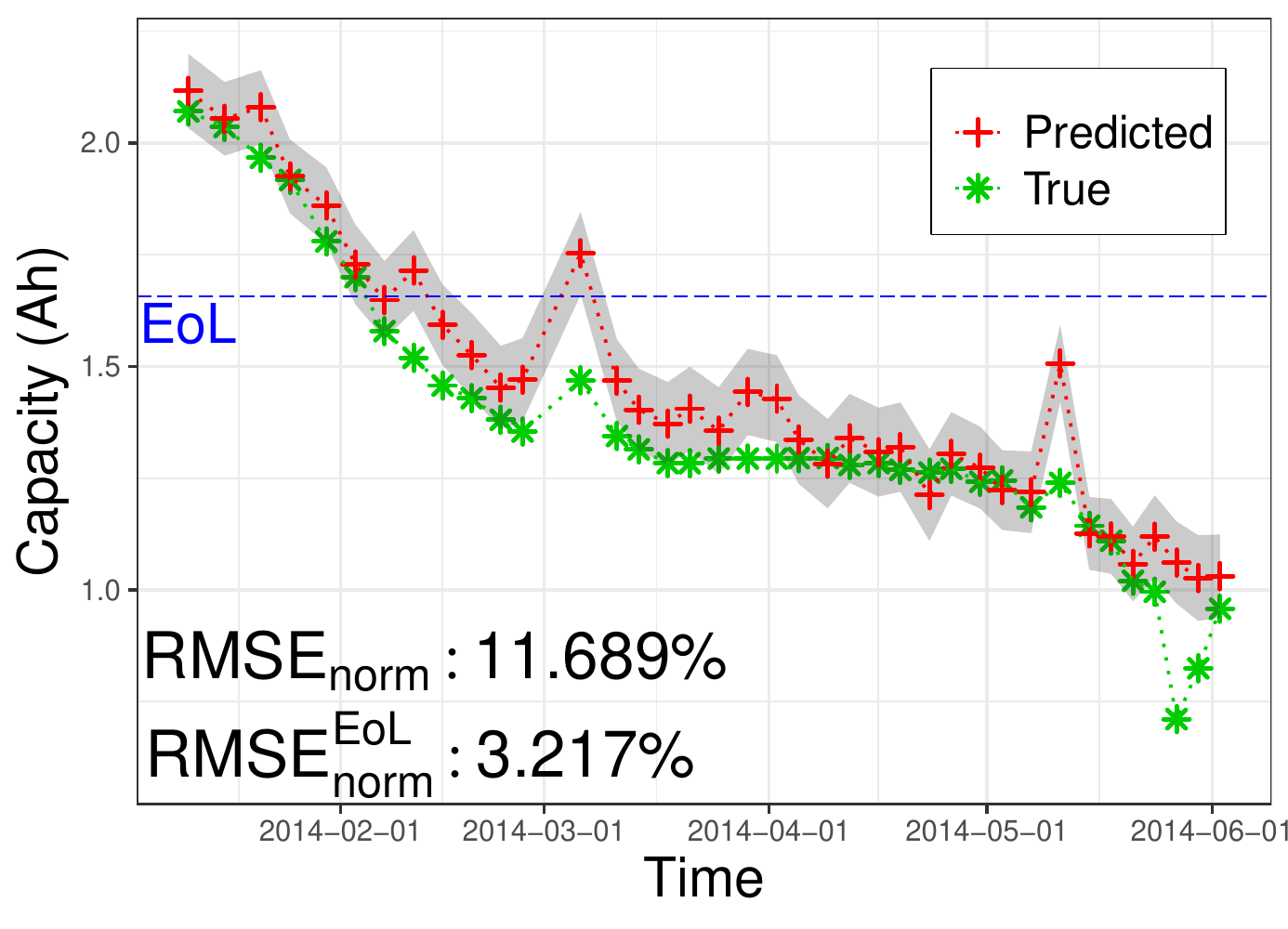}
        \hfill\includegraphics[width=0.315\textwidth, valign=c]{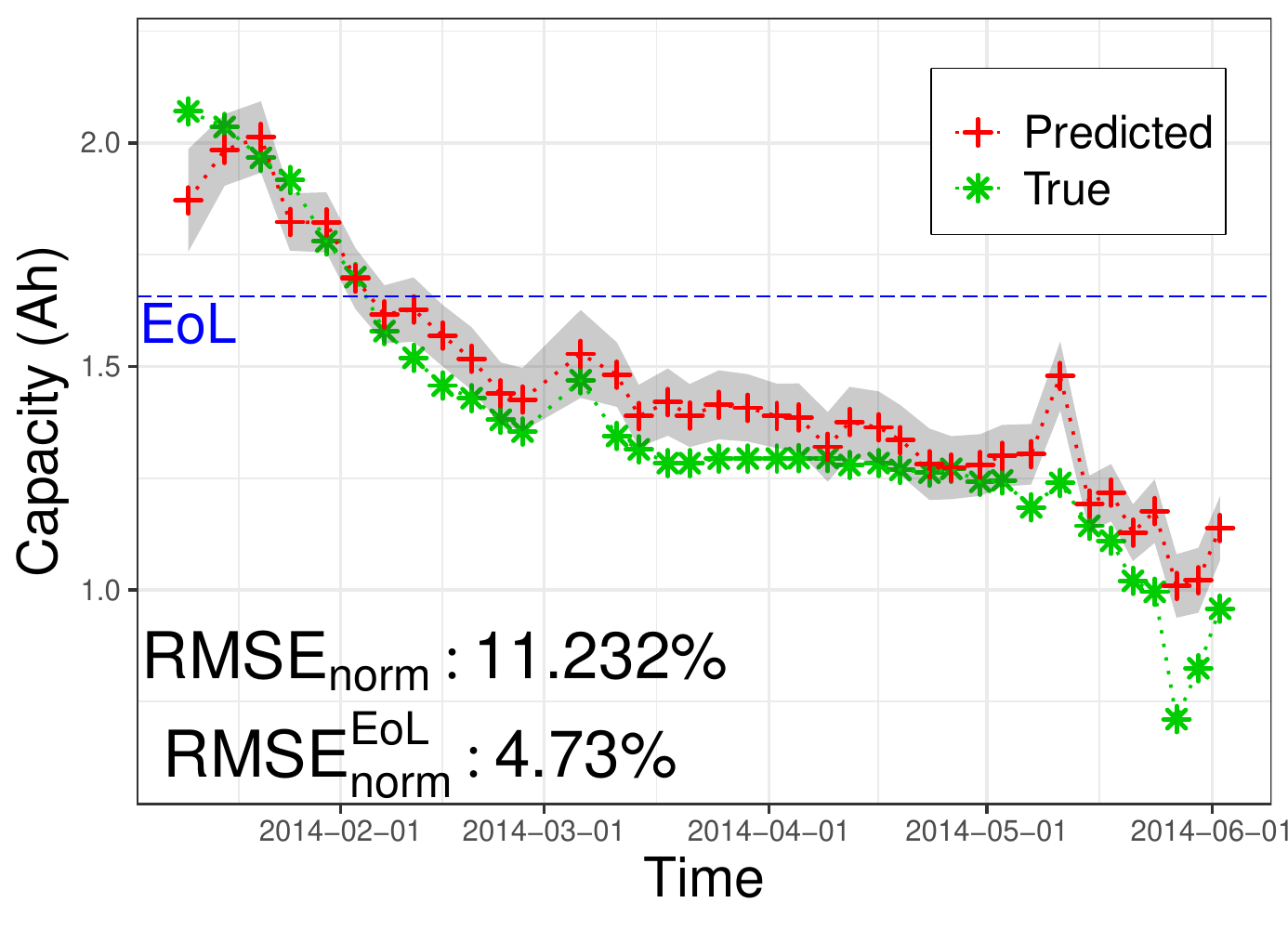}
        \hfill\includegraphics[width=0.315\textwidth, valign=c]{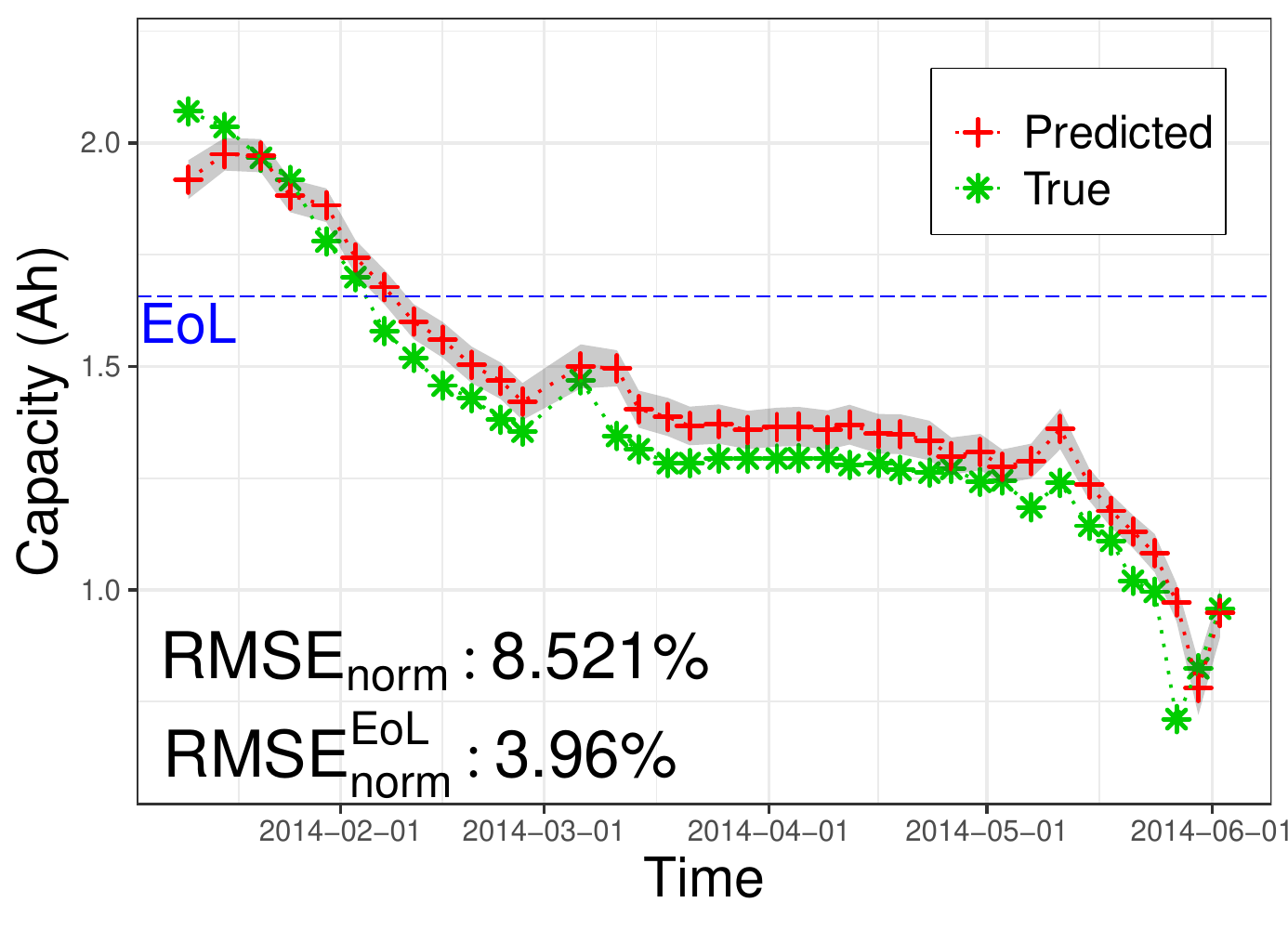}      
        
        {\rotatebox{90}{\textbf{RW12}}} \hspace{0cm} \hfill\includegraphics[width=0.315\textwidth, valign=c]{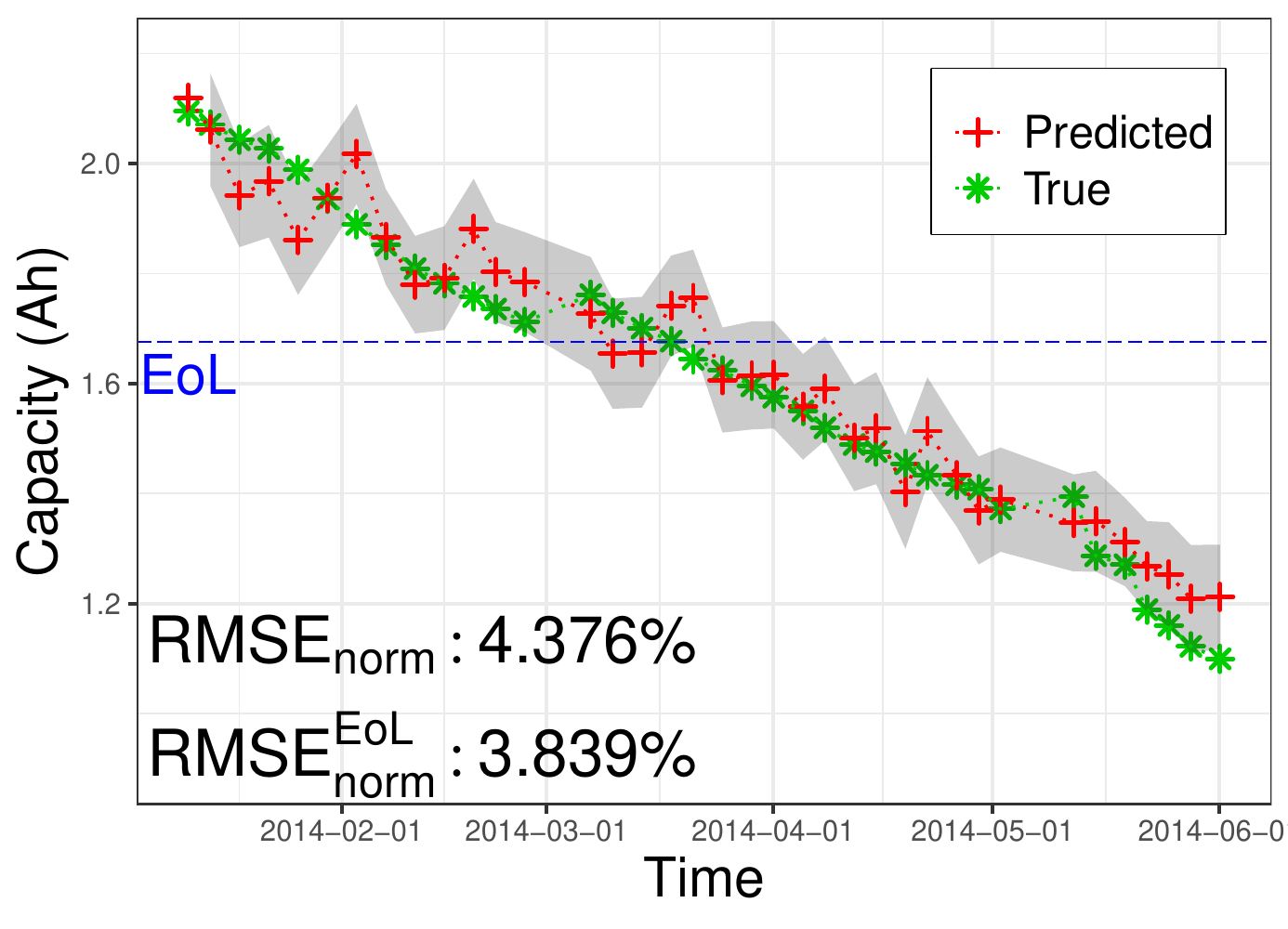}
        \hfill\includegraphics[width=0.315\textwidth, valign=c]{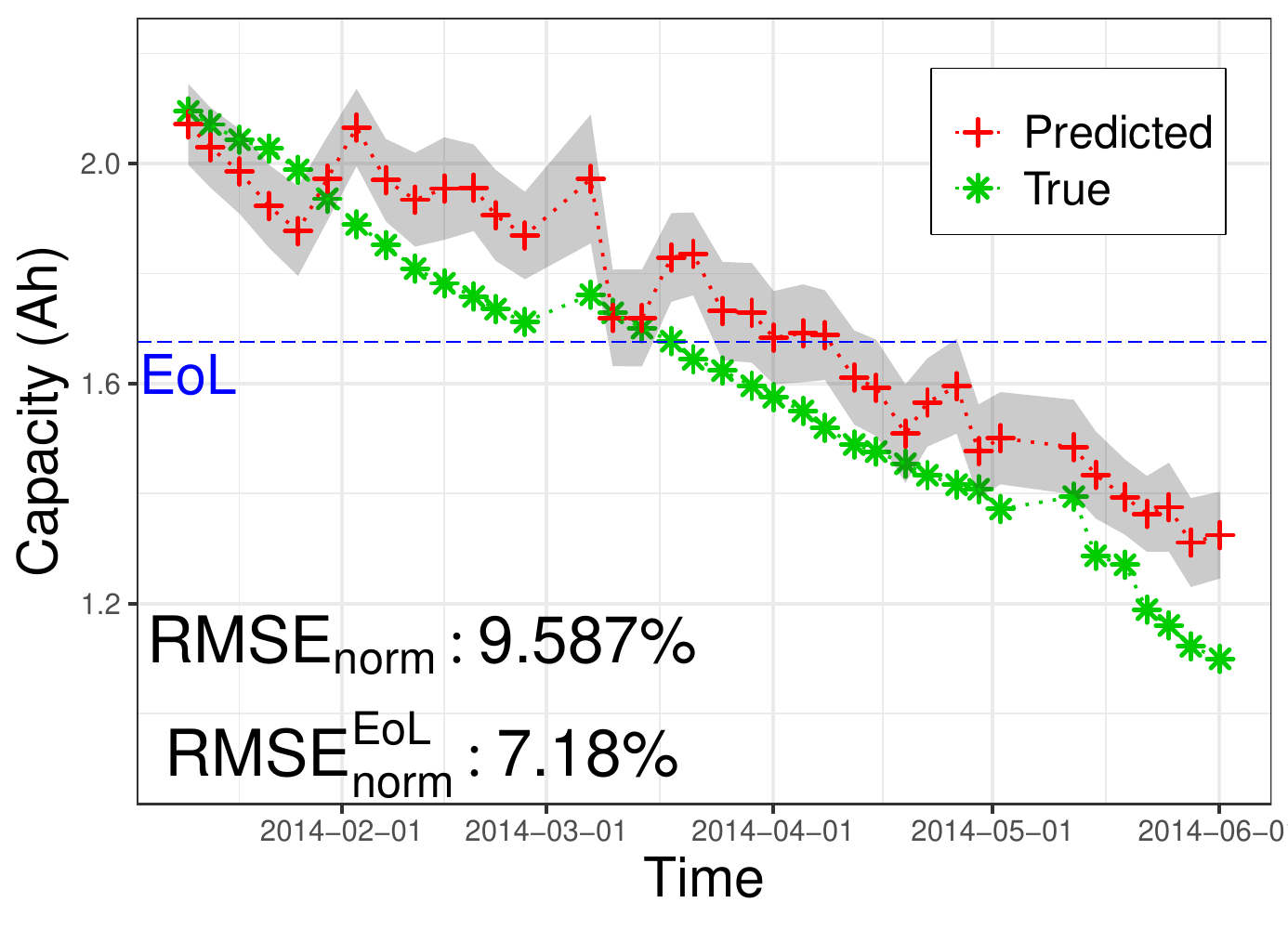}
        \hfill\includegraphics[width=0.315\textwidth, valign=c]{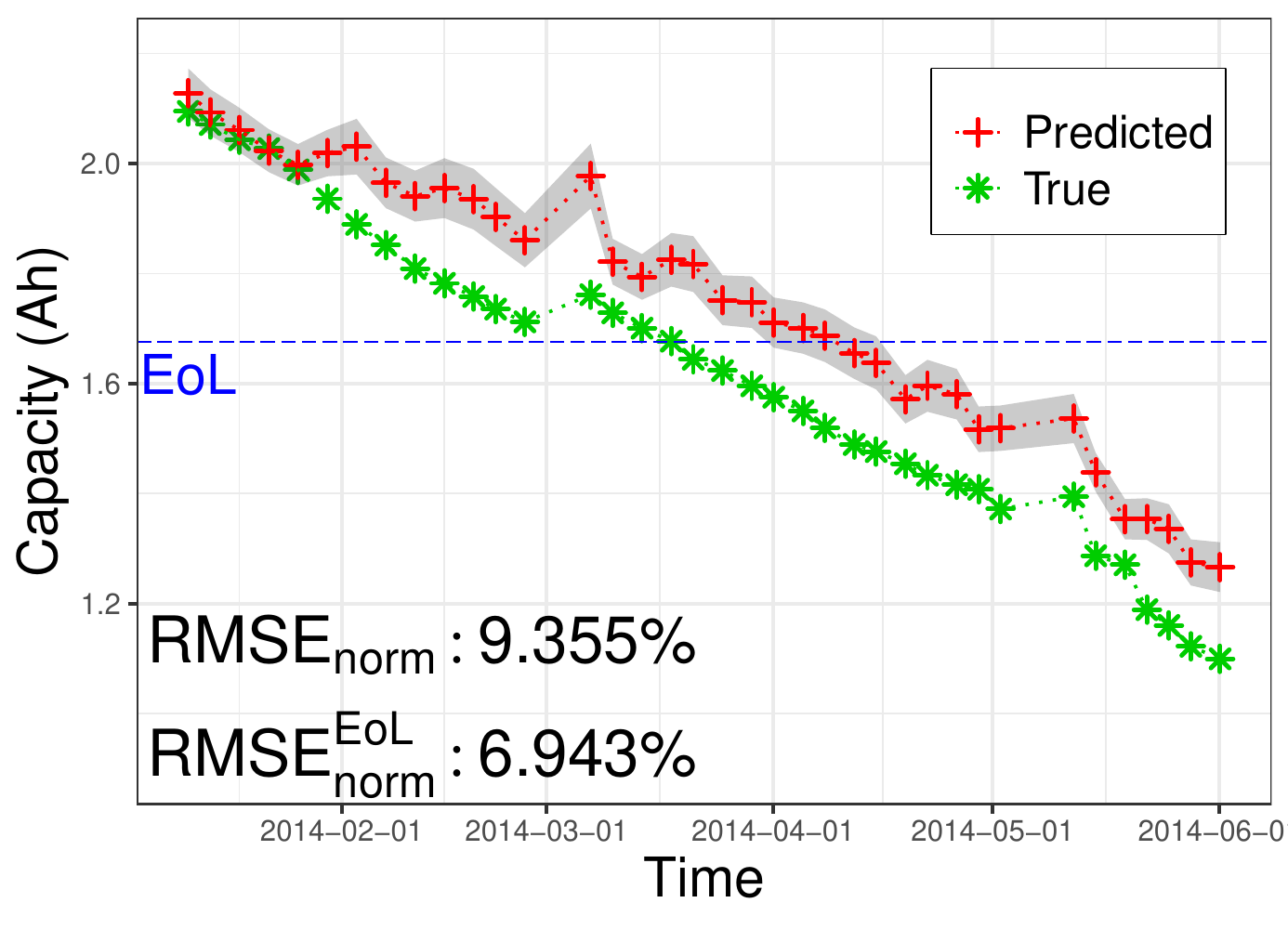}        
\caption{Capacity fade prediction results using the three models MFPa (left column), MFPb (middle column), and MFPc (right column), together with the considered error metrics. The two apparent increases in each capacity curve, produced by a prolonged rest of the cells, are in most cases predicted by the models. All models have good predictive accuracy with RMSE$_{\tiny \mbox{norm}}$  and RMSE$^{\tiny \mbox{EoL}}_{\tiny\mbox{norm}}$ ranging from 2.22\% to 11.69\% and from 3.21\% to 7.18\%, respectively.  The higher errors affecting cells RW11 and RW12 are due to the different profiles of these cells (Section \ref{Prob_formulation_dataset}), but are in line with the existing literature (Section \ref{discussion}). Other error metrics can be found in Section S.5 in the Supplementary Material.}   
\label{results}
\end{figure}

When it comes to accuracy in prediction, the normalised RMSEs are presented in Figure \ref{results}, where the predicted capacity fade according to each model is compared to the true values. The grey area represents the 90\% prediction interval, which has been computed using basic asymptotic results at almost no additional computational cost.
The first row (cell RW9) reports the training error of each model. The results show that the difference in the performances of the three considered models is not huge: all of them have good predictive accuracy with RMSE$_{\tiny \mbox{norm}}$  and RMSE$^{\tiny \mbox{EoL}}_{\tiny\mbox{norm}}$ ranging respectively from 2.22\% to 11.69\% and from 3.21\% to 7.18\%. The errors reflect the similarities and dissimilarities in the production phase and operational history of the four cells, as shown in Figure \ref{capacity_all_batteries} in Section \ref{Prob_formulation_dataset}. Considering RMSE$_{\tiny \mbox{norm}}$, there is a consistent improvement going from model MFPa to MFPb and MFPc for batteries RW10 and RW11, while the minimum RMSE$^{\tiny \mbox{EoL}}_{\tiny \mbox{norm}}$ is obtained with MFPa; for RW12, remarkably, we obtain better results with model MFPa according to both the error metrics.

\subsubsection{Groups 1, 2, 4: method solidity} \label{other_groups}

\begin{figure}[H]
\centering
        \textbf{\hspace{.6cm} MFPa \hspace{3.5cm} MFPb \hspace{3.5cm} MFPc}
        
        {\rotatebox{90}{\textbf{RW1}}} \hspace{0cm} \hfill\includegraphics[width=0.315\textwidth, valign=c]{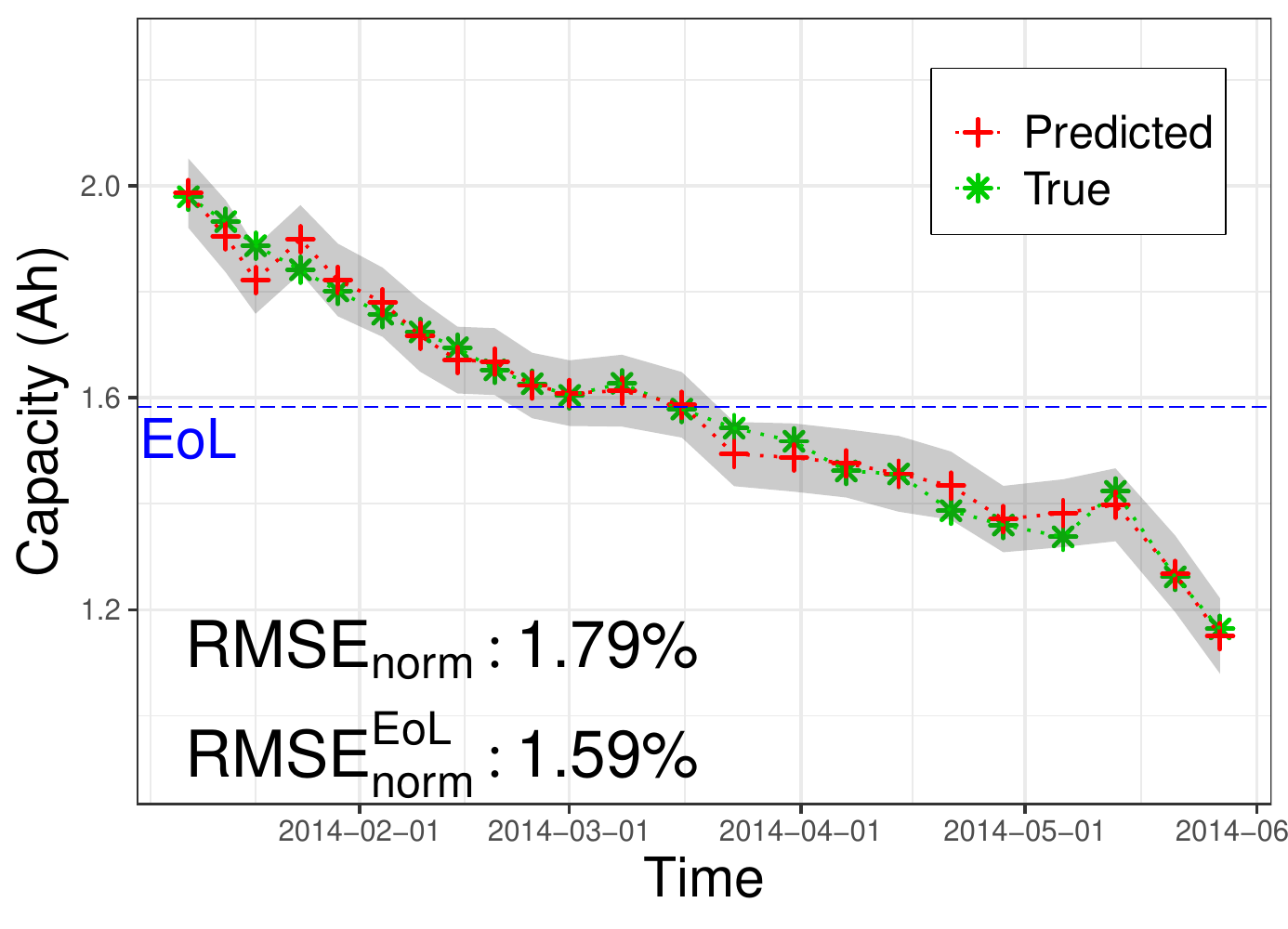}
        \hfill\includegraphics[width=0.315\textwidth, valign=c]{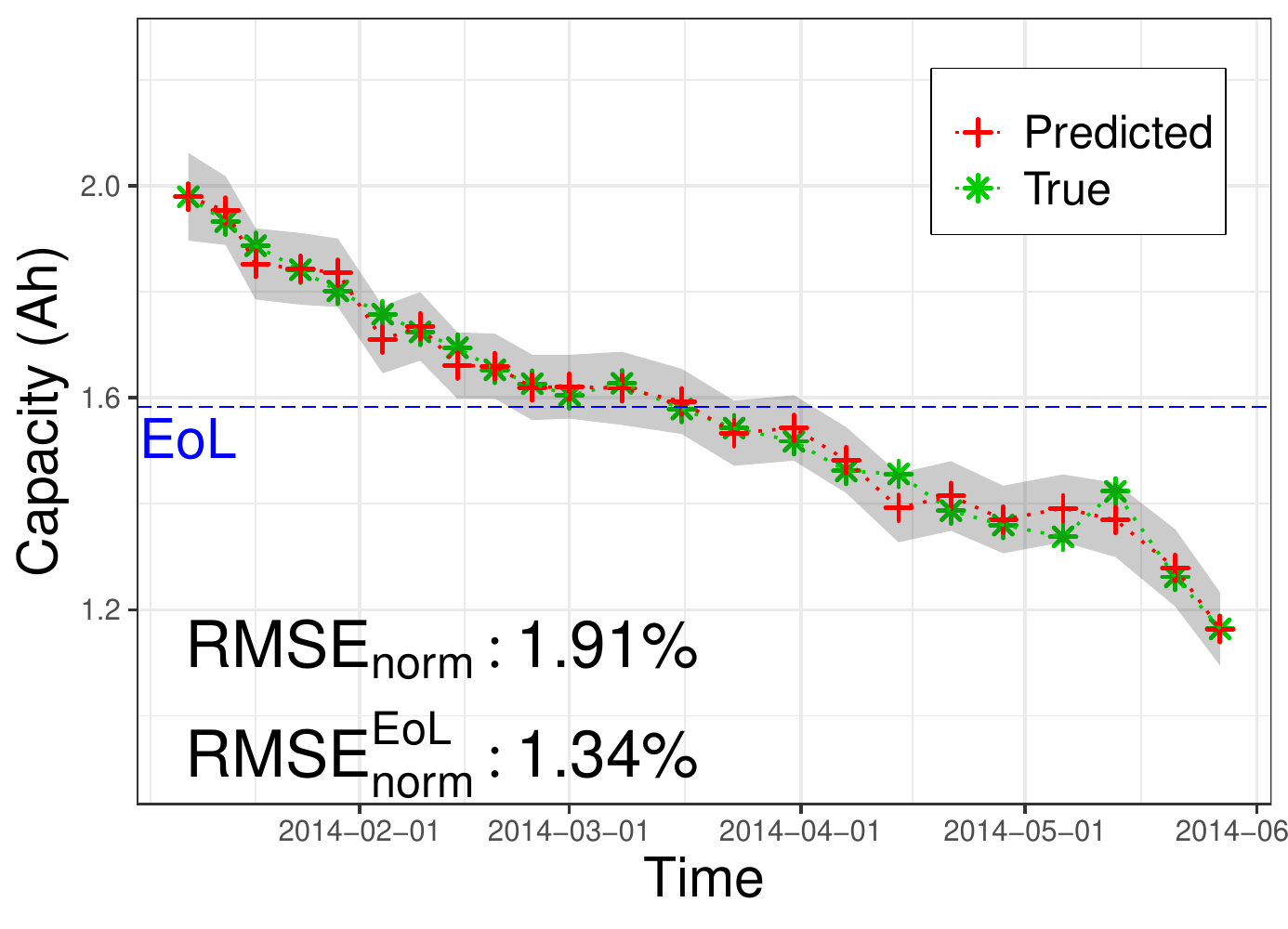}
        \hfill\includegraphics[width=0.315\textwidth, valign=c]{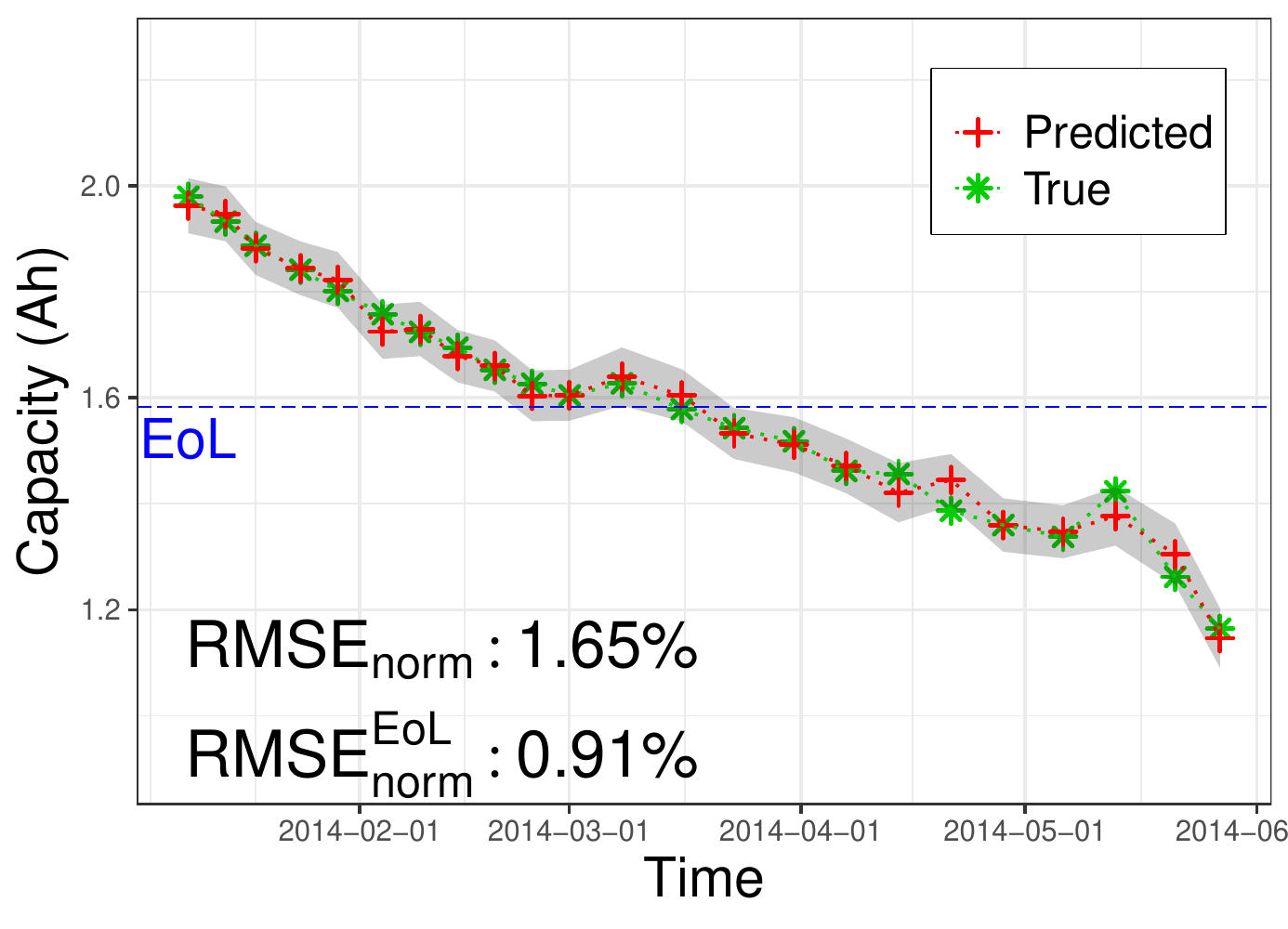}
        
        {\rotatebox{90}{\textbf{RW7}}} \hspace{0cm} \hfill\includegraphics[width=0.315\textwidth, valign=c]{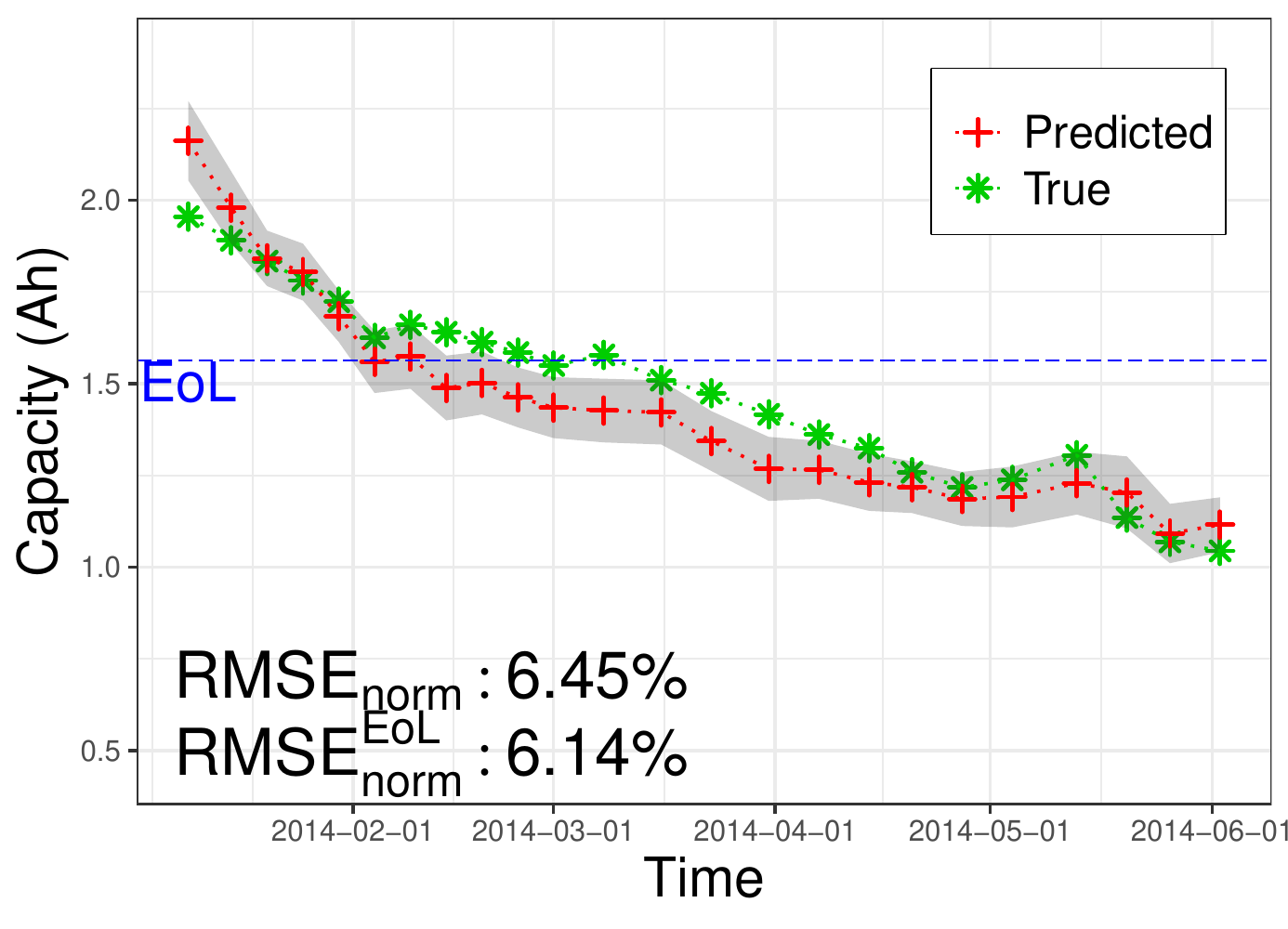}
        \hfill\includegraphics[width=0.315\textwidth, valign=c]{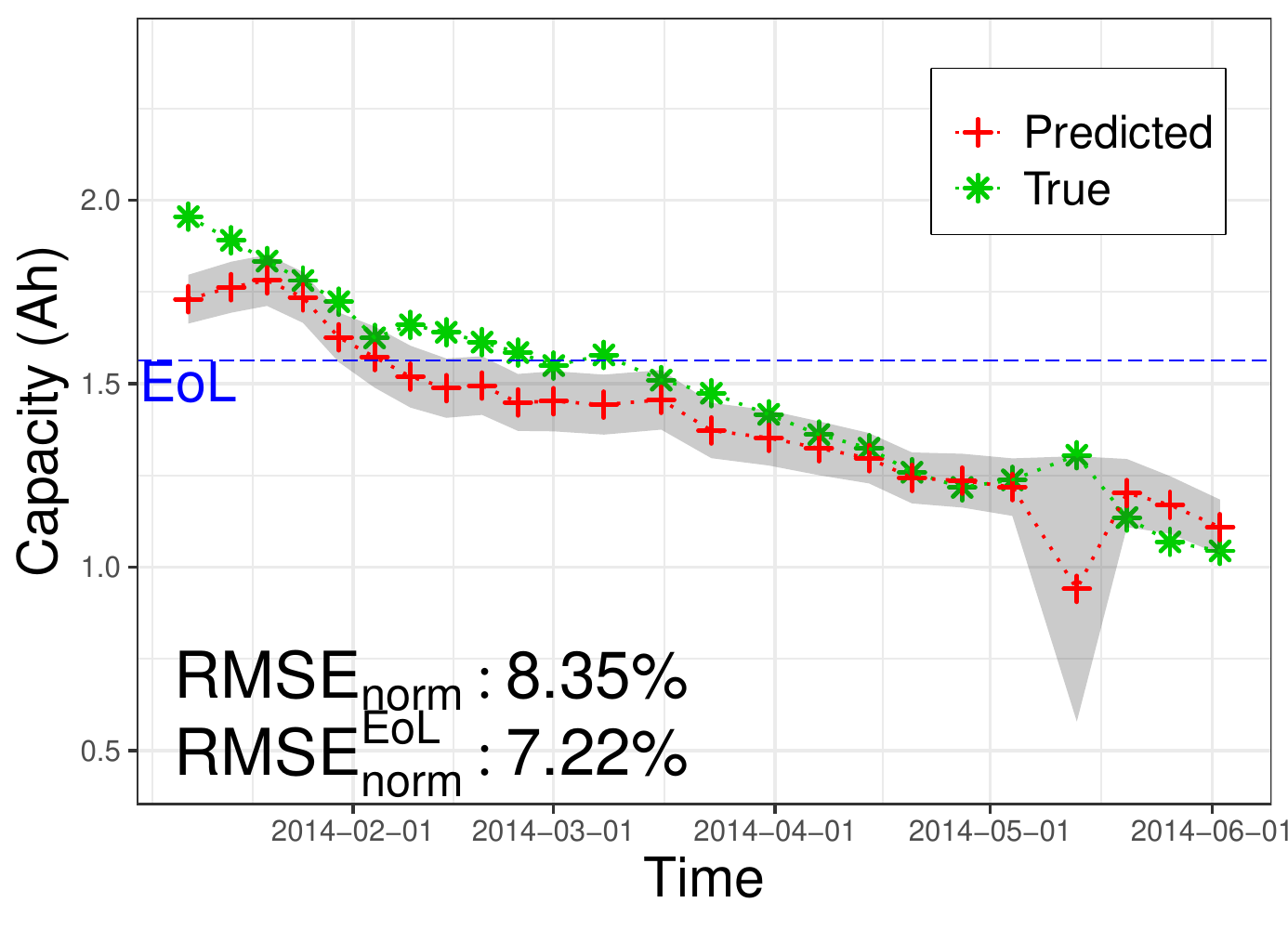}
        \hfill\includegraphics[width=0.315\textwidth, valign=c]{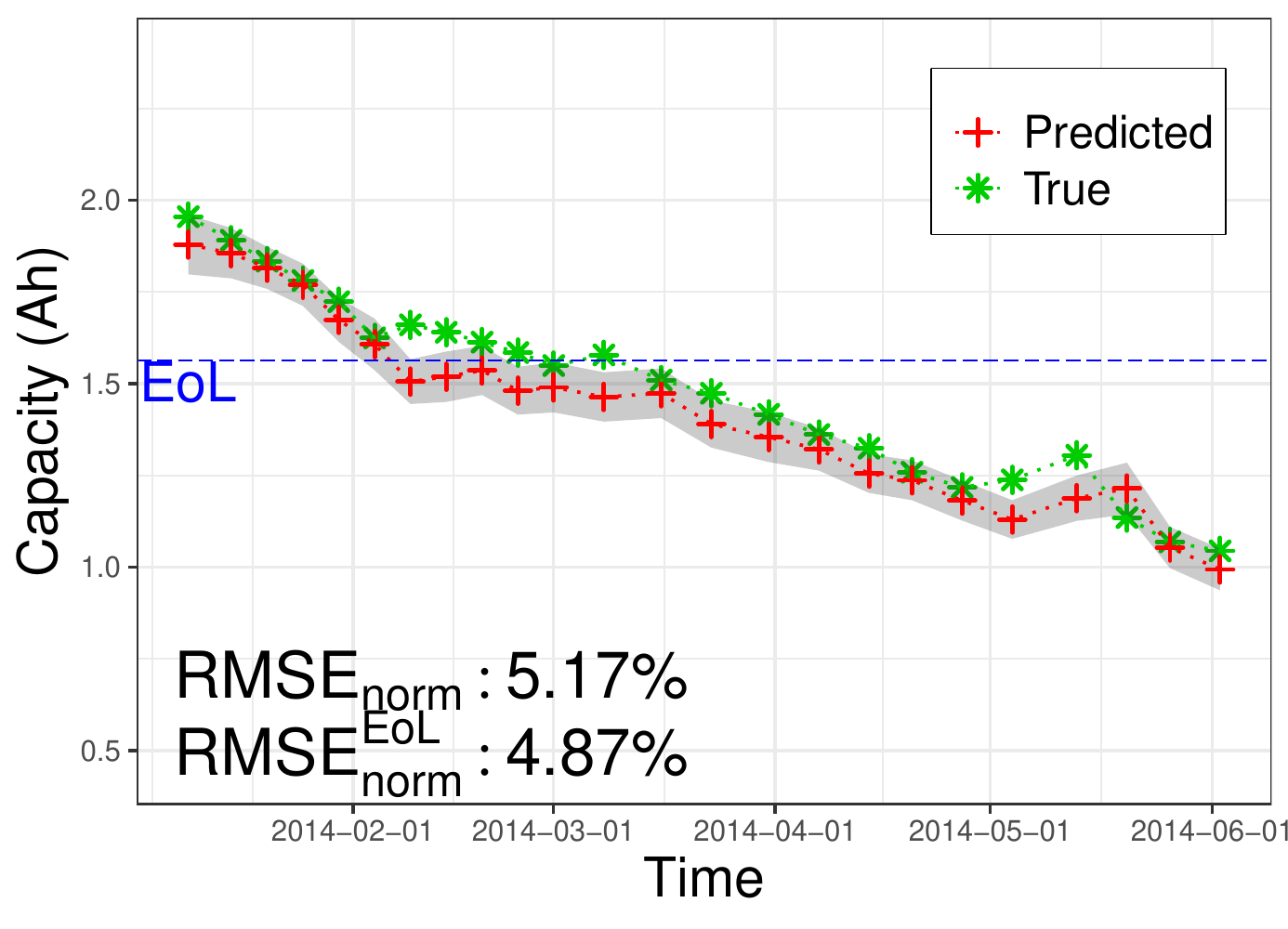}
        
        {\rotatebox{90}{\textbf{RW8}}} \hspace{0cm} \hfill\includegraphics[width=0.315\textwidth, valign=c]{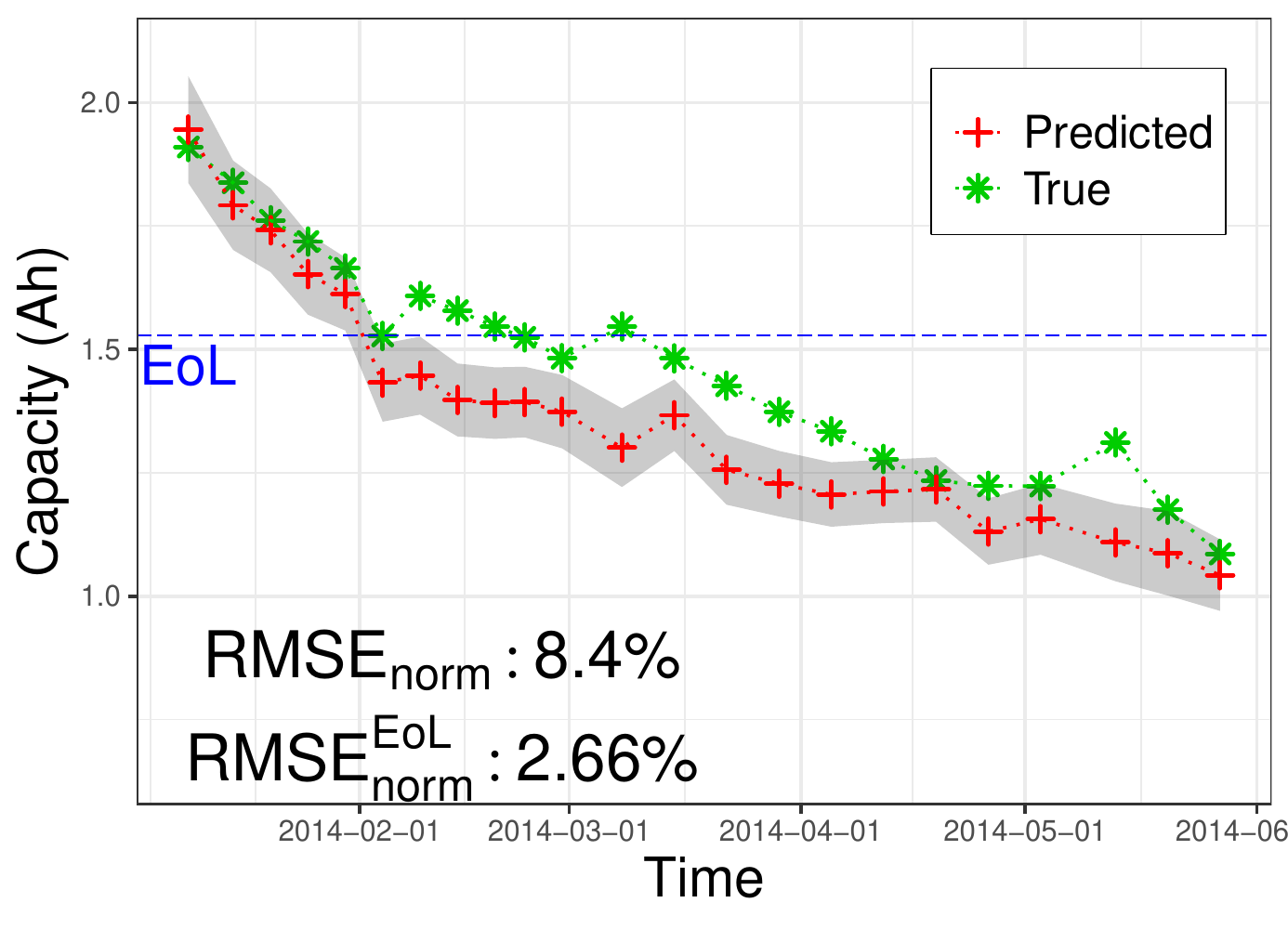}
        \hfill\includegraphics[width=0.315\textwidth, valign=c]{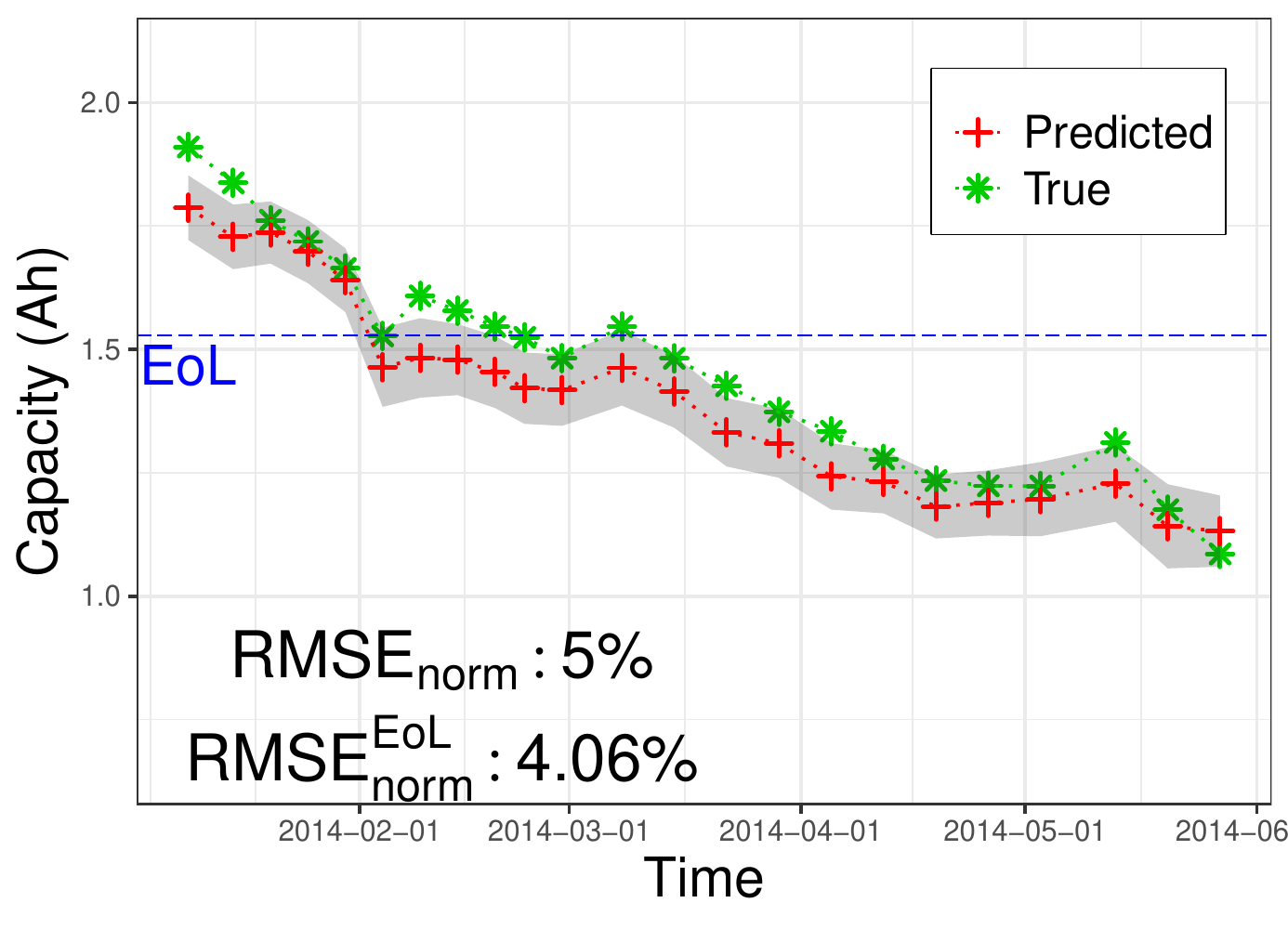}
        \hfill\includegraphics[width=0.315\textwidth, valign=c]{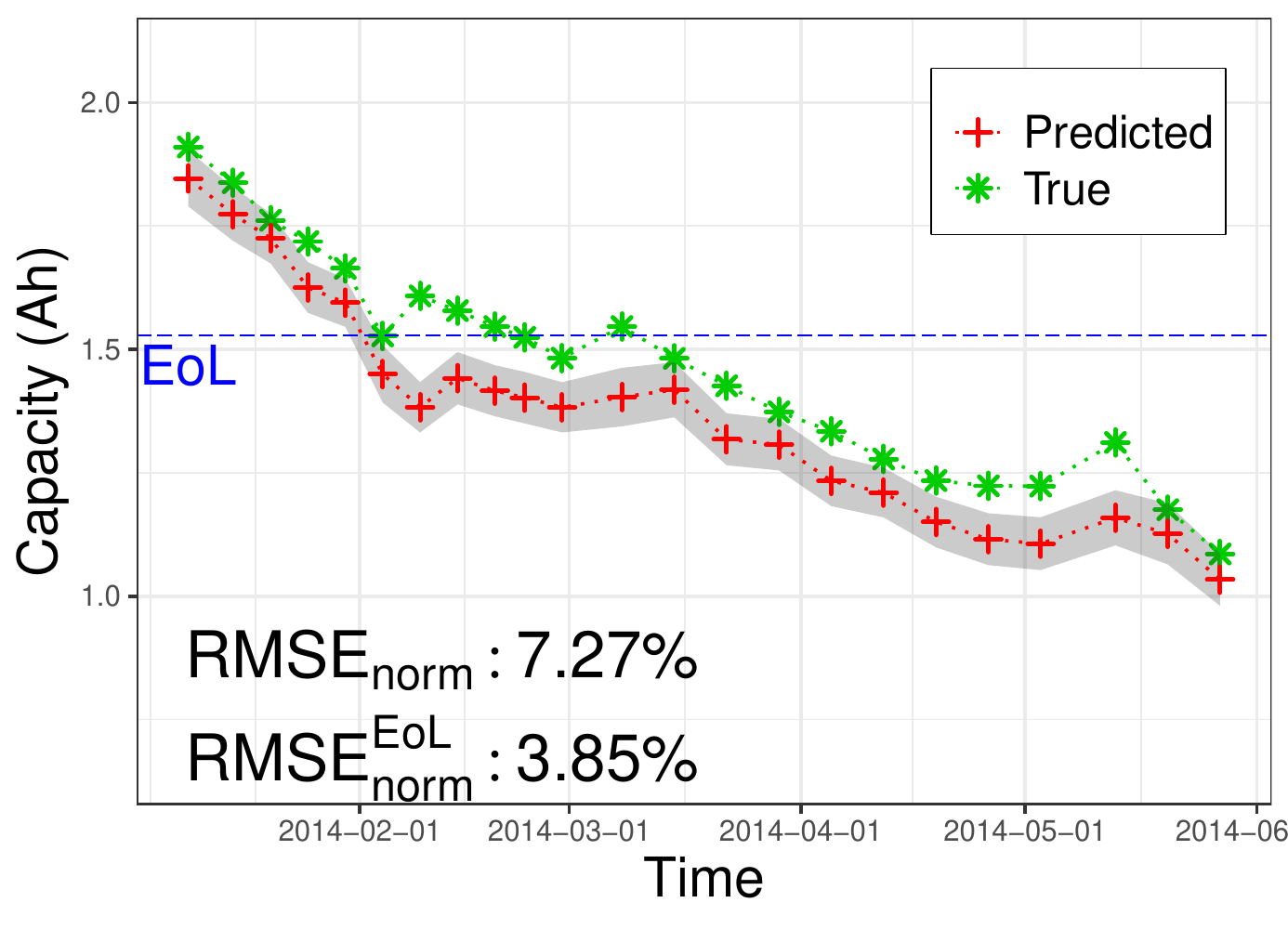}      
        
\caption{Capacity fade prediction results for the cells in group 1 using the three models MFPa (left column), MFPb (middle column), and MFPc (right column), together with the considered error metrics. Cell RW1 is used for training, while RW7 and RW8 are used for test. The range in the prediction errors is 5\% - 8.4\% for  RMSE$_{\tiny \mbox{norm}}$  and 2.66\%-7.22\% for RMSE$^{\tiny \mbox{EoL}}_{\tiny\mbox{norm}}$.}   
\label{results_group1}
\end{figure}

\begin{figure}[H]
\centering
        \textbf{\hspace{.6cm} MFPa \hspace{3.5cm} MFPb \hspace{3.5cm} MFPc}
        
        {\rotatebox{90}{\textbf{RW4}}} \hspace{0cm} \hfill\includegraphics[width=0.315\textwidth, valign=c]{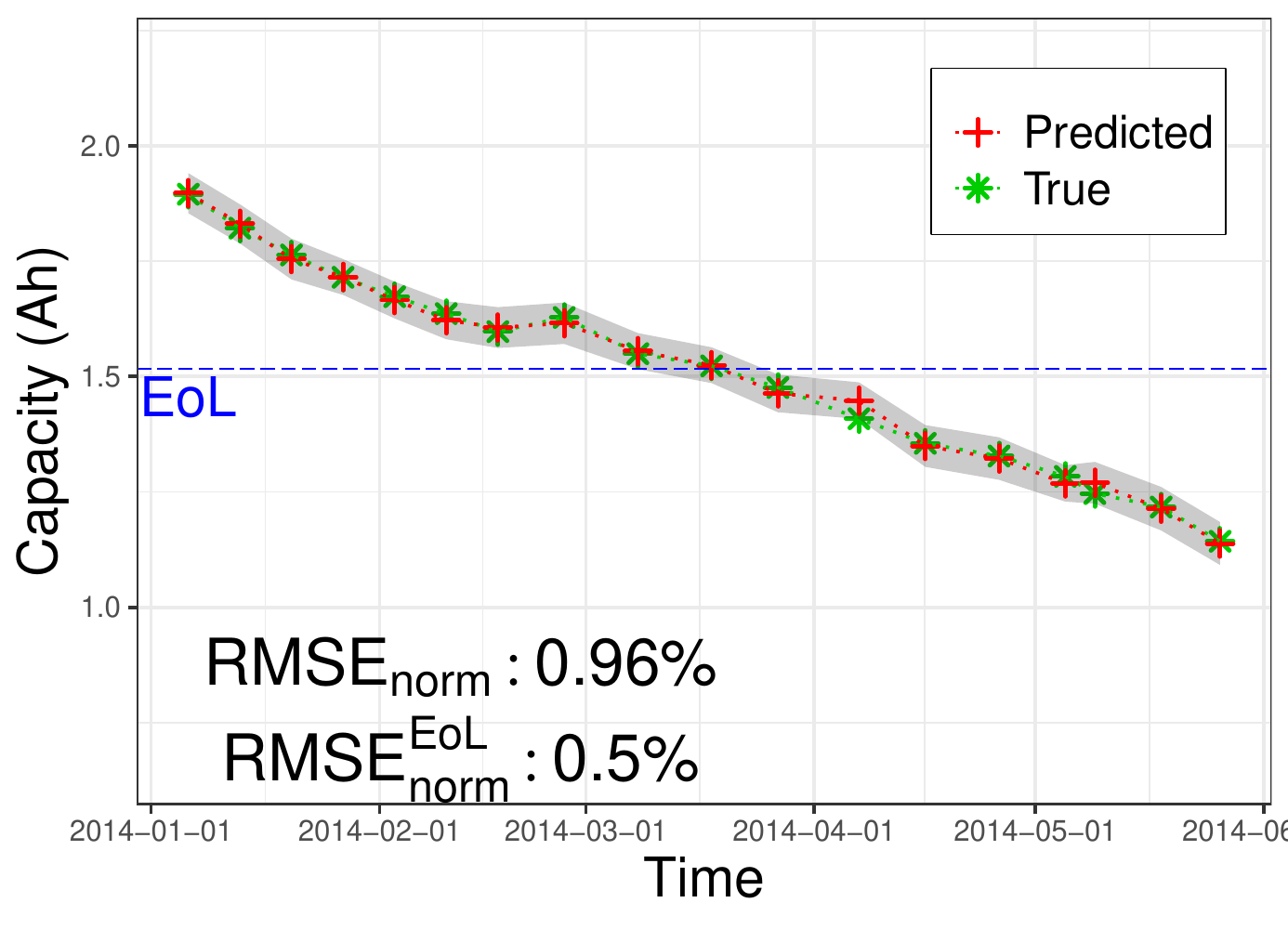}
        \hfill\includegraphics[width=0.315\textwidth, valign=c]{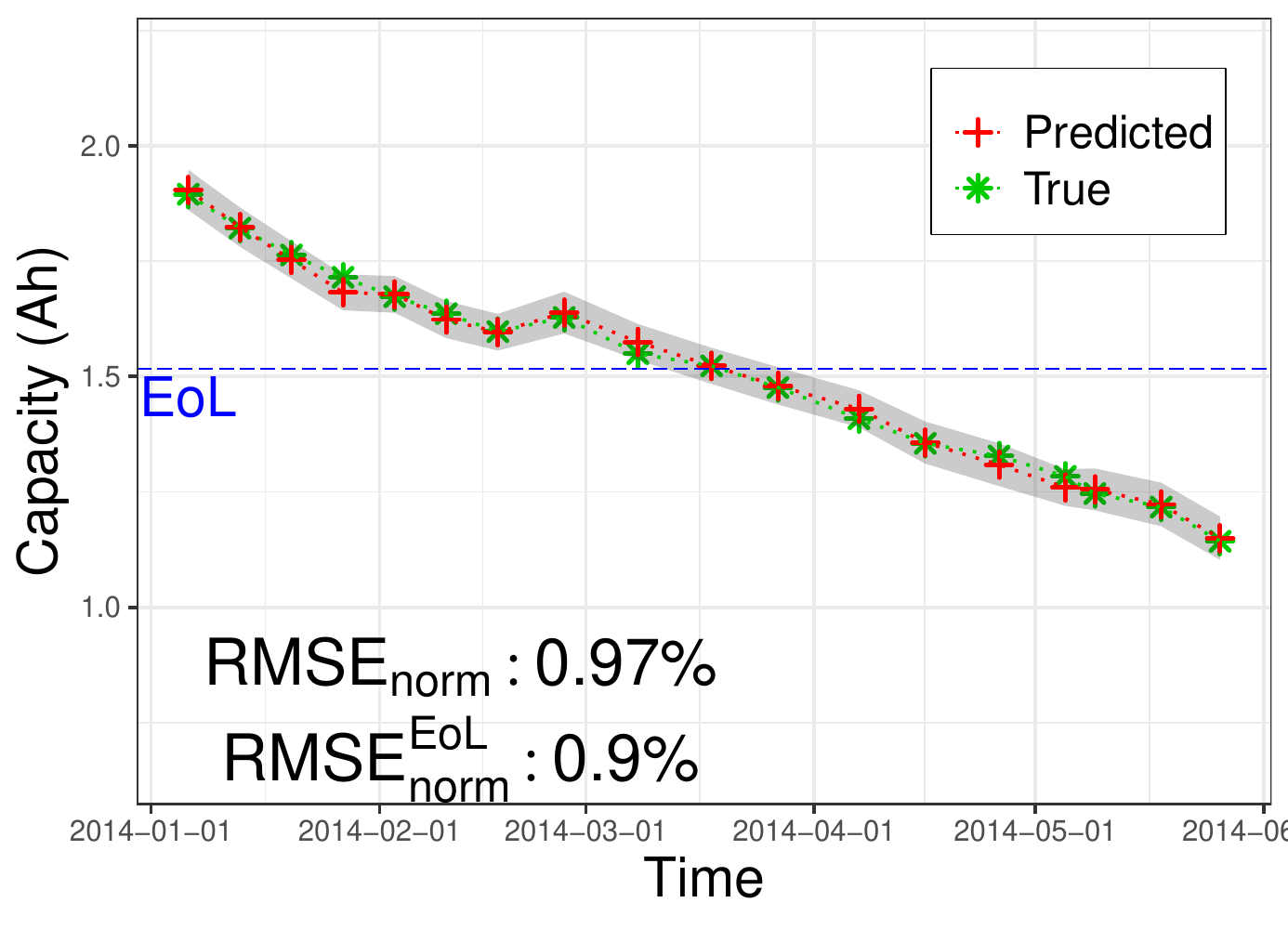}
        \hfill\includegraphics[width=0.315\textwidth, valign=c]{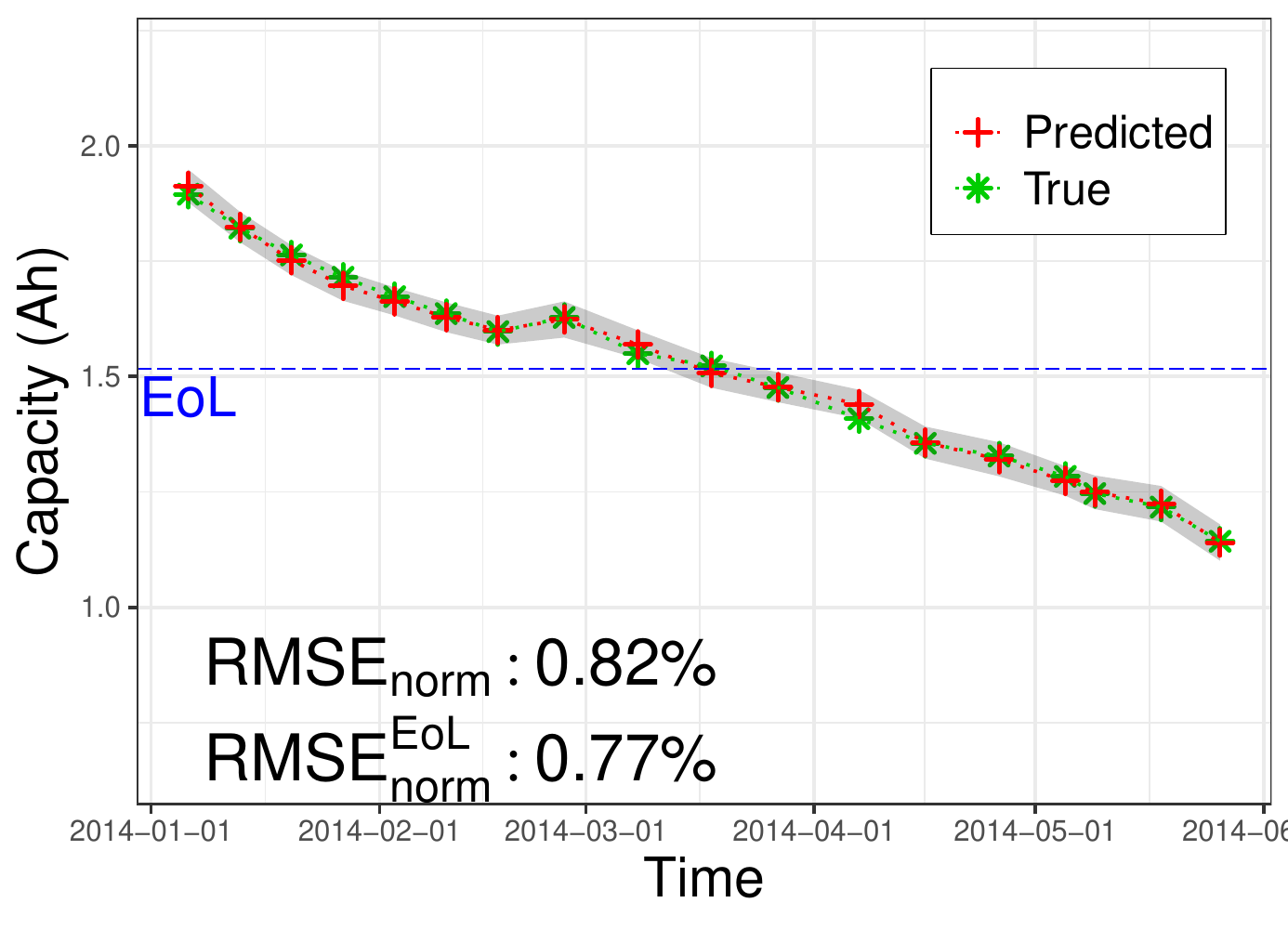}
        
        {\rotatebox{90}{\textbf{RW5}}} \hspace{0cm} \hfill\includegraphics[width=0.315\textwidth, valign=c]{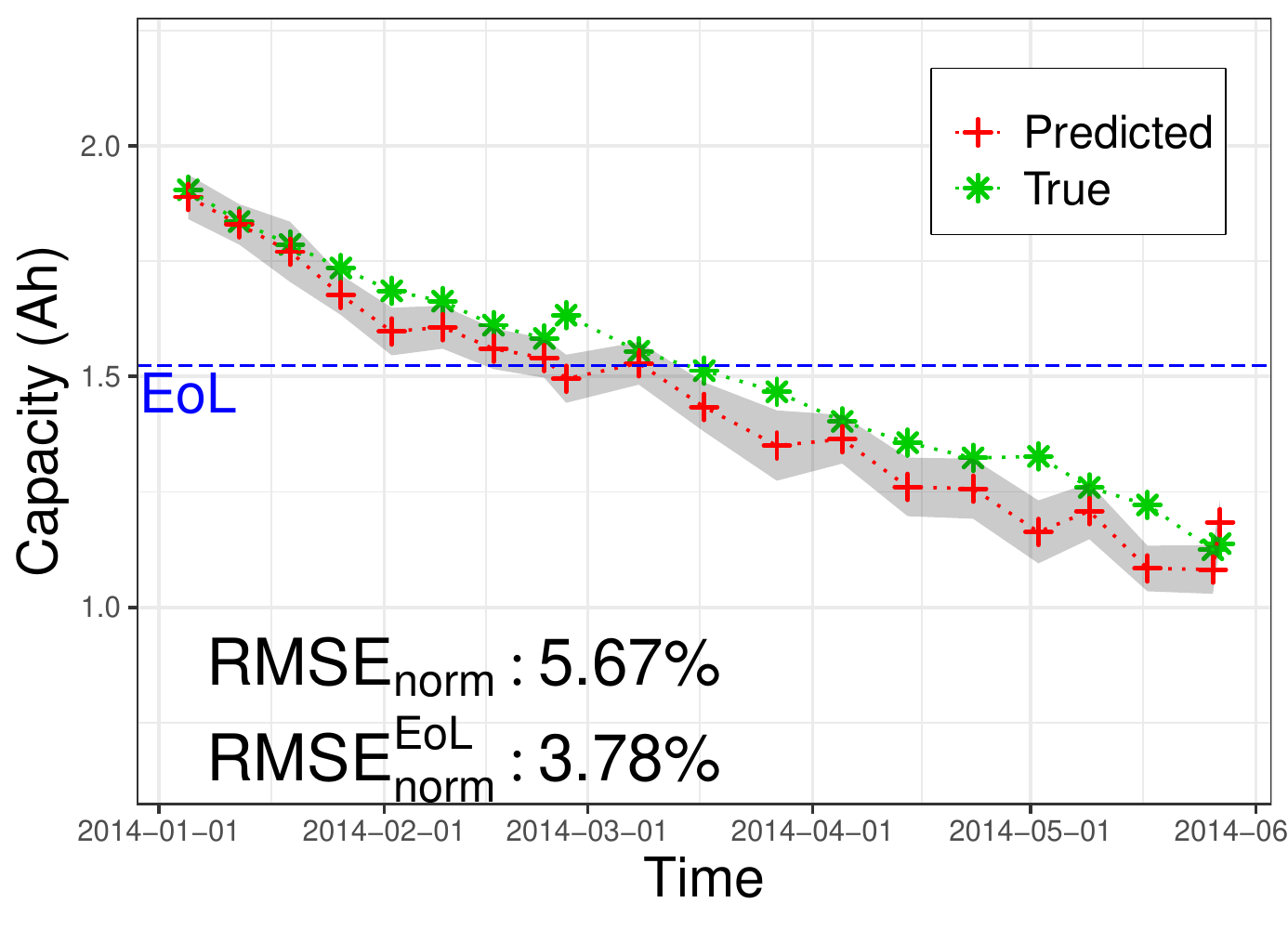}
        \hfill\includegraphics[width=0.315\textwidth, valign=c]{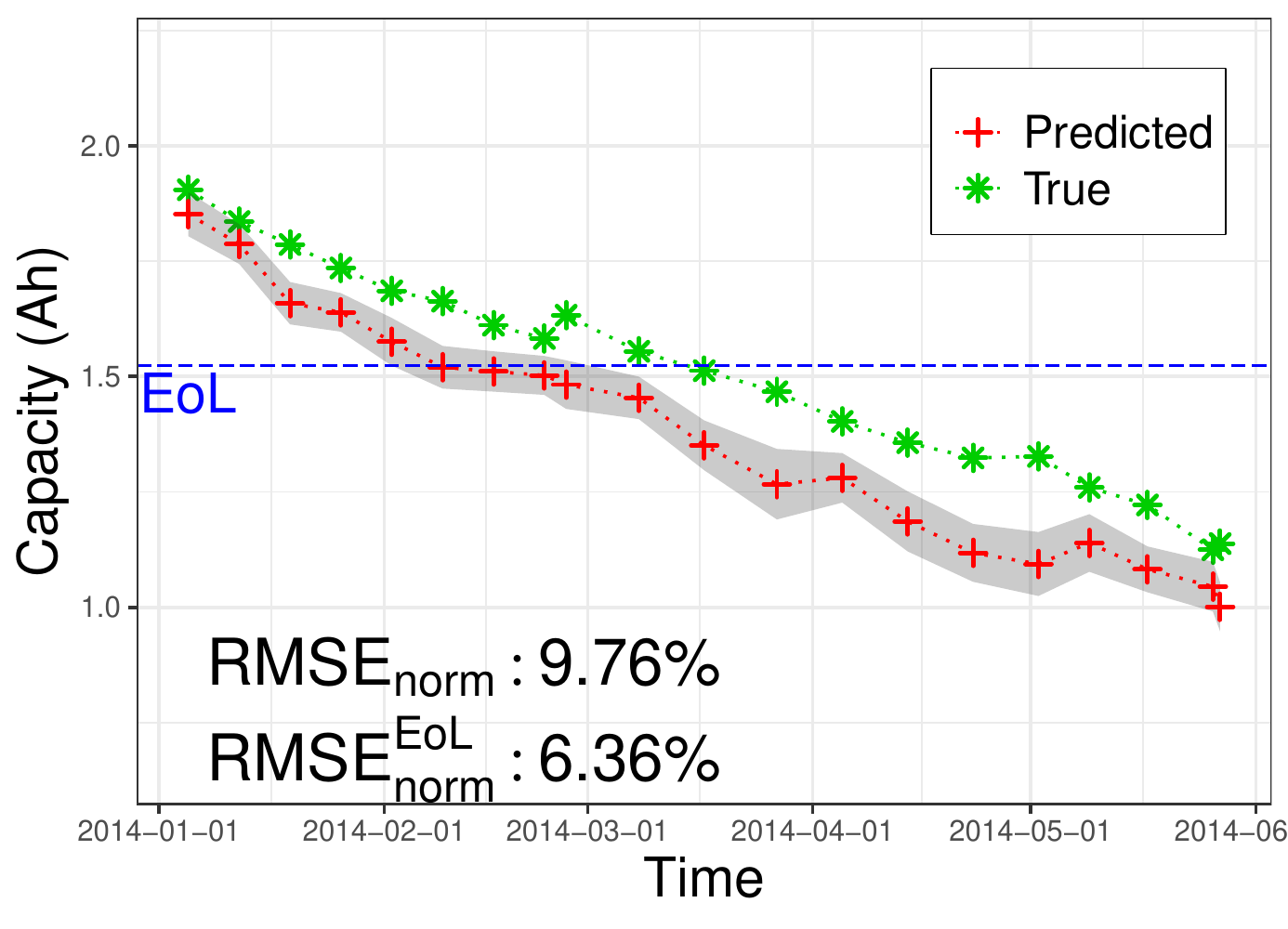}
        \hfill\includegraphics[width=0.315\textwidth, valign=c]{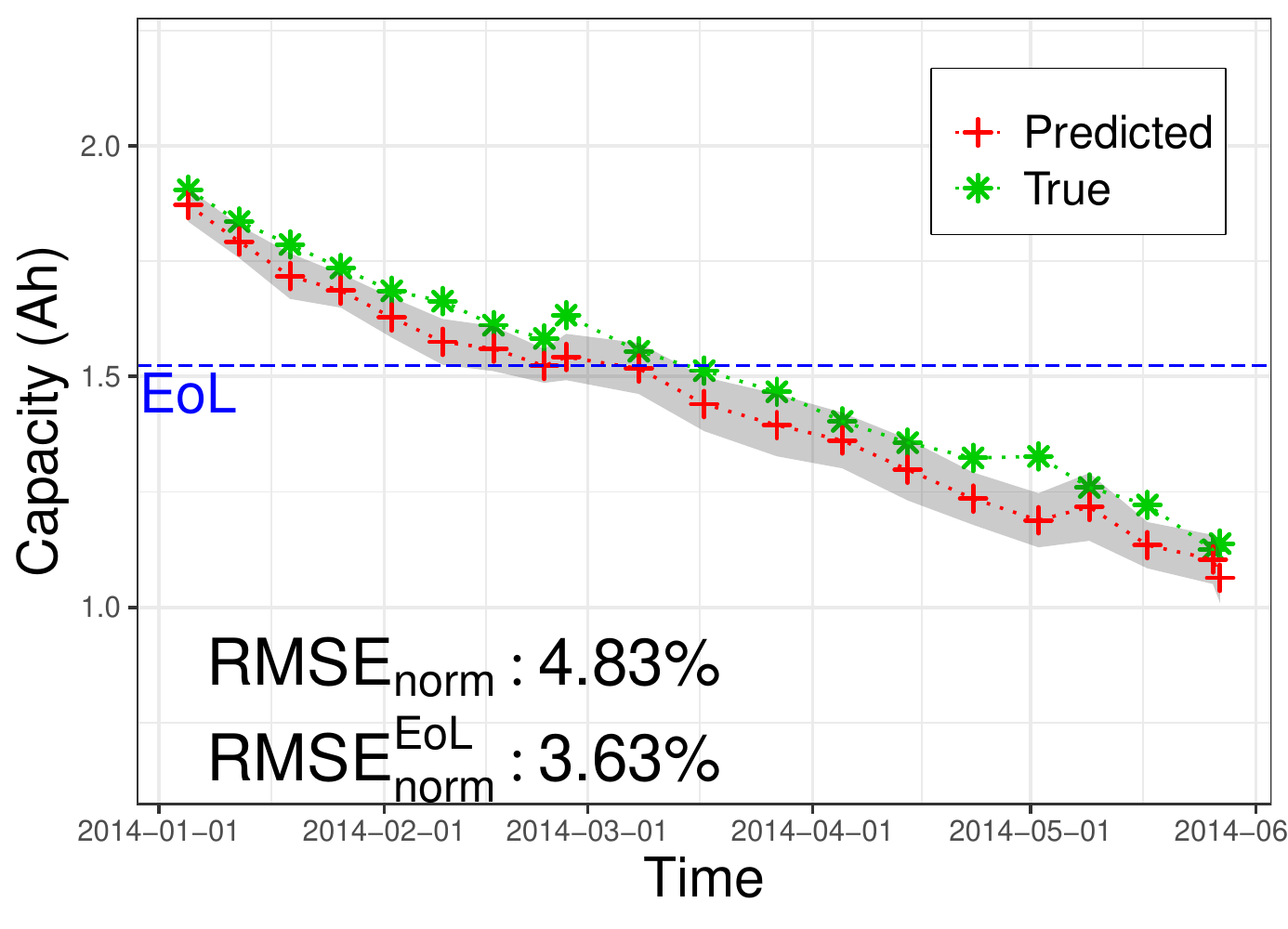}
        
\caption{Capacity fade prediction results for the cells in group 2 using the three models MFPa (left column), MFPb (middle column), and MFPc (right column), together with the considered error metrics. Cell RW4 is used for training, while RW5 is used for test. The range in the prediction errors is 4.83\% - 9.76\% for  RMSE$_{\tiny \mbox{norm}}$  and 3.63\%-6.36\% for RMSE$^{\tiny \mbox{EoL}}_{\tiny\mbox{norm}}$.}   
\label{results_group2}
\end{figure}

\begin{figure}[H]
\centering
        \textbf{\hspace{.6cm} MFPa \hspace{3.5cm} MFPb \hspace{3.5cm} MFPc}
        
        {\rotatebox{90}{\textbf{RW19}}} \hspace{0cm} \hfill\includegraphics[width=0.315\textwidth, valign=c]{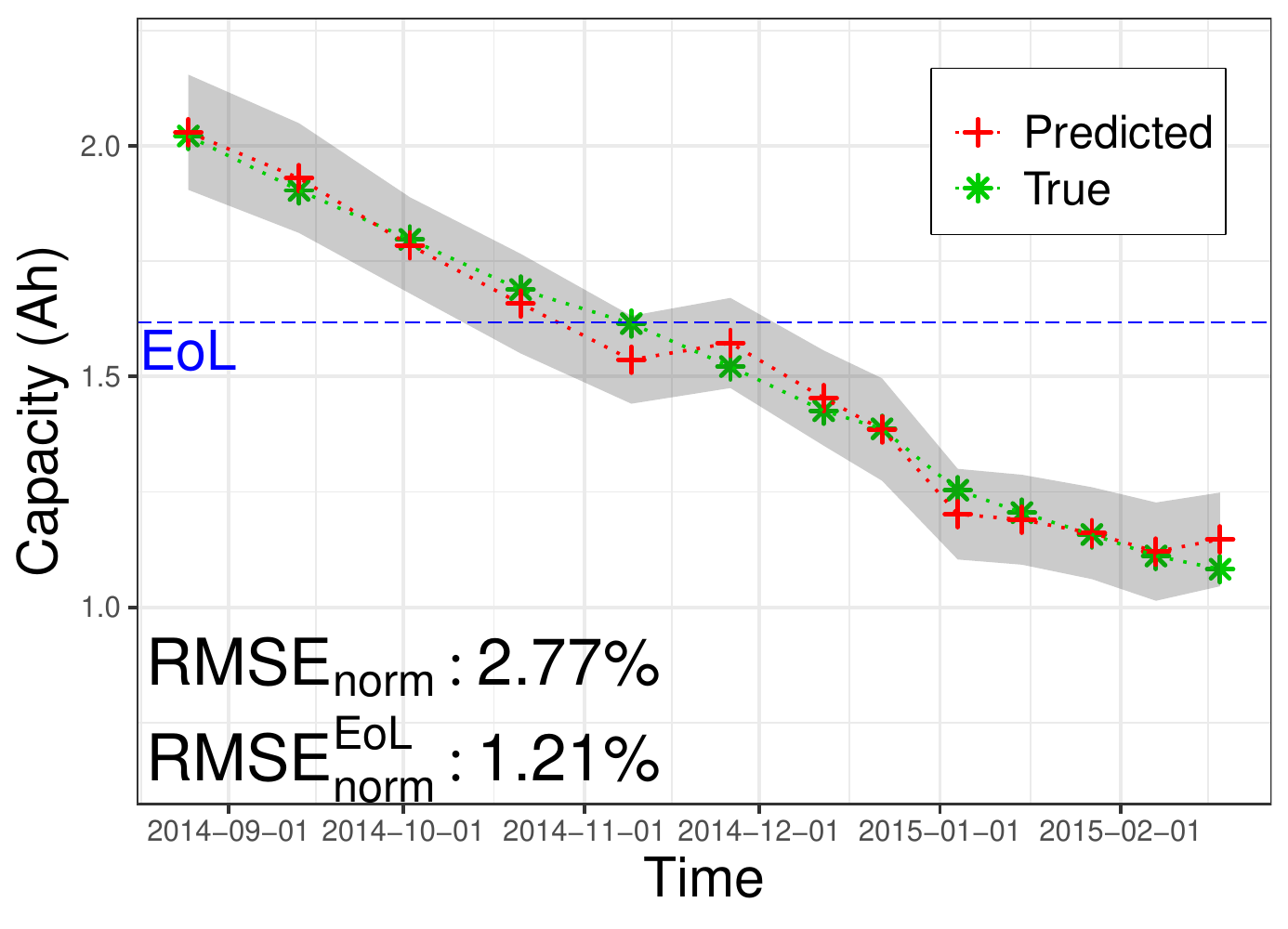}
        \hfill\includegraphics[width=0.315\textwidth, valign=c]{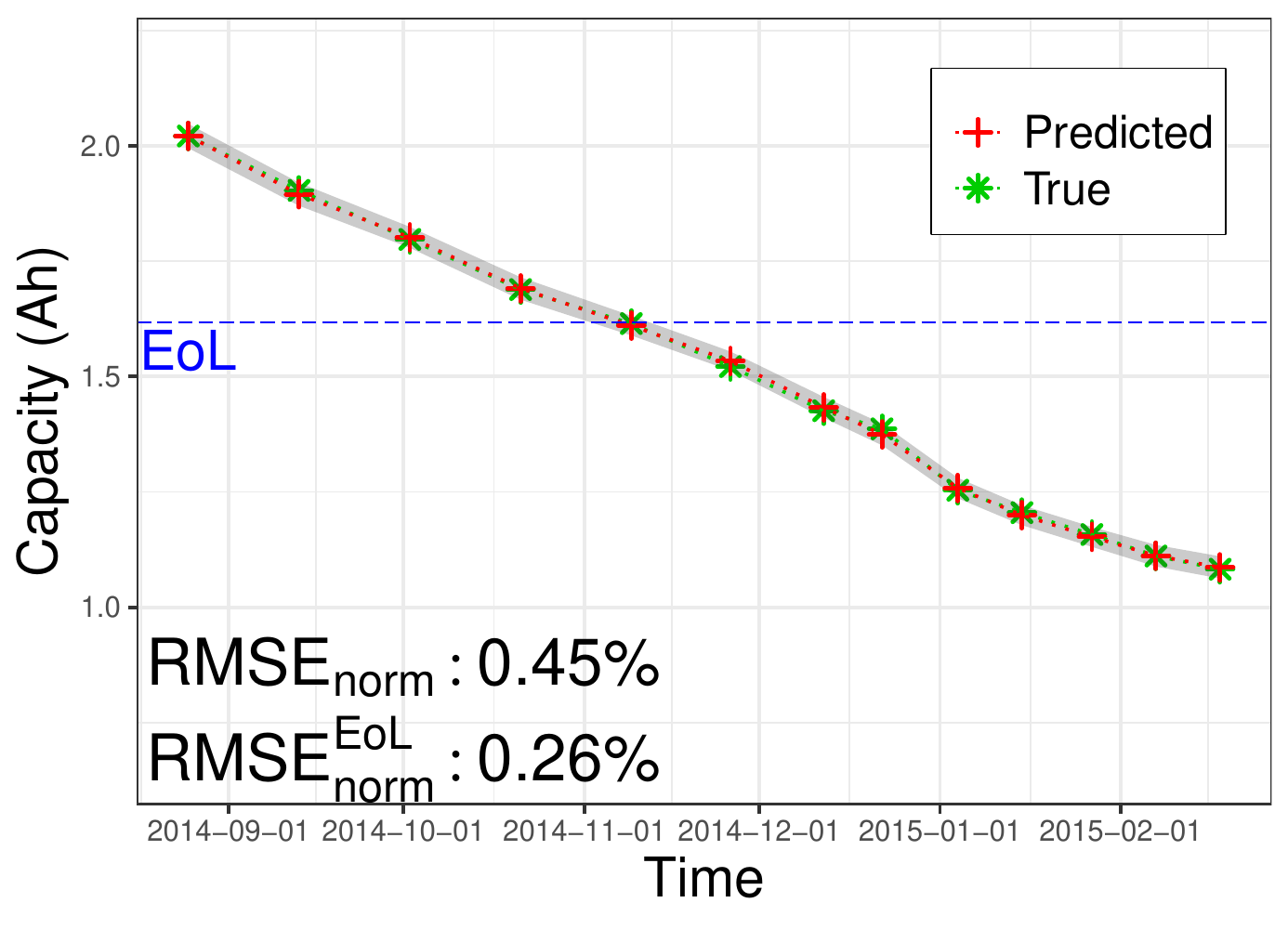}
        \hfill\includegraphics[width=0.315\textwidth, valign=c]{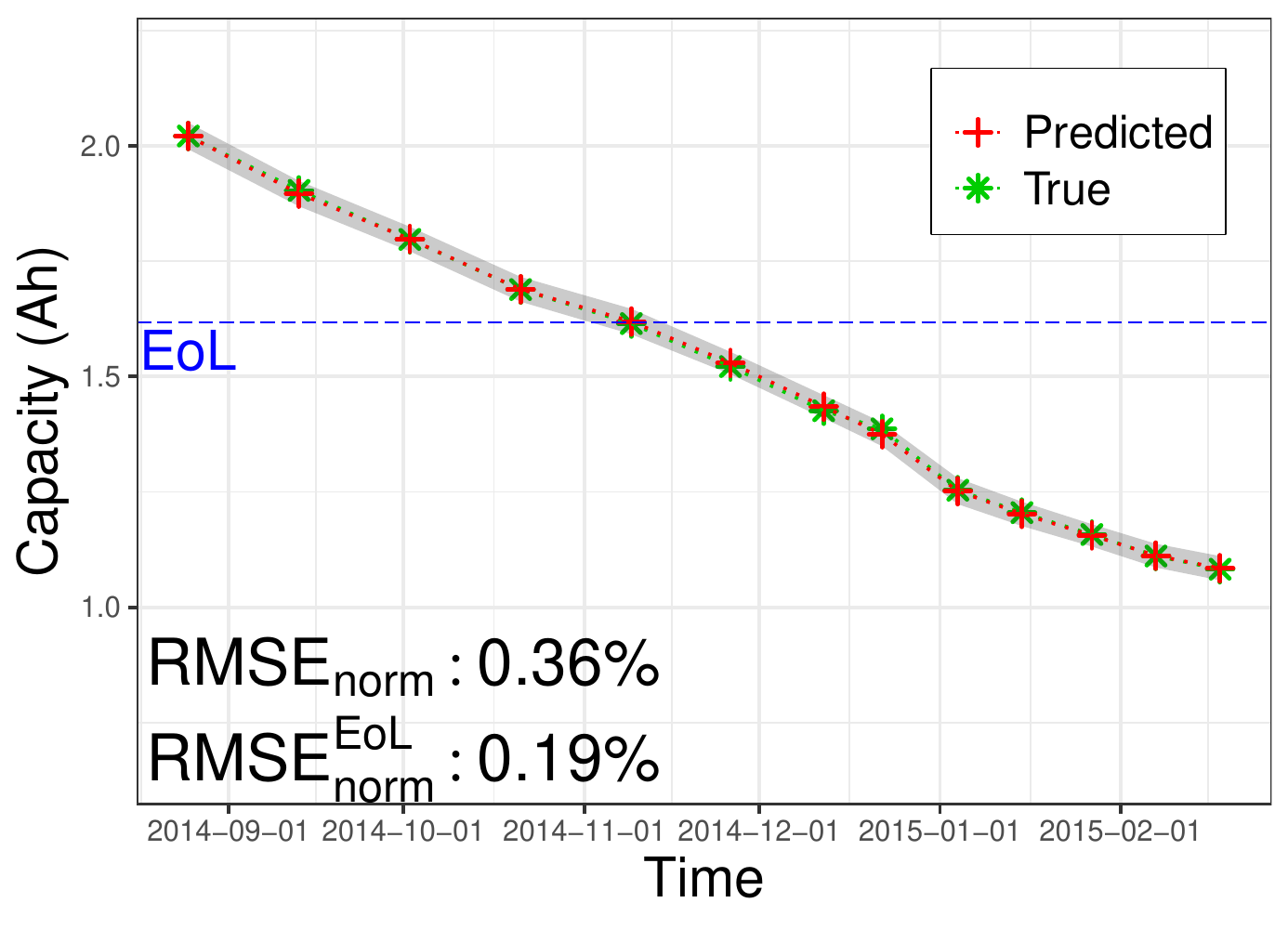}
        
        {\rotatebox{90}{\textbf{RW20}}} \hspace{0cm} \hfill\includegraphics[width=0.315\textwidth, valign=c]{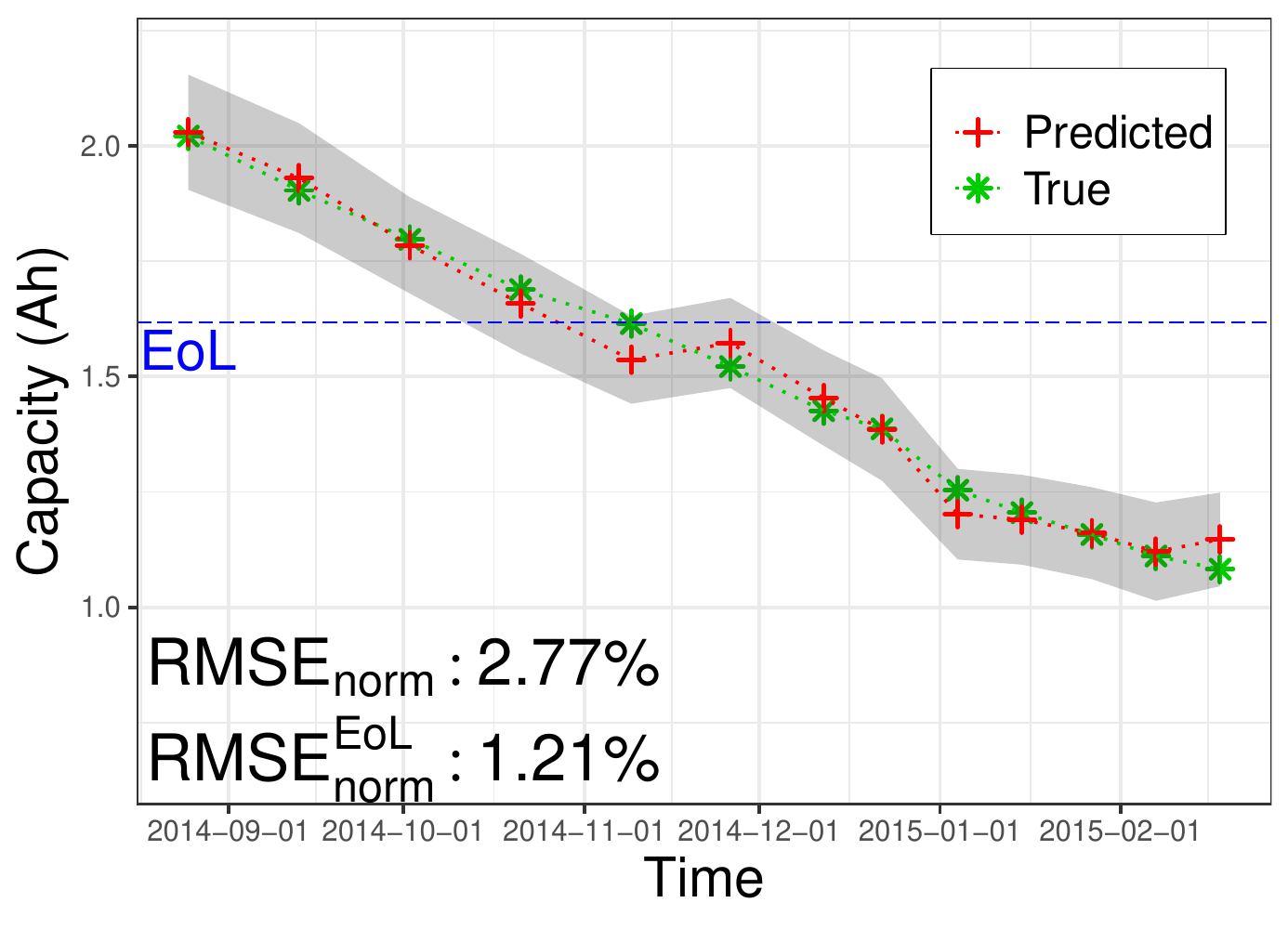}
        \hfill\includegraphics[width=0.315\textwidth, valign=c]{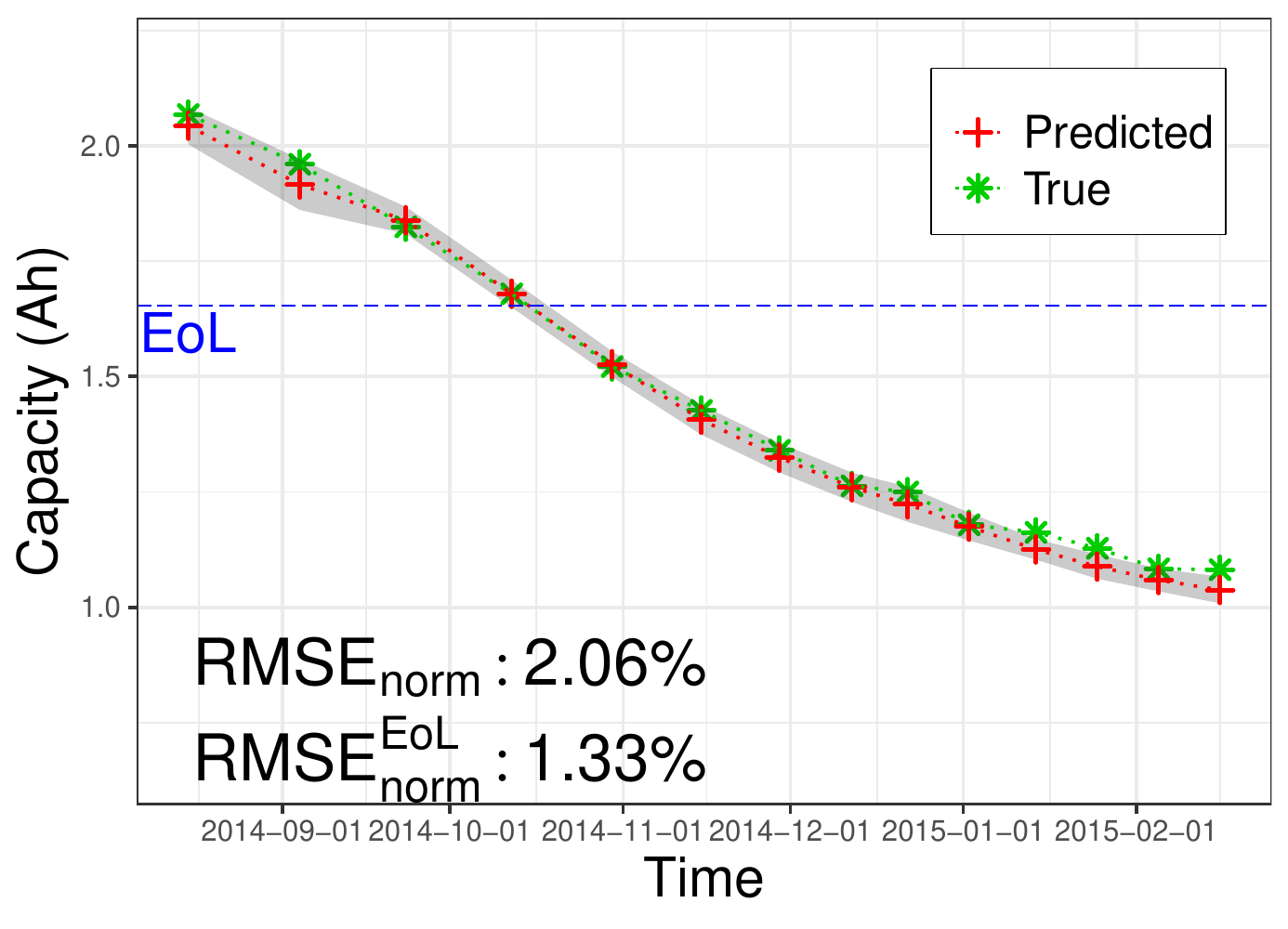}
        \hfill\includegraphics[width=0.315\textwidth, valign=c]{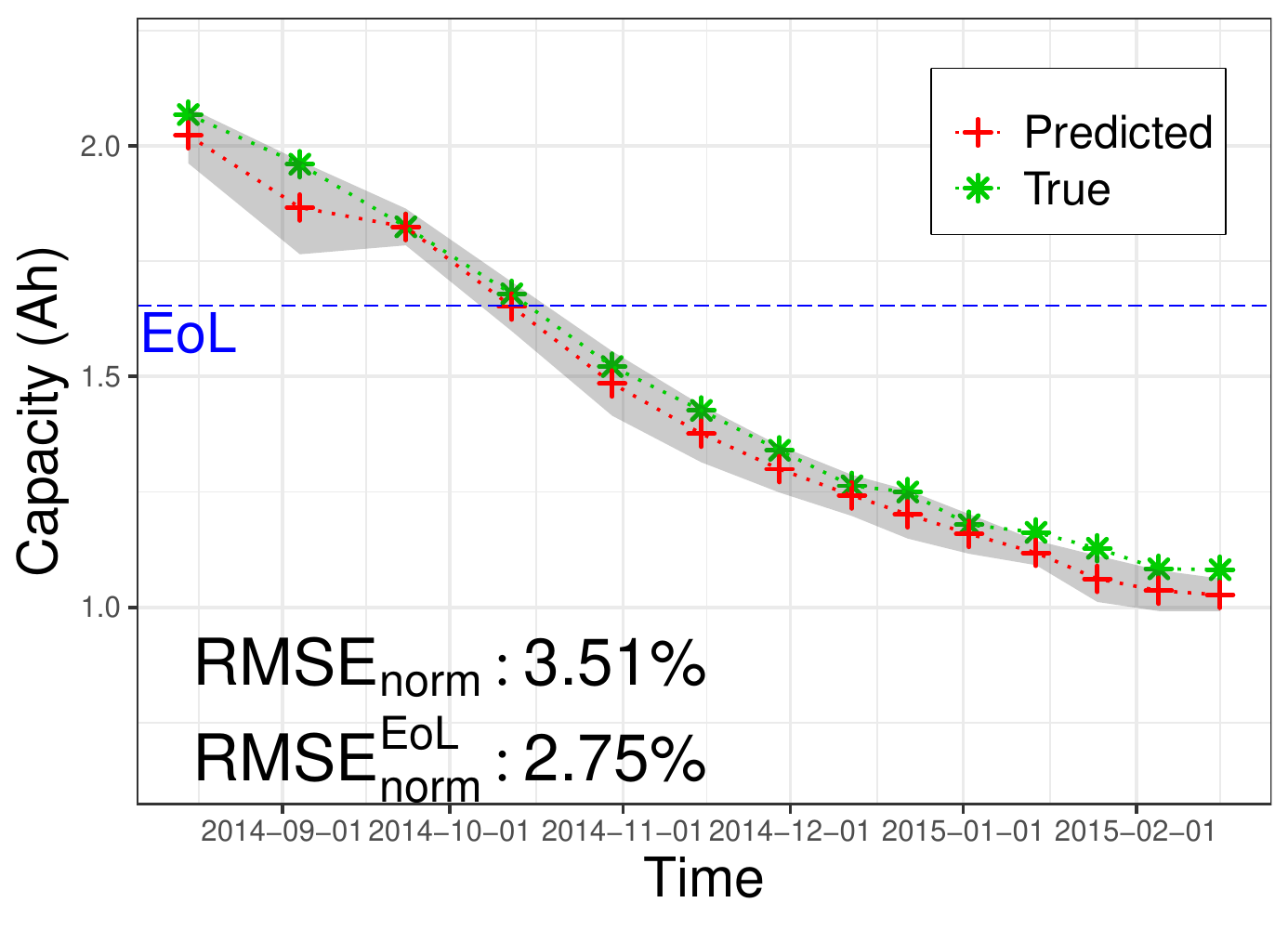}
        
\caption{Capacity fade prediction results for the cells in group 4 using the three models MFPa (left column), MFPb (middle column), and MFPc (right column), together with the considered error metrics. Cell RW19 is used for training, while RW20 is used for test. The range in the prediction errors is 2.06\% - 3.51\% for  RMSE$_{\tiny \mbox{norm}}$  and 1.21\%-2.75\% for RMSE$^{\tiny \mbox{EoL}}_{\tiny\mbox{norm}}$.}   
\label{results_group4}
\end{figure}

Figures \ref{results_group1}, \ref{results_group2} and \ref{results_group4} show the predictions obtained respectively for the first, second and fourth cell groups. The first row reports the training error of each model, since the first cell degradation path in each group is used as training set. The results, with errors never above 10\% and mostly within 5\%, confirm that the models can effectively predict the capacity drop (or apparent increase) of a cell on the basis of data from one exhausted cell that has been cycled under similar conditions. Note that the features input to the models were firstly designed with respect to the cycling routine of group 3, and were not changed when modelling group 1, 2 and 4.

\section{Comparison with other methods} \label{discussion}

\subsection{Comparison with Deep Long Short-Term Memory Regression Network}

In this subsection, we compare the proposed MFP approach with a Deep Long Short-Term Memory Regression Network (D-LSTM-RN). The regression network has been trained on the battery cells of group 3, using RW9 as training set and RW10, RW11 and RW12 as test data. It is composed of: an input layer through which the features are normalised and input to the model; two hidden layers of $200$ LSTM cells each; a fully connected layer; and a regression layer that predicts the capacity fade. Background information and further details about the structure, input and implementation of the D-LSTM-RN can be found in the Supplementary Material, sections S.3 and S.4. Note that the capacity-fade prediction problem is considered as a \say{multivariate sequence to one scalar} regression problem, differently from what was done with MFP which could not handle input in the form of multivariate sequence. The outcome is also different as D-LSTM-RN enables to consider the variation in capacity compared to the last capacity prediction, whereas MFPs allowed the comparison to the known nominal capacity only,

\begin{equation}
    \Delta \, C(t_p, t_n)  \, = \, \hat{C}(t_p) - C(t_n), 
\end{equation}

\noindent where $p$ and $n$ are two consecutive reference cycles, and $\hat{C}(t_p)$ is the capacity evaluated by the D-LSTM-RN at cycle $p$.

\medskip

\begin{figure}[H]
\centering
        {\rotatebox{90}{\textbf{RW9}}}  
        \includegraphics[width=0.41\textwidth, valign=c]{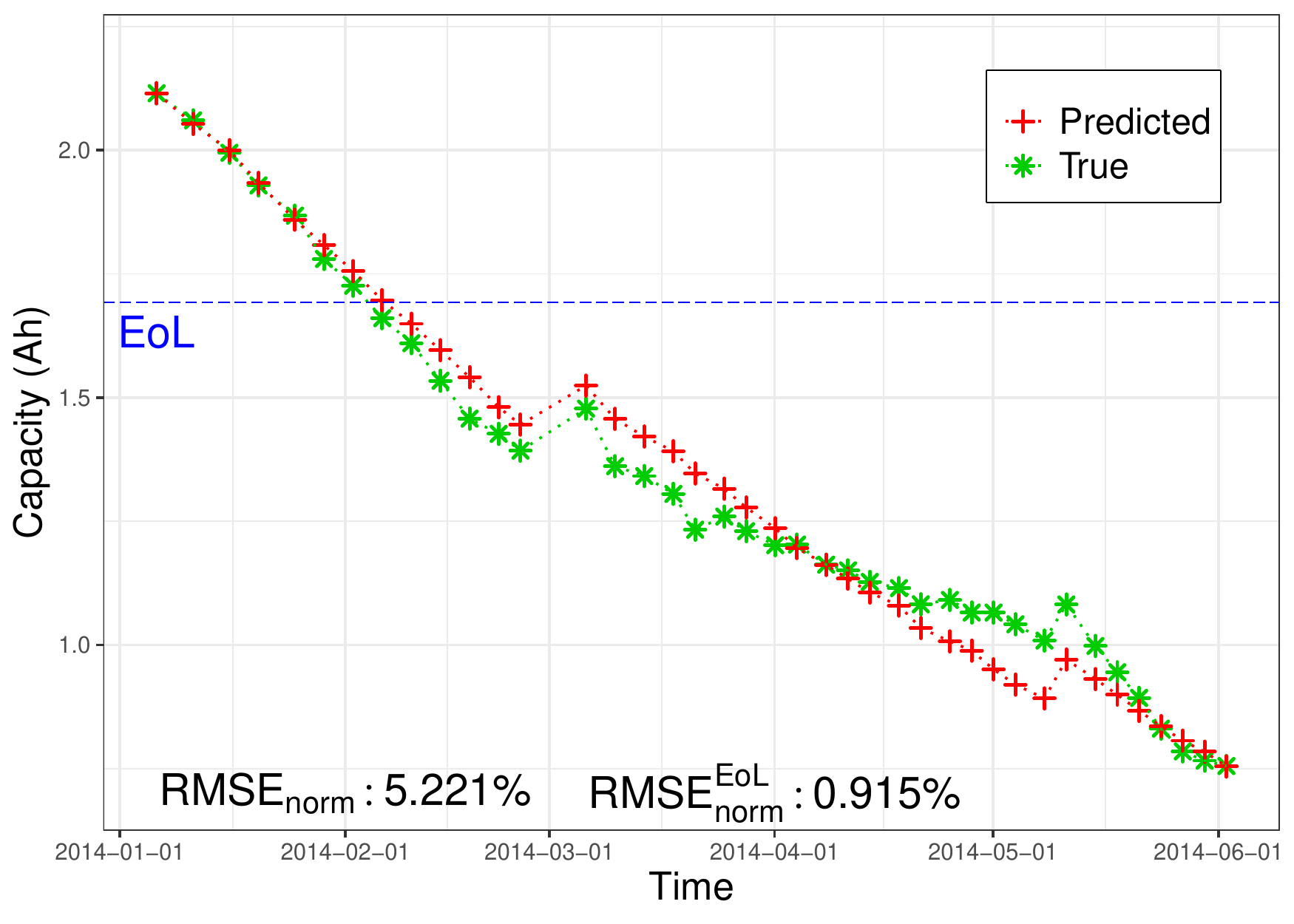}\hfill
        {\rotatebox{90}{\textbf{RW10}}}  
        \includegraphics[width=0.41\textwidth, valign=c]{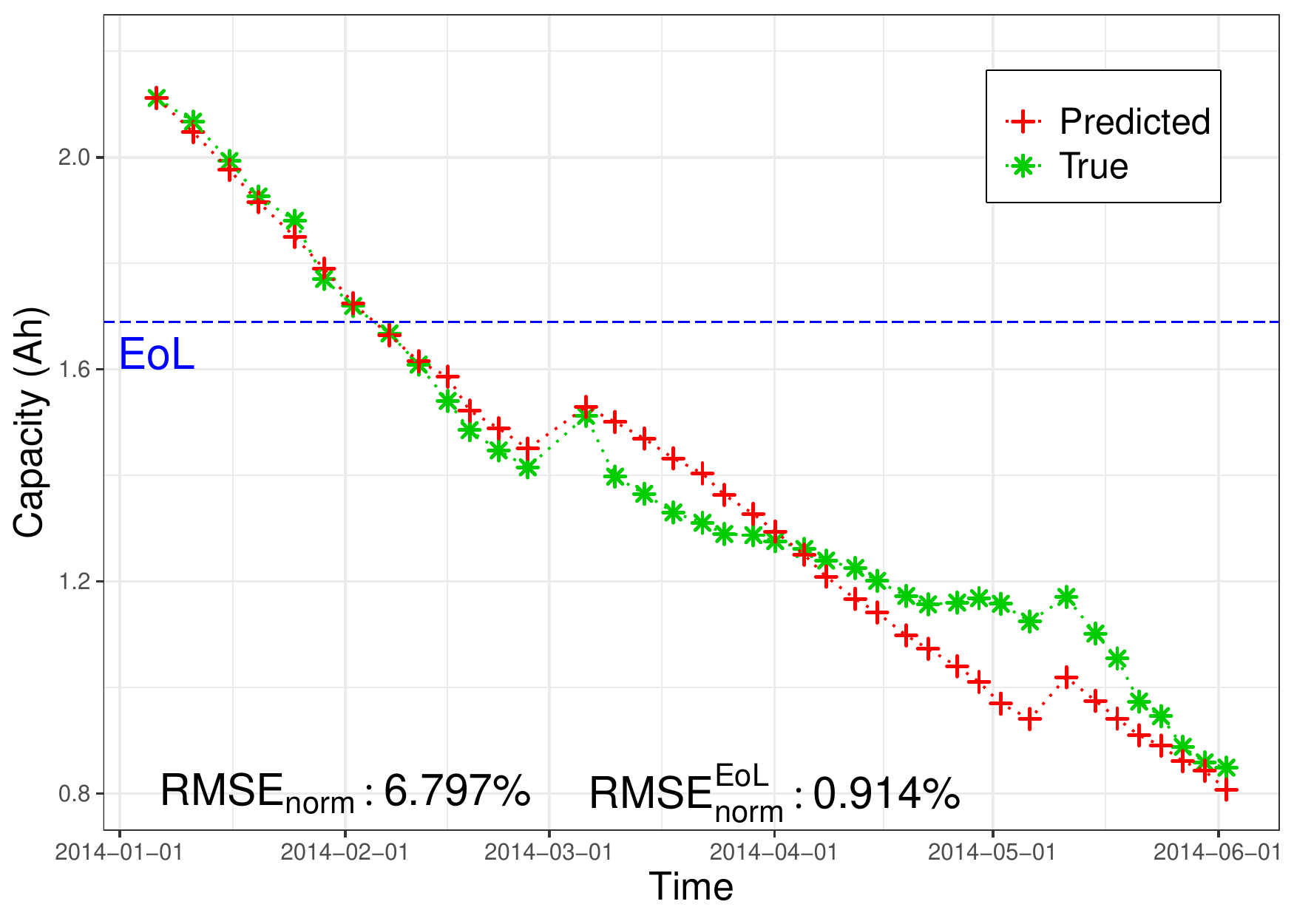}
        
        {\rotatebox{90}{\textbf{RW11}}}
        \includegraphics[width=0.41\textwidth, valign=c]{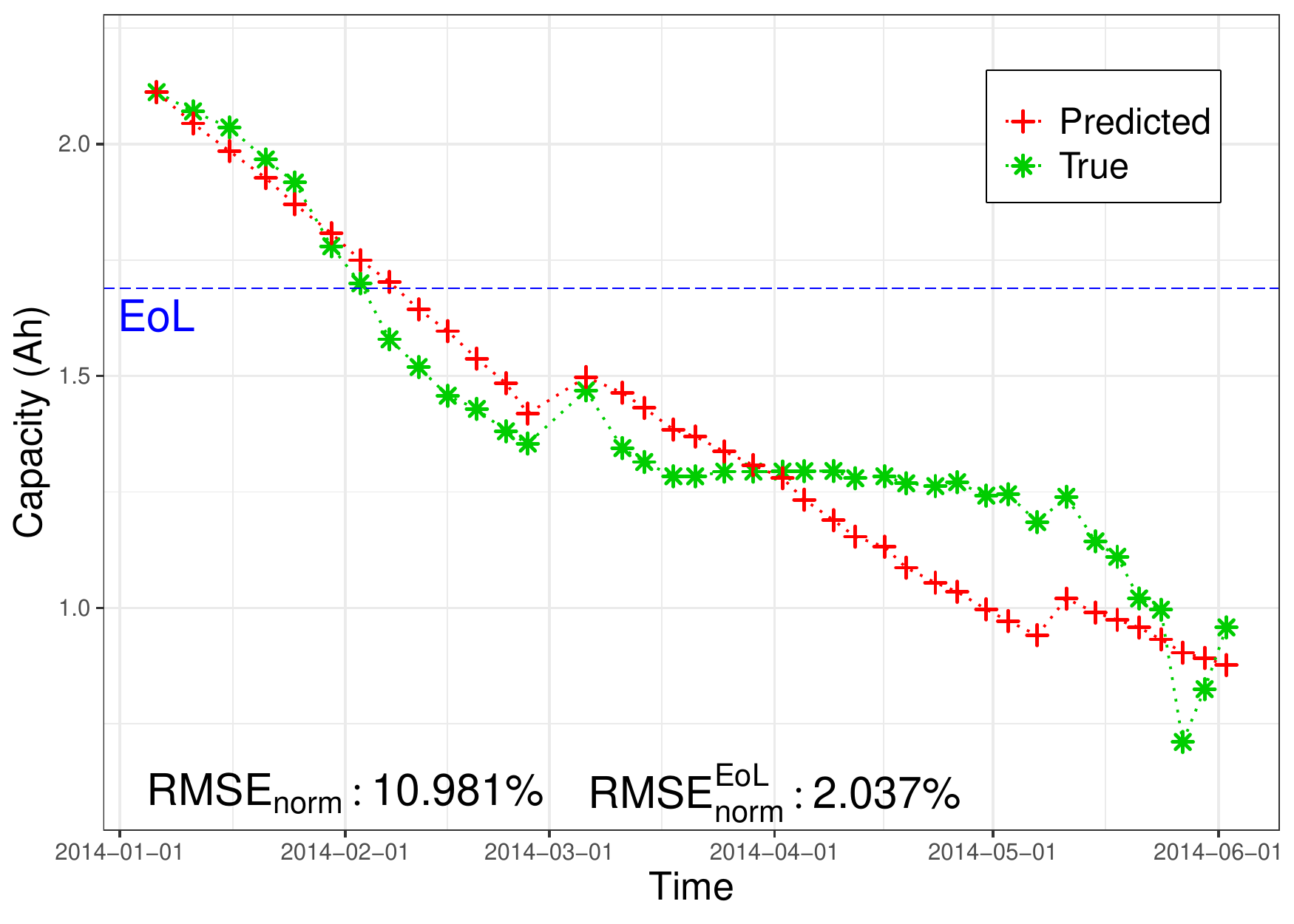}\hfill
        {\rotatebox{90}{\textbf{RW12}}}  
        \includegraphics[width=0.41\textwidth, valign=c]{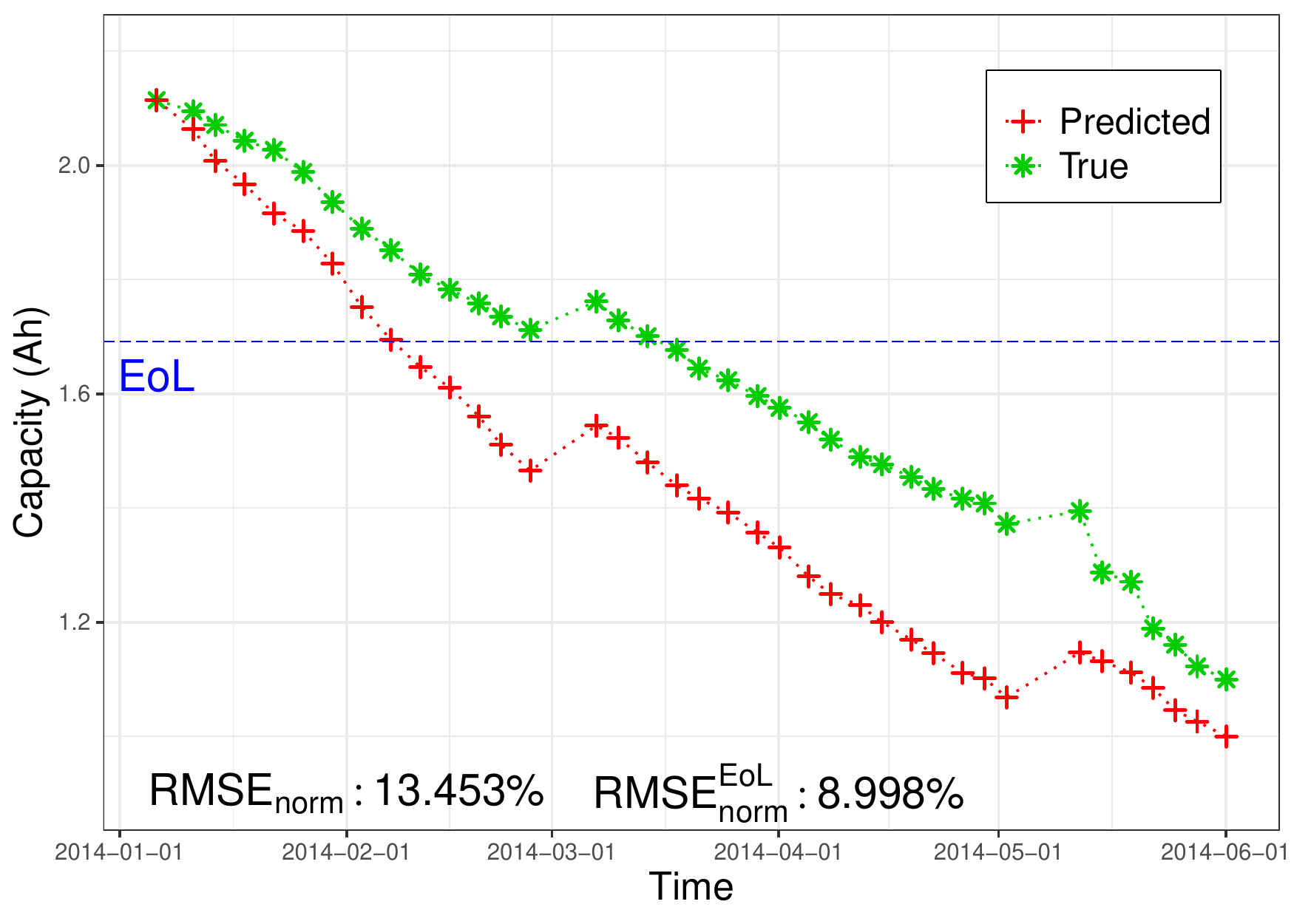}
        
\caption{Capacity fade prediction results using D-LSTM-RN, together with the considered error metrics. The prediction suffers from cumulative error, with RMSE$_{\mbox{\tiny norm}}$ ranging from 6.8$\%$ to 13.45$\%$, but RMSE$^{\tiny \mbox{EoL}}_{\tiny \mbox{norm}}$ from 0.9$\%$ to 9$\%$. Other error metrics can be found in section S.5 in the Supplementary Material.}
\label{Figure_LSTM_Results}
\end{figure}

Figure \ref{Figure_LSTM_Results} depicts the capacity fade predictions obtained with the D-LSTM-RN. It is clear that the predictions suffer from cumulative error and the normalised RMSEs tend to be slightly higher than the errors obtained with the MFP models; however, when considering the capacity fade only up to the cell EoL the prediction errors are remarkably low (except for RW12 which, as discussed before, has a different degradation profile). Nevertheless, the closeness of the prediction error achieved by the MFP models (range 2.2$\%$ - 11.7$\%$) and those of the D-LSTM-RN (range 5.2$\%$ - 13.5$\%$) implies that both methods have good performance. It is remarkable that our simple linear regression models achieve results that are comparable to those of the deep learning architecture, and they offer several key advantages that make them absolutely worthy of consideration: they have a much higher computational efficiency (for this study, the computational time was about 10 seconds for training MFP against 1 hour for D-LSTM-RN with comparable computing resources \footnote{Hardware/software configurations for MFP: Intel \textregistered $\hspace{0.05cm}$ Core\texttrademark $\hspace{0.05cm}$ i7-8550U CPU; Memory 16 G, Programming Language R 3.6.3. Hardware/software configurations for D-LSTM-RN: Intel \textregistered $\hspace{0.05cm}$ 6 Core\texttrademark $\hspace{0.05cm}$ i7 CPU; Memory 16 G, Programming Language MATLAB R2018a.}); they give the chance to interpret the results and discuss the effect of each feature, while allowing to rank the most important ones; they easily provide prediction intervals by using fundamental results from the classic asymptotic theory; they allow feature selection, to the benefit of prediction accuracy, interpretability and portability; they do not require the tuning of many hyperparameters, as opposed not only to D-LSTM-RN, but to most of the ML methods commonly used for capacity prediction of LIBs; and their much smaller number of parameters (4-6 against $ > 3\times10^5$ for D-LSTM-RN) grants a crucial practical advantage for PHM  as an MFP algorithm can be easily implemented on the BMS to monitor and optimise the battery performance.

\subsection{Comparison with contemporary works}

Table \ref{contemporary_studies_table} lists the results of the most recently published methodologies chosen to model the capacity curve for the NASA Randomized dataset. A one-to-one comparison would not take into consideration several differences across the studies, including the data-split into training and test and, more importantly, fundamental dissimilarities in the aim and methodology (estimation vs prediction, using particular cycles or segments vs the whole historical sequence): concerning the predictive objective of our study, in fact, the only work close enough is \cite{richardson2019} (though different in methodology), that reports much higher errors than ours with a RMSE range from 2.48\% to 21.95\%. The models proposed by \cite{venugopal} have very low prediction errors, but those results were achieved having included a transformation of the response among the inputs, and hence cannot be employed for a fair comparison.  

\medskip

\begin{table}[h]
\renewcommand\arraystretch{2}
\centering
\begin{tabular}{p{0.15\textwidth}p{0.15\textwidth}p{0.17\textwidth}p{0.15\textwidth}p{0.15\textwidth}}
\toprule
Method & Dataset & Train - test & Evaluation metric & Range \\[-.1em]
\midrule
\mmm{DCNN \\ DCNN-TL \cite{shen2020} \\ DCNN-ETL \\ }  & \mmm{10-year source \\ dataset + \\ NASA \\ first 20 cells}   & \mmm{Train: 16 cells \\ Test: 4 cells \\  \\ } & \mmm{RMSE$_{\mbox{\tiny norm}}$\\MaxE$_ {\tiny \mbox{norm}}$ \\ \\ } & \mmm{ 1.5\% - 3.68\% \\ 9.505\% - 21.778\% \\  \\ }  \\

\hline

\m{GP \cite{tagade} \\ } &  --  & --  & \m{MAE\\R$^2$} & \m{$\sim$ 0.045 - 0.090 \\ $>0.90$}  \\

\hline

\m{GP \cite{richardson2019} \\ } & \m{All cells but \\ 16 and 17 }  & \m{Train: even cells\\Test: odd cells} & \m{ RMSE \\ } & \m{ 0.070 - 0.642 \\ } \\

\hline

\m{GP \cite{richardson2017bis} \\ } & \m{Cells RW9, \\ RW10, RW11}  & \m{Train: RW9-10 \\ Test: RW11} & \m{RMSE \\ } & \m{$<$ 0.025 - $<$ 0.05  \\ } \\

\hline

GP-ICE \cite{richardson2017} & First 20 cells  & -- & RMSE$_{\mbox{\tiny norm}}$ &  2.48\%-21.95\%  \\[1ex]

\hline

IndRNN \cite{venugopal} & \mm{Cells RW9, \\ RW10, RW11, \\ RW12} & \mm{Train: RW9, \\ RW10, RW11 \\ Test: RW12} & \s{RMSE$_ {\tiny \mbox{norm}}$ \\[.5ex] MAE$_ {\tiny \mbox{norm}}$ \\[.5ex] MaxE$_{\tiny \mbox{norm}}$ \\[.5ex] RMSE$^{\tiny \mbox{EoL}}_{\tiny \mbox{norm}}$ \\[.5ex] MAE$^{\tiny \mbox{EoL}}_{\tiny \mbox{norm}}$ \\[.5ex] MaxE$^{\tiny \mbox{EoL}}_{\tiny \mbox{norm}}$ } & \mm{1.736\% - 3.015\% \\[.5ex] 1.380\% - 2.462\% \\[.5ex] 3.780\% - 6.840\% \\[.5ex] 1.337\% - 2.664\% \\[.5ex] 1.140\% - 2.255\% \\[.5ex] 2.594\% - 5.152\%}  \\

\bottomrule
\end{tabular}
\caption{Summary of the most recent publications and results for SoH/capacity analysis with the NASA Randomized Battery Usage Data Set.}
\label{contemporary_studies_table}
\end{table}

\newpage

\noindent Table \ref{contemporary_studies_table} gives an overview of general performances in contemporary works, and an indirect comparison shows the good performance of our models, which are competitive to most recent studies. A detailed report of the most commonly considered error metrics obtained in our analysis can be found in Table \ref{errors_supplementary} in the Supplementary Material (section S.5).

\section{Conclusions} \label{conclusion}

This paper has proposed a simple Multivariable Fractional Polynomial regression to predict the SoH of lithium-ion batteries under randomised load conditions recreated in the NASA Ames Prognostics Center of Excellence. The State of Health capacity-related definition has been considered, and an adjustment to the capacity values obtained by integration of the discharge current has been introduced to account for the battery transient effects. The degradation modelling is based on historical data from LIBs and rely on sensor data solely. It is shown that the degradation behaviour of the batteries under examination is influenced by their historical data, as supported by the low prediction errors achieved, in a multi-factor perspective which allows to study the impact of different factor combinations. We have shown that the use of simple statistical models which linearly include suitably-transformed inputs has competitive performance, while offering additional advantages such as: the opportunity to easily select the most significant inputs, and the interpretability of their coefficients; the capability of producing prediction intervals around the outcome; the absence of large numbers of hyperparameters; the great computational efficiency and consequent implementability on the BMS, which enables the adjustment of operating conditions for potential increase of longevity and safety. Machine learning methods such as neural networks have become the dominant data-driven methodology in modelling the degradation of lithium-ion batteries, and there is consensus on the fact that they are highly successful in a variety of applications; however, classical methods should not be overlooked just as such. It is worth mentioning that while machine learning methods are less interpretable, MFP might have difficulties in handling strong correlations. Moreover, as any data-driven methods (including machine learning techniques), it requires representative and sufficient training data to learn the relationship between features and response. Besides, as it is based on the whole operating history of the battery cells, problems might arise in case of periods of missing data. 

A shortcoming of this study might be that the considered cells are all characterised by the same chemical composition, though they underwent different cycling routines and show different degradation patterns: it should be investigated whether this could result in poor generalisation properties, therefore it could be interesting to extend the study to different cells. However, in real applications, it is often easy to find cells of the same composition treated under similar conditions: hence, training the model on one exhausted cell to predict the State of Health evolution of the similar ones remains a valuable possibility. Different sets of inputs could also have been selected, where additional care could have been posed on the correlations existing between the various stress factors. Besides, more efficient  methods could have been adopted for the capacity adjustment, as opposed to splines which can be poor in extrapolating. The next sensible step would be to extend the methods, developed on laboratory data (albeit randomised), to the more complex and challenging situations that characterise real applications.

\section*{Acknowledgments}

This research is funded by the Norwegian Research Council research-based innovation center \\ BigInsight, project no 237718.

\newpage
\section*{Supplementary Material}
\beginsupplement

\subsection*{S.1 \hspace{0.05cm} Difference between original and adjusted capacity values (group 3)}

\begin{figure}[H] 
\centering
    \includegraphics[width=0.41\linewidth, valign=c]{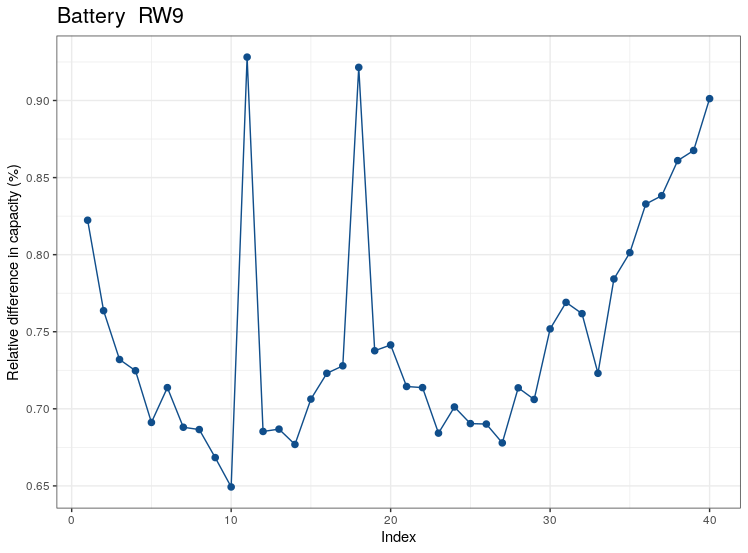} 
    \includegraphics[width=0.41\linewidth, valign=c]{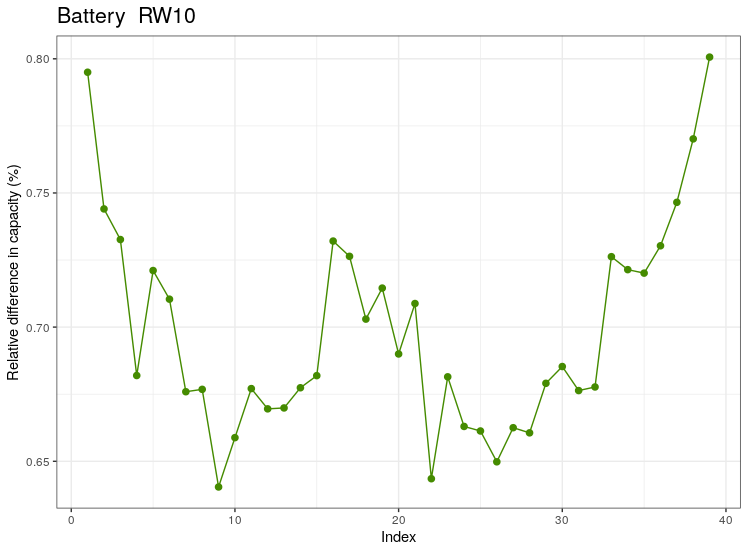} 
    \includegraphics[width=0.41\linewidth, valign=c]{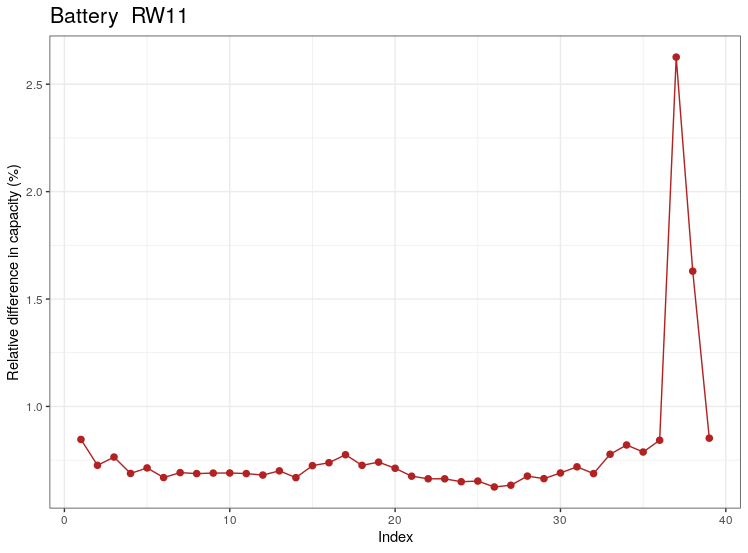} 
    \includegraphics[width=0.41\linewidth, valign=c]{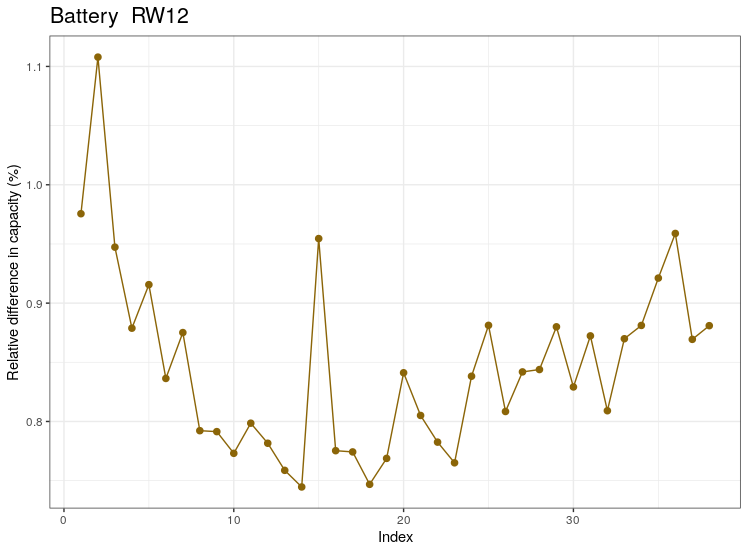} 
\caption{Relative difference between capacity values before and after introducing a correction accounting for the uncertain initial voltage of the reference cycle.}
  \label{corrected_capacity} 
\end{figure}

\bigskip

\subsection*{S.2 \hspace{0.05cm} Preliminary linear model for MFP}

The models MFPb and MFPc described in Section \ref{section_models} include the feature 

\begin{equation}
C_{\mbox{\footnotesize approx}} \, = \, \frac{1}{m} \sum_{k=1}^m (\hat{C}_{k}) \tag{S1}
\end{equation}

\noindent where $\hat{C}_{k} = I_k \cdot \hat{t}_{\mbox{\footnotesize dis},k}$ are rough capacity predictions at each RW step, obtained multiplying the current intensity $I_k$ by the discharge time $t_{\mbox{\footnotesize dis},k}$ estimated from a linear regression model:

\begin{equation} 
\begin{split}
\hat{t}_{\mbox{\footnotesize dis},k} \, = \, 
    \hat{\gamma}_0 \, &+ \, 
    \hat{\gamma}_1 \, \widebar{T}_k \, + \, 
    \hat{\gamma}_2 \, I_k \, + \, 
    \hat{\gamma}_3 \, \widebar{V}_k \, + \, 
    \hat{\gamma}_4 \, \Delta V_k \, + \, 
    \hat{\gamma}_5 \, \Delta T_k  \, + \, 
    \hat{\gamma}_6 \, t_{\mbox{\footnotesize triangle},k} \\[1ex] 
    &+ \mbox{two-way interactions} \, + \,  \mbox{three-way interactions.}
\end{split}
\label{preliminary_model} \tag{S2}
\end{equation}

\noindent In particular, $\hat{t}_{\mbox{\footnotesize dis},k}$ is an estimate of the time that we would have observed if the corresponding $k$-th RW step had been a complete reference discharge, rather than a small portion of discharge occurring under a randomly selected current load. To train the model, we used piecewise linear interpolation between each pair of capacity measurements, considering it an approximate response at each RW step, since we lack capacity measurements during the RW phase of the cycles. The inputs included in model (\ref{preliminary_model}) are, for each RW discharge step or reference discharge profile $k$:

\begin{itemize}
 \setlength\itemsep{1em}

\item[$\cdot$] $\widebar{T}_k$, the average capacity of the discharge process;

\item[$\cdot$] $I_k$, the current intensity. It constant and equal to 1 A during the reference cycles; for the RW steps, it is a constant and randomly selected value in the set 

\begin{center} \{0.75~A,~1.5~A,~2.25~A,~3~A,~3.75~A,~4.5~A\}. \end{center}

\item[$\cdot$] $\widebar{V}_k$, the average voltage of the discharge process

\item[$\cdot$] $\Delta V_k$, the signed difference between the beginning and final voltage values of the discharge process

\item[$\cdot$] $\Delta T_k = T_{\mbox{\footnotesize max}} - T_{\mbox{\footnotesize min}}$, the temperature range during the discharge process

\item[$\cdot$] $t_{\mbox{\footnotesize triangle},k} = (t_k \cdot A) / b_k$, where $t_k$ is the duration of the discharge, $A$ is the voltage drop characterising the previous reference cycle, and $b$ is the voltage drop of the current discharge.

\end{itemize}

\begin{figure}[h]
        \centering
        \includegraphics[scale=.3]{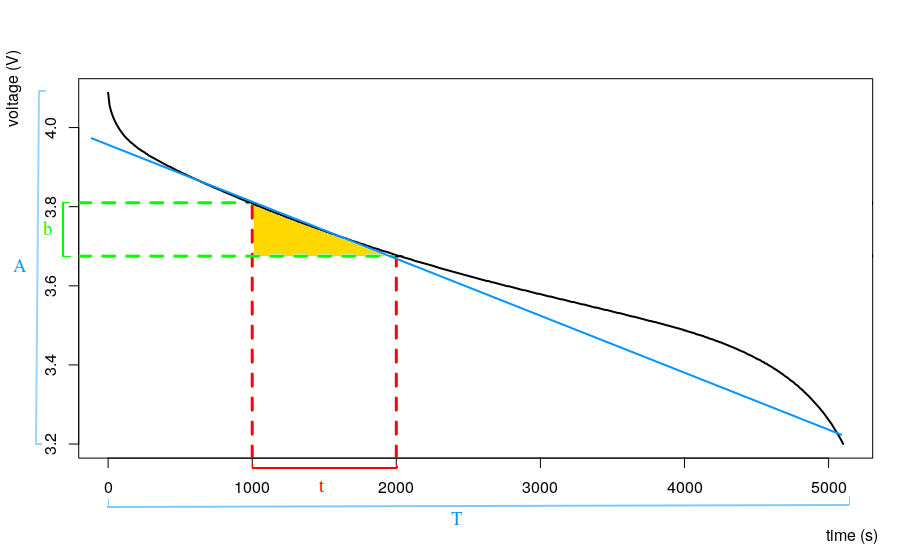}
        \caption{Illustration of the geometrical consideration behind $t_{triangle}$. Black curve: voltage curve of a reference discharge profile; light blue line: linear approximation of the voltage curve; $T$: total duration of the reference discharge; $t$: duration of a discharge RW step; $A$: voltage drop of the reference discharge; $b$: voltage drop of the RW discharge.}
        \label{Tprop}
    \end{figure}

\noindent The input $t_{\mbox{\footnotesize triangle},k}$ is motivated by a simple geometrical consideration: looking at the voltage-vs-time curve of a reference cycle (Figure \ref{Tprop}), we interpret a RW step as a portion of such a curve. Approximating the curve with a straight line, we get two similar rectangular triangles having, respectively, the duration of the reference cycle $T$ and the duration of the RW step $t$ as longer cathetus. The proportion

\begin{equation}
T : t \, = \, A:b \tag{S3} 
\end{equation}

\noindent is then valid under the approximation, and it implies $t_{\mbox{\footnotesize triangle}} = (t \cdot A) / b$. Note that, for this input to be defined and positive, the dataset was filtered in order to exclude all the RW steps having $b = \Delta V \leq 0$.

\bigskip
    
\subsection*{S.3 \hspace{0.05cm} D-LSTM-RM: background information}

Long short-term memory (LSTM) is a type of Recurrent Neural Network (RNN), i.e. a multi-layer NN.  The LSTM architecture was originally introduced by Hochreiter and Schmidhuber \cite{LSTM} with the purpose of overcoming the vanishing or exploding gradients problem \cite{gradient_problem},  by allowing constant error flow through self-connected units embedded in the LMST cell. This key feature of LSTM makes it capable of learning long-term dependencies, as opposed to vanilla-RNN.

The computational unit of a NN is the neuron, often called node or cell. The LSTM-NN has a particular neuron, called LSTM cell or memory cell, which will be explained in the following based on the description in \cite{mathWorks_LSTM} and \cite{blog_LSTM}. The state of the network at the $k$-th LSTM cell, $(\boldsymbol{c}_k, H_k)$, is composed of: the cell state $\boldsymbol{c}_k$, which encloses the learnt information up to step $k$; and the hidden state $H_k$, which is the output of the cell. The network state $(\boldsymbol{c}_k, H_k)$ is fed back as input by the cell at the next step, $k+1$, which can optionally modify the state by adding or removing information. The learnable weights of an LSTM layer are: the input weights $WI$, the recurrent weights $WR$, and the bias $B$. Inside the LSTM cell, the network state is modified through three main steps that are controlled by gates:

\begin{itemize}
 \setlength\itemsep{2em}
    \item \emph{Forget}: a \say{forget gate} decides how much information to keep from the previous cell state, through a sigmoid activation function $\sigma(x) = [1+\exp(-x)]^{-1}$,
    
    \begin{equation}
    f_{k+1} \, = \, \sigma(WI_f \, \boldsymbol{x}_{k+1} + WR_f \, H_{k} + B_f). 
    \end{equation}
    
\item \emph{Update}: an \say{input gate} decides which are the values that shall be updated, again through a sigmoid function,

    \begin{equation}
    i_{k+1} \, = \, \sigma(WI_i \, \boldsymbol{x}_{k+1} + WR_i \, H_{k} + B_i); 
    \end{equation}

\noindent in addition, another sigmoid or hyperbolic tangent layer produces new candidate values $\tilde{c}_{t+1}$ that could be used to update the cell state,

    \begin{equation}
    \tilde{c}_{k+1} \, = \, \sigma(WI_{\tilde{c}} \, \boldsymbol{x}_{k+1} + WR_{\tilde{c}} \, H_{k} + B_{\tilde{c}}) \quad \quad \mbox{or} \quad \quad  \tilde{c}_{k+1} \, = \, \tanh(WI_{\tilde{c}} \, \boldsymbol{x}_{k+1} + WR_{\tilde{c}} \, H_{k} + B_{\tilde{c}}).
    \end{equation}

\noindent The combination of these two procedures, added up to the product of the previous cell state by the forget gate, creates an update to the cell state,

\begin{equation}
    \boldsymbol{c}_{k+1} \, = \, f_{k+1} \odot \boldsymbol{c}_k + i_{k+1} \odot \tilde{c}_{k+1},
\end{equation}

\noindent where $\odot$ indicates the element-wise multiplication of vectors.

\item \emph{Output}: Lastly, the sigmoidal \say{output gate}

    \begin{equation}
    o_{k+1} \, = \, \sigma(WI_o \, \boldsymbol{x}_{k+1} + WR_o \, H_{k} + B_o)
    \end{equation}

\noindent decides how much of the information carried by the newly updated cell state ought to be added to the hidden state,

\begin{equation}
    H_{k+1} \, = \, o_{k+1} \odot \sigma(\boldsymbol{c}_{k+1}) \quad \quad \mbox{or} \quad \quad    H_{k+1} \, = \, o_{k+1} \odot \tanh(\boldsymbol{c}_{k+1}),
\end{equation}    

\noindent where a sigmoid or $\tanh$ function has been applied to the updated cell state $\boldsymbol{c}_{k+1}$. 
\end{itemize}

\noindent At the end of the process, the weights $WI$, $WR$ and the biases $B$ of a LSTM cell are concatenations of each gate's weights and biases:

\begin{equation}
WI \, = \,  \begin{bmatrix}
WI_i \\[1ex]
WI_f \\[1ex]
WI_{\tilde{c}} \\[1ex]
WI_o
\end{bmatrix}, \quad \quad 
WR \, = \,  \begin{bmatrix}
WR_i \\[1ex]
WR_f \\[1ex]
WR_{\tilde{c}} \\[1ex]
WR_o
\end{bmatrix}, \quad \quad
B \, = \,  \begin{bmatrix}
B_i \\[1ex]
B_f \\[1ex]
B_{\tilde{c}} \\[1ex]
B_o
\end{bmatrix}.
\end{equation}

\textbf{}

\subsection*{S.4 \hspace{0.05cm} D-LSTM-RM: implementation} 

A minimum and sufficient set of input features for the Deep LSTM regression network (D–LSTM–RN) accounts for all the degradation acceleration factors described in the article, using exclusively sensor data and without need of State of Charge (SoC) estimation. For each RW step $j$, we input: 

\begin{itemize}
	\item[$\cdot$] $V_j$, initial voltage at the beginning of the cycle;
	\item[$\cdot$] $\Delta V_j$, cycle signed difference between initial and final voltage values;  
	\item[$\cdot$] $\Delta t_j$, duration of the cycle;
	\item[$\cdot$] $T_{\mbox{\footnotesize min},j}$, minimum temperature during the cycle;
	\item[$\cdot$] $T_{\mbox{\footnotesize max},j}$, maximum temperature during the cycle;
	\item[$\cdot$] $\bar I_j$, mean current during the cycle.
\end{itemize}

\begin{figure}[h]
\centering
\includegraphics[width=\textwidth]{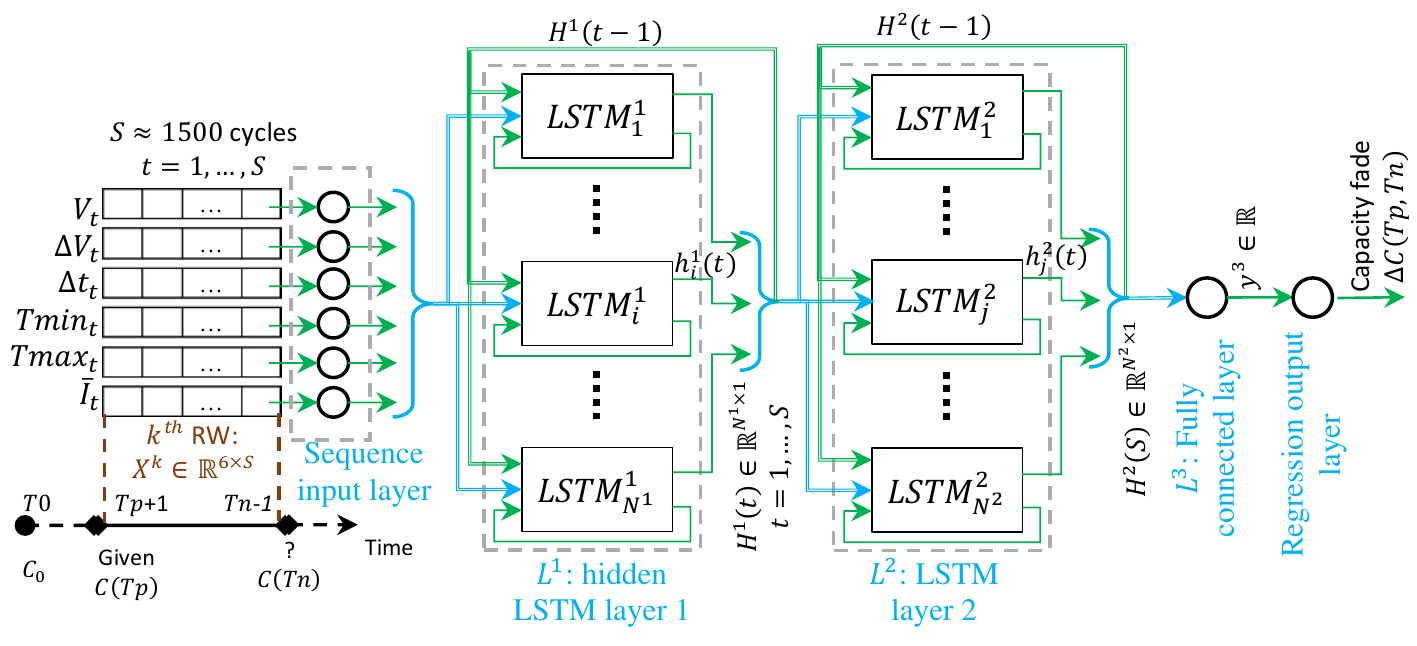}
\caption{Structure of the deep LSTM regression network with one RW example sequence of cycles.}
\label{Figure_LSTM_Structure}
\end{figure}

The structure of the deep learning model is illustrated in Figure \ref{Figure_LSTM_Structure} with one training example. The model consists of: an input layer through which the six features are normalised and the multivariate sequence of $S\approx1500$ cycles is input to the model; a hidden LSTM layer $L^1$ of $N^1=200$ LSTM cells that outputs the whole sequence of hidden states $H^1(t)$ $(t=1,\ldots,S)$; a second LSTM layer $L^2$ of $N^2=200$ LSTM cells that outputs the last hidden state $H^2(S)$; a fully connected layer $L^3$;  and a regression layer that predicts the capacity fade. 

Training the network consists of optimizing the network parameters to minimise the loss function, here the mean squared error between the true and predicted target values. The network parameters are the input weights $WI^1\in \R ^{4N^1\times6}$, $WI^2\in \R^{4N^2\times N^1}$, $WI^3\in \R^{1\times N^2}$ and biases $B^1\in \R^{4N^1\times1}$, $B^2\in \R^{4N^2\times1}$, $B^3\in \R$ for $L^1$, $L^2$, and $L^3$, respectively; and the recurrent weights $WR^1\in \R^{4N^1\times N^1}$ and $WR^2\in \R^{4N^2\times N^2}$ of LSTM layers $L^1$, and $L^2$, respectively. The model is trained using back propagation through time and stochastic gradient descent method with batch size $b=10$, gradient threshold $\tau=10$, and constant learning rate $\gamma=0.03$.

\newpage

\subsection*{S.5 \hspace{0.05cm} Supplementary prediction error results for cell group 3}

\begin{table}[H]
\renewcommand\arraystretch{2}
\setlength{\tabcolsep}{3.4pt}
\footnotesize
\begin{tabular}{llllllllll}
\toprule
Model & Test cell & RMSE & RMSE$_{\tiny \mbox{norm}}$ & MAE & MAE$_{\tiny \mbox{norm}}$ & MaxE$_ {\tiny \mbox{norm}}$ & RMSE$^{\tiny \mbox{EoL}}_{\tiny \mbox{norm}}$ & MAE$^{\tiny \mbox{EoL}}_{\tiny \mbox{norm}}$ & MaxE$^{\tiny \mbox{EoL}}_{\tiny \mbox{norm}}$ \\
\midrule

MFPa & RW9  & 0.0417 &  3.88\% & 0.0339 & 2.93\% & 12.14\% & 1.71\% & 1.61\% &  2.43\%  \\

 & RW10  & 0.0816  &  6.49\% & 0.0672 & 5.20\% & 16.31\% & 4.02\% & 3.19\% & 7.97\%   \\

 & RW11  & 0.1208  &  11.69\% & 0.0913  & 7.65\% & 49.25\% & 3.22\% & 2.57\% &  5.75\%  \\

 & RW12  & 0.0661  &  4.38\% & 0.0549  & 3.58\% & 10.24\% & 3.84\% & 3.15\% & 7.00\%   \\[3ex]

MFPb & RW9  & 0.0303 &  2.62\% & 0.0241 & 2.06\% & 6.41\% & 1.43\% & 0.97\% & 3.47\%  \\

 & RW10  & 0.0614  & 4.50\%  & 0.0487  & 3.67\% & 9.62\%  & 4.39\%  & 3.52\%  & 9.13\%   \\

 & RW11  & 0.1163  & 11.23\%  & 0.0975  & 8.23\% & 41.88\%  & 4.73\%  & 3.66\% & 9.61\%   \\

 & RW12  & 0.1392  & 9.59\%   & 0.1269  & 8.41\% & 20.45\%  & 7.18\%  & 6.10\% & 11.95\%  \\[3ex]

MFPc & RW9  & 0.0183 & 1.49\% & 0.0147 & 1.19\% & 3.52\% & 0.96\% & 0.79\% & 1.72\%  \\

 & RW10  & 0.0352  & 2.22\%  & 0.0222 & 1.60\%  & 7.38\%  & 3.37\%  & 2.38\% & 7.38\%  \\

 & RW11  & 0.0907 & 8.52\%  & 0.0795 & 6.46\% & 36.71\%  & 3.96\% & 3.26\% & 7.37\%   \\

 & RW12  & 0.1383  & 9.35\%  & 0.1281 & 8.49\% & 15.15\% & 6.94\% & 5.84\% & 12.24\%  \\[3ex]
 
 \hline \hline
 
 D-LSTM-RN & RW9  & 0.0603  & 5.22\%  & 0.0477  & 4.02\%  & 11.70\%  & 0.92\%  & 0.66\%  & 1.71\%  \\

 & RW10  & 0.0797   & 6.80\%   & 0.0612  & 5.09\%  & 16.36\%   & 0.91\%   & 0.77\%  & 1.62\%  \\

 & RW11  & 0.1303  & 10.98\%   & 0.1087  & 8.82\%  & 27.04\%   & 2.04\%  & 1.82\%  &  2.93\%  \\

 & RW12  & 0.2052   & 13.45\%   & 0.1884  & 12.18\%  & 22.14\%  & 9.00\%  & 7.93\% & 14.41\%  \\[3ex]
 
\bottomrule
\end{tabular}
\caption{Capacity fade prediction errors for each model considered in the machine learning approach and in the classical statistics approach for the batteries RW9 (training), RW10, RW11, and RW12.}
\label{errors_supplementary}
\end{table}


\newpage

\bibliographystyle{ieeetr}
\bibliography{refs}

\end{document}